\documentclass[aps,amsmath,amssymb,floatfix,twocolumn,prb]{revtex4-1}
\usepackage[]{standalone}
\usepackage{times}
\usepackage{graphicx}
\usepackage{color}
\usepackage{bm}
\usepackage{braket}
\usepackage{mathtools}
\usepackage{mathcomp}
\usepackage[caption=false,justification=raggedright,position=top,singlelinecheck=off]{subfig}
\usepackage{soul}
\usepackage[mathscr]{euscript}

\newcommand{\enni}{\noindent}
\newcommand{\enbe}{\begin{equation}}
\newcommand{\enee}{\end{equation}}

\newcommand{\en}[1]{#1}
\newcommand{\note}[1]{#1}

\begin{document}
\title{Charge transport in graphene-based mesoscopic realizations of Sachdev-Ye-Kitaev models}

\author{Oguzhan Can}
\email[Corresponding author: ]{ocan@phas.ubc.ca}

\author{Emilian M.\ Nica}

\author{Marcel Franz}
\affiliation{Department of Physics and Astronomy and Stewart Blusson Quantum Matter Institute,
University of British Columbia, Vancouver, B.C., V6T 1Z1, Canada}

\date{\today}

\begin{abstract}
We consider a recent proposal for a physical realization of the
Sachdev-Ye-Kitaev (SYK) model in the zeroth-Landau-level sector of an
irregularly-shaped graphene flake. We study in detail charge transport
signatures of the unique non-Fermi liquid state  of such a quantum dot
coupled to non-interacting leads. The properties of this setup depend essentially on the
ratio $p$ between the number of transverse modes in the lead $M$ and the
number of the fermion degrees of freedom $N$ on the SYK dot. This
ratio can be tuned via the magnetic field applied to the dot. Our
proposed setup gives access to the non-trivial conformal-invariant
regime associated with the SYK model as well as a more conventional
Fermi-liquid regime via tuning the field. The dimensionless
linear response 
conductance acquires distinct $\sqrt{p}$ and $1/\sqrt{p}$ dependencies
for the two phases respectively in the low-temperature limit, with a
universal jump at the transition. We find that corrections scale
linearly and quadratically in either temperature or frequency on the
two sides of the transition. In the weak tunneling regime we find
differential conductance proportional to the inverse square root of
the applied voltage bias $U$. This dependence is replaced by a conventional Ohmic
behavior with constant conductance proportional to $1/\sqrt{T}$ for
bias energy $eU$ smaller than temperature scale $k_BT$. 
 We also describe the out-of-equilibrium current-bias characteristics
 and discuss various crossovers between the limiting behaviors
 mentioned above.  
\end{abstract}

\maketitle

\section{Introduction}

SYK is an exactly solvable quantum mechanical model describing $N$
fermions with random all-to-all interactions.\cite{SY1996,Kitaev2015} The model is 
connected to black hole physics in AdS$_{2}$ space-time gravity
theories through holographic principle. \cite{Sachdev2015,Maldacena2016}
It exhibits a host of remarkable properties such as non-vanishing residual
entropy \cite{PhysRevLett.105.151602}  and saturating the
universal chaos bound \cite{boundonchaos} which are also properties of
quantum  black holes. SYK and its variants
\cite{Xu2016,Polchinski2016,Verbaar2016,Fu2016,Altman2016,Gu2016,Berkooz2016,Hosur2016,Liu2017,Huang2017,Balents2017,Bi2017,Lantagne2018} 
are important examples of
holographic quantum matter where non-Fermi liquid (NFL) behaviour is
observed in the presence of strong correlations and strong
disorder. In a non-Fermi liquid, elementary excitations of the system
can not be associated with non-interacting electronic excitations
through adiabatic continuity arguments. This means that the familiar
quasiparticle description fails, making theoretical considerations
difficult. Nevertheless, SYK model is special: despite the strong
correlations it can be solved in the large $N$ limit and many
observable quantities can be analytically obtained. 

The
distinct non-Fermi liquid behaviour of the SYK model remains to be
experimentally observed. Recently, various realizations of the model have been proposed (see ref. ~\onlinecite{marcelmoshereview} for a recent review.) involving ultracold atoms \cite{ultracoldrealization}, Majorana modes on the surface of a topological insulator\cite{mzmsyk}, semiconductor quantum wires attached to a quantum dot \cite{majoranawiresyk}, and finally a graphene flake in external magnetic field\cite{Chen} which will be the focus in this paper. Remarkably, this 
relatively simple setup contains all of the essential ingredients 
of the SYK model. More specifically, the low-energy sector of this 
system involves electrons in the zeroth Landau level with 
virtually no kinetic energy. For the chemical potential $\mu$ in the
zeroth Landau level, the irregular 
boundary of the flake ensures that the electronic wavefunctions acquire 
a random spatial structure. A quasi-degeneracy is maintained 
via the preserved chiral symmetry. Correspondingly, 
the Coulomb interactions projected onto the 
lowest Landau level reflect the disorder and are likewise random and
all-to-all, as required by the SYK model.  
\noindent \begin{figure}[t!]
\includegraphics[width=1\columnwidth]{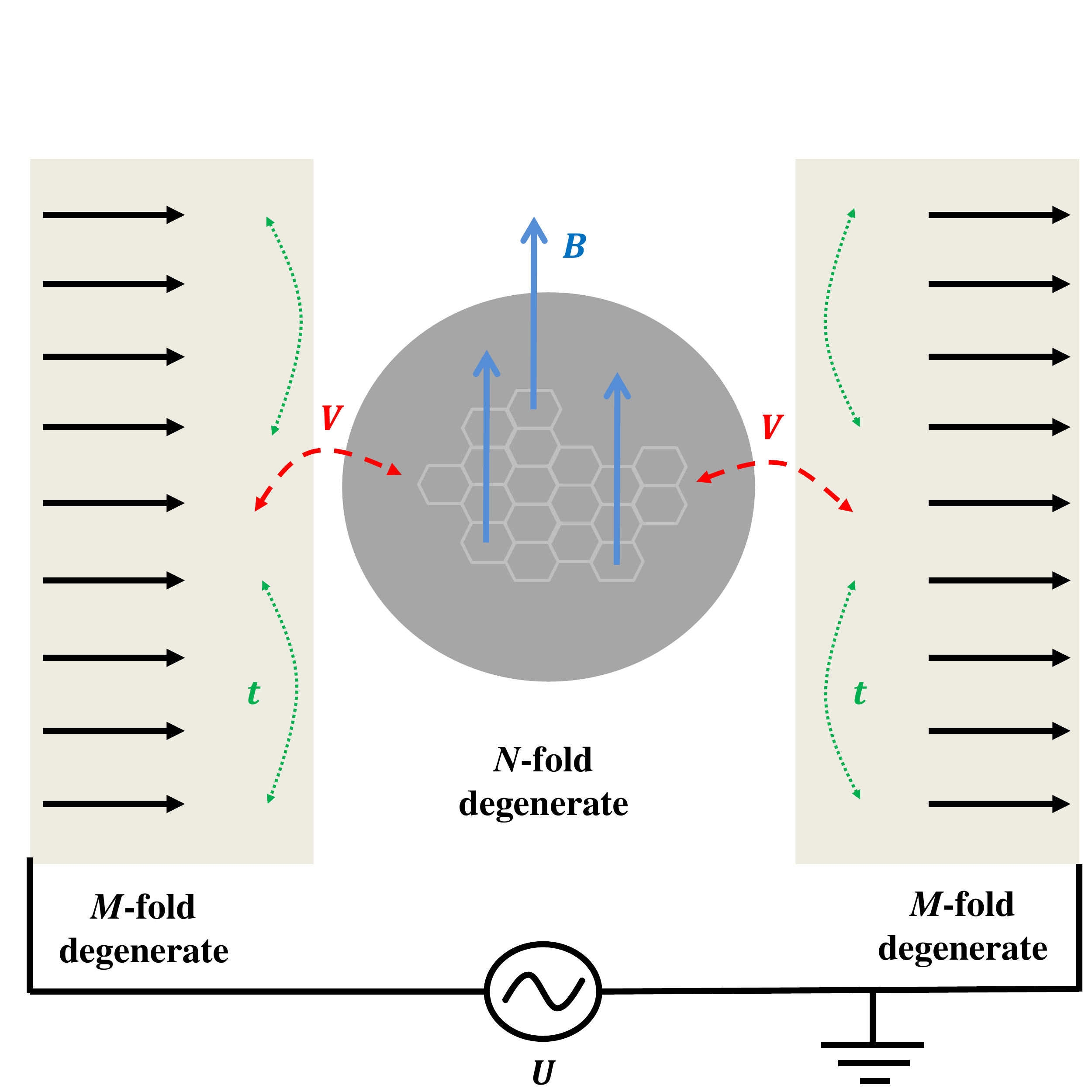}
\caption{Sketch of the proposed 
experimental setup for measurements of 
tunneling conductance of a graphene-based 
$\text{SYK}_4$ model. A graphene dot with an irregular boundary is 
placed under a perpendicular magnetic field $B$. The Coulomb interactions 
projected onto the zeroth Landau level of degeneracy $N$ provides 
an effective realization of an $\text{SYK}_4$ model\cite{Chen}. The dot is coupled 
to identical, quasi-one dimensional, ballistic leads each with $M$ transverse modes. 
We consider general models which also allow for the effects of disorder on leads in the vicinity of the junction $(t)$.
In addition to the random, all-to-all interactions on the dot, which 
are specific to $\text{SYK}_4$ models, we also include 
disordered, all-to-all scattering between dot and lead end points $V$. 
The applied bias is labeled by $U$.}
\label{Fig:Setup}
\end{figure}


In this paper, we study the tunneling conductance and current-voltage characteristics of a disordered graphene-flake 
realization\cite{Chen}  of the complex-fermion version of the SYK
model~\cite{Sachdev2015} in a setup shown in Fig.~\ref{Fig:Setup}. The transport properties 
are obtained via analytical and numerical solutions in the limit 
of large degeneracy of ballistic channels in the leads and of the zeroth Landau level on the graphene flake quantum dot. Our aim is to provide 
clear signatures of the non-trivial, conformal-invariant regime
of the SYK model which can be readily observed in 
a charge transport experiment. 

\noindent \begin{figure}[t!]
\includegraphics[width=1\columnwidth]{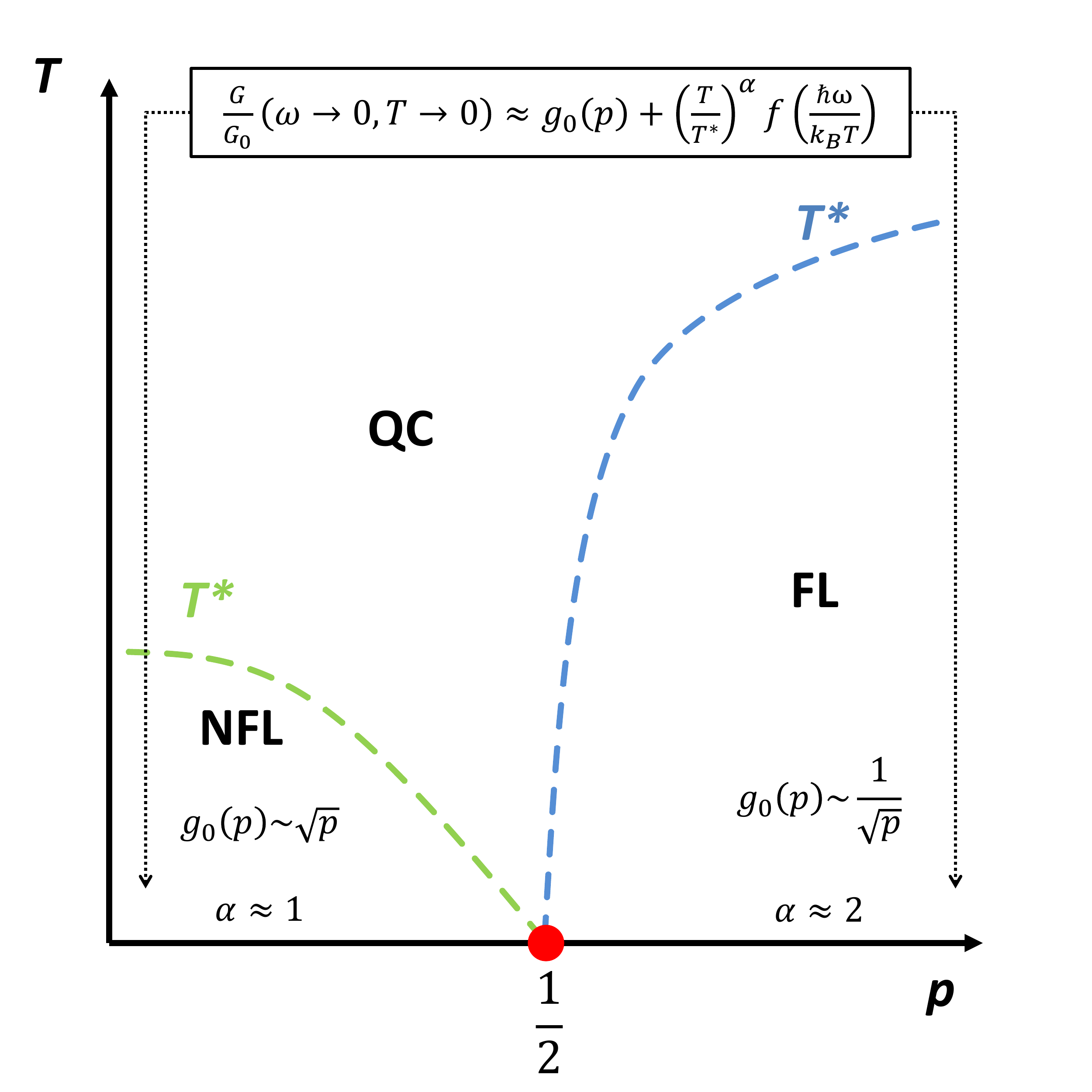}
\caption{Sketch of the expected $T-p$ phase diagram of the 
graphene dot in contact with leads corresponding 
to the setup in Fig.~\ref{Fig:Setup}. $p=M/N$ is the 
ratio between the number of transverse channels in each lead $M$ and 
the degeneracy of the zeroth Landau level on the dot $N$, as 
defined in Eqs.~\ref{Eq_N}-\ref{Eq:p}. For fixed 
$M$, $N$ and $p$ can be tuned via the applied magnetic field $B$. 
At $T=0$, a second-order quantum phase transition separates an emergent, 
conformal-invariant NFL regime from a 
more conventional FL phase, as originally discussed in Ref.~\onlinecite{Altman2016}.
We expect that both regimes survive for finite temperatures below 
 the cross-over scales labeled by $T^{*}$. The two phases 
are separated via a quantum-critical (QC) regime, which is not 
discussed in this work. }
\label{Fig:Phase_diagram}
\end{figure}
Our setup Fig.~\ref{Fig:Setup} is reminiscent of well-known
quantum-impurity systems, such as the multi-channel Kondo
model~\cite{Parcollet_Kondo}. Although the analogy is not exact, it is
natural to expect that the low-temperature properties of the junction
are essentially controlled by the ratio of the number of channels in
the leads to the effective degeneracy on the dot, $p=M/N$. While $M$
is typically fixed by the lead geometry, $N$ can be tuned in our setup
via the applied magnetic field on the dot. Therefore, our proposed setup naturally allows for quantum phase 
transitions as a function of the 
magnetic field on the graphene flake.
Our results, presented below, are in agreement with these expectations. 

Our model for the junction is very similar 
to the model introduced by Banerjee and Altman (BA)  in
Ref.~\onlinecite{Altman2016}. The BA model consists of $N$ fermions described by
the SYK$_4$ Hamiltonian  (Eq.~\ref{Eq:Model} below) coupled to $M$
`peripheral' fermions described by an SYK$_2$ model.  Here SYK$_q$
refers to an SYK model with $q$-fermion interactions. Because for
large enough $M$ the coupling
to peripheral non-interacting fermions is a relevant perturbation the BA model exhibits a 
second-order quantum phase transition at $p=1$ from an SYK-like NFL
phase at small $p$ to a Fermi liquid at large $p$. In our setup
peripheral fermions describe electrons in the leads. In analogy to the
BA results coupling to the leads can destabilize the SYK state on the
dot which makes the transport properties of the junction highly non-trivial.

Our results are summarized in Fig.~\ref{Fig:Phase_diagram}. 
For sub-critical fields ($p < p_{c}=1/2$) a phase with emergent
conformal invariance is realized on the dot well below a cross-over
scale ~\cite{Altman2016} $\hbar \omega^{*}(p) \propto k_{B}
T^{*}$. This regime is characterized by a leading
spectral density for the dot electrons which exhibits non-trivial $\omega^{-1/2}$ and $T^{-1/2}$ scaling, 
as predicted for $\text{SYK}_4$ model in the absence of the leads~\cite{Sachdev2015}. 
Following BA~\cite{Altman2016}, we refer  to this as the non-Fermi liquid phase. 
Upon approaching the transition we expect the cross-over scale $T^{*}$
to vanish\cite{Altman2016} as $\sqrt{p_c -p}$. For fields above the
critical value ($p > p_{c}$), the effects of the random interactions
on the graphene flake become sub-leading at low temperatures. At
frequencies and temperatures below $T^{*}(p)$, the spectral density
of the dot develops a resonance peak with corrections which scale as
$\omega^{2}$ and $T^{2}$, as in conventional Fermi-liquid (FL)
regimes. The cross-over scale $T^{*}$ is expected to decay to zero as $(p-p_c)$ as we approach the transition from this side. ~\cite{Altman2016}

In the particle-hole symmetric case at $T=0$, we find that the linear-response
dimensionless dc tunneling conductance (Eq.~\ref{Eq:Dmns_cndc} below)
has a distinct  dependence on parameter $p$:
\begin{align}
g_{0}  =  \biggl\{
\begin{array}{cc}
\pi \sqrt{p}, \ \ \  & p < p_c, \\
2/\sqrt{p}, & p> p_c.
\end{array}
\label{Eq:DC_cndc}
\end{align}
At the transition, $g_0$ undergoes a \emph{universal jump} from $\pi /
\sqrt{2}$ to $2\sqrt{2}$.  At nonzero temperatures, the sharp transition with 
increasing $p$ is broadened into 
a smooth crossover. We find corrections 
to the dimensionless conductance which scale linearly and
quadratically with either temperature or frequency on the NFL and FL
sides, respectively. Crossovers to quantum critical and
high-temperature regimes are observed with increasing temperature. 

In the weak tunneling regime on the NFL side (i.e., when the dot-lead
coupling is the smallest energy scale) we
predict the tunneling differential conductance of the form
\begin{align}
g(U)  \propto  \biggl\{
\begin{array}{cc}
1/\sqrt{U}, \ \  & eU \gg k_BT, \\
1/\sqrt{T}, \ \ & eU \ll k_BT.
\end{array}
\label{Eq:DC_tun}
\end{align}
At low temperature $T$ compared to the applied bias $eU$ the behavior is
highly non-Ohmic reflecting the divergent spectral density of the SYK
dot at low energy. At higher temperature the divergence is cut off and
a more conventional Ohmic dependence prevails albeit with a highly unusual temperature dependence.

Aside from the linear response and weak tunneling regimes we consider
also fully non-equilibrium situations with no natural small parameter
in which one can perturb. In
this case we employ the Keldysh formulation of the transport theory.
We match these results to the simple limiting cases mentioned above and obtain interesting crossover behaviors as a function of
temperature, voltage bias and lead-dot coupling. 
Throughout, we focus on the tunneling conductance deep within each of the two stable 
phases and do not address the behavior in the quantum-critical regime in great detail.

We note that Ref.~\onlinecite{Beenakker2018} discussed charge
transport in a similar setup involving an irregular graphene flake in
the presence of an applied field. The authors consider the limit of
few tunneling channels corresponding to $M \rightarrow 0$ in our
terminology. In the conformal regime for the dot, they find a
"duality" in the zero-temperature differential conductance which
scales as the square root and inverse of the square root of the bias
in the limit of small and large biases, respectively. These results
are in effect complementary to those found for our setup, which are
valid in the limit of large degeneracy of both leads and dot.    

In Section ~\ref{Sec:Setup}
we describe our model. Section ~\ref{Sec:Results}
presents our main results for the frequency and temperature-dependent 
conductance obtained 
in the linear-response regime and the $I-V$ characteristics 
for arbitrary static biases. Our conclusions are 
presented in Sec.~\ref{Sec:Conclusions}. Detailed discussions of the 
theory and calculations are available in the Appendix.

\section{Model}
\label{Sec:Setup}

We now describe the setup shown in Fig.~\ref{Fig:Setup} in greater detail. 
As previously mentioned, it consists of an irregularly-shaped graphene flake, under a perpendicular magnetic field,
in proximity to the end points of two leads each with $M$  quasi-one dimensional, ballistic modes. We consider  leads which are sufficiently 
long such that effects due to 
coupling to large reservoirs can be ignored. 
The low-energy sector of the graphene flake is described by the effectively random interactions within the zeroth Landau level manifold of degeneracy $N$.
\en{For a detailed discussion on the realization of the SYK model in the zeroth LL sector of the irregularly-shaped graphene flake, we refer the reader to Ref.~\onlinecite{Chen}.}

Due to the disorder inherent to the irregularly-shaped graphene flake, we expect 
that the matrix elements for tunneling to and from the end points of the leads are essentially random. We expect that our predictions are valid for 
systems where both the number of transverse modes $M$ and the
degeneracy of the zeroth Landau level $N$ are large which in practice
means at least of $O(10)$. 
We also assume that the filling of the system can be tuned 
via applied gate voltages. In this work, we only consider
statistically identical left and right leads, with equivalent random hoppings to the dot. 
In addition, the leads remain in
thermal equilibrium with large reservoirs. In the following, we shall refer to the graphene flake and the random disordered end points of the left and right leads as the dot and the leads for simplicity. Note that we ignore the electron spins as the external magnetic field on the dot results in large spin splitting, allowing us to consider only one spin sector\cite{Chen}.

    We first consider a setup where the lead end points in the vicinity of the junction are modeled 
by an \emph{effective local} $\text{SYK}_2$ model, implying that the
low-energy dynamics of the lead end point is dominated by disorder scattering. This amounts to ignoring the effects of coupling to the bulk of the \emph{non-interacting leads} to leading order. 
Equivalently, the neglected couplings are assumed to be marginal or irrelevant in the RG sense.
We stress that this approximation is not an essential part of our model and we 
show that the two phases and the respective conductances are essentially unchanged when the local 
disorder on the leads is neglected altogether in favor of a coupling to 
non-interacting extended leads more typical of quantum-impurity
models~\cite{Ludwig}. 

The situation described above is modeled by the BA-type
Hamiltonian~\cite{Altman2016} with two flavors of peripheral fermions corresponding to
the two leads, 
\begin{align}\label{Eq:Model}
H_I = H_D+H_{L} + H_{R}  + H_{LD} + H_{RD}.
\end{align}
The dot is described by the SYK$_4$ Hamiltonian 
\begin{align}\label{hdot}
H_{D} = \frac{1}{(2N)^{3/2}}  \sum_{ij;kl} J_{ijkl} c^{\dagger}_{i} c^{\dagger}_{j} c_{k} c_{l} - \mu \sum_{i} c^{\dagger}_{i}  c_{i},
\end{align}
where $i \in \{1, \hdots, N \}$ labels the degenerate, randomized zeroth Landau level 
states. 
\en{In the absence of any symmetry, we use the indices $i,j,k,l$ to label four distinct fermions in the zeroth LL. As discussed in Ref.~\onlinecite{Chen}, the effective vertices $J_{ij;kl}$ are computed by projecting the Coulomb interaction onto the zeroth LL sector. They result from the spatial average of the spatially-random zeroth LL wave functions of the four electrons. Consequently, they are also randomized in the second-quantized form used here.}
The antisymmetrized vertex $J_{ij;kl}= - J_{ji;kl} = -J_{ij;lk}$, obeys 
a Gaussian distribution with zero mean and $\overline{|J_{ij;kl}|^2} =
J^2$ variance. 
Based on the previous proposal  \cite{Chen} for a realization of the
$\text{SYK}_4$ model, it is estimated that $J \approx 25$ meV in this
setup. \en{We refer the reader to Ref.~\onlinecite{Chen} for an in-depth discussion of the emergence of the SYK model shown in Eq.~\eqref{hdot} in the zeroth LL sector of an irregularly-shaped graphene flake under an applied magnetic field.}

The end points of the two
leads are modeled by a pair of SYK$_2$ Hamiltonians
\begin{align}\label{hlead}
H_{a} = & \sum_{\alpha \beta} \frac{t_{a, \alpha \beta}}{M^{1/2}} \psi^{\dagger}_{a \alpha} \psi_{a \beta} + \text{h.c.} 
  -\mu \sum_{\alpha} \psi^{\dagger}_{a \alpha} \psi_{a\alpha},
\end{align}
with $a=L,R$ labeling the left and the right lead, respectively. The index $\alpha \in \{1, \hdots, M \}$ corresponds to
 transverse channels in the bulk of the lead. We 
assume that the local couplings are drawn from a Gaussian random 
distribution with zero mean and variance
$\overline{|t_{\alpha\beta}|^2} = t^2$. The $1/(2N)^{3/2}$ and
  $1/\sqrt{M}$ factors are chosen so that the Hamiltonians exhibit
  sensible scaling in the thermodynamic limit. Coupling between the
  dot and the leads is effected by  
\begin{align}\label{hdot-lead}
H_{a D} = & \sum_{i \alpha} \frac{V_{a i \alpha}}{(NM)^{1/4}} c^{\dagger}_{i} \psi_{a \alpha} + \text{h.c.}, 
\end{align}
where the tunneling matrix elements $V_{a \alpha i}$ are chosen as
random-Gaussian with $\overline{|V_{a \alpha i}|^2} = V^2$ variance. 

Except for two flavors of peripheral fermions $\psi_{a \beta}$
corresponding to two leads $H_I$ is essentially the BA Hamiltonian of
Ref.\ \onlinecite{Altman2016} and we may thus adopt results of that
work with only minimal modifications. Specifically, we will make an
extensive use of  the
expressions derived by BA for the fermion propagators in the FL and
NFL phases. These will be reviewed below as needed. Here we record
for future use the expression for the crossover temperature indicated
in Fig. ~\ref{Fig:Phase_diagram},
\begin{align}
T^*(p)  \simeq  \biggl\{
\begin{array}{cc}
(V^4/t^2 J)\sqrt{p_c-p}/p, \ \ \  & p < p_c, \\
(V^2/t)\sqrt{p(p-p_c)}, & p> p_c,
\end{array}
\label{Eq:T*}
\end{align}
derived in Ref.~\onlinecite{Altman2016} for $p$ close to $p_c$.

In a realistic experimental setup the leads will be spatially extended which we
model by connecting the lead end points to semi-infinite ballistic
chains for each transverse channel $\alpha$. This is represented by an
`extended lead' Hamiltonian $H^{\rm ext}=H_I+H_E$ with 
\begin{align}\label{hE}
H_{E} = & \sum_{ |\tilde{i}| >1, \alpha}  \left[ t_{E} \psi^{\dagger}_{\tilde{i} \alpha} \psi_{\tilde{i}+1, \alpha}  + \text{h.c.} \right] - \mu \sum_{ |\tilde{i}| >1, \alpha} \psi^{\dagger}_{\tilde{i} \alpha} \psi_{\tilde{i} \alpha} \notag \\
& + \sum_{\alpha} \left[ t_{-1, \alpha} \psi^{\dagger}_{-1 \alpha} \psi_{L \alpha} + t_{1,\alpha} \psi^{\dagger}_{1 \alpha} \psi_{R \alpha} + \text{h.c.} \right].
\end{align}
Electrons in the bulk of the leads are annihilated by $\psi_{\tilde{i} \alpha}, |\tilde{i}| \ge 1$
and are not subject to either disorder nor interactions. 
$t_{1/-1, \alpha}$ are the couplings between the bulk and end point states.
To conserve the large $M$ degeneracy, we approximate these to be independent 
of the index $\alpha$ and ignore any randomness.
Likewise, we assume that the $M$ transverse channels in the bulk are quasi-degenerate on a scale set by 
the variance of the interactions on the dot $J$. 
Throughout, we ignore any source of asymmetry between left and right leads. 
In our calculations, we set $t_{1/-1} > t = V =J/2 > 0$ unless otherwise
stated. 

In considering an effective local model for the junction, we ignore the coupling to the bulk of the leads which are given by $H_{EL/R}$. As previously mentioned and supported by numerical results in Sec.~\ref{Sec:Results}, including these terms and/or ignoring any local disorder on the lead end points does not modify our main results. 

We estimate \en{the degeneracy of the zeroth LLs} $N$ (from Ref.\onlinecite{Chen}) and \en{the number of quasi-one dimensional, ballistic modes in each lead} $M$ in our setup as 

\enni \begin{align}
N = & \frac{SB}{\Phi_{0}} \label{Eq_N} \\
M = & \frac{h G_{L}}{e^{2}},
\label{Eq_M}
\end{align}

\enni where $S$ is the area of the graphene flake, $B$ is the applied field, and 
$\Phi_{0}= hc/e$ is the quantum of flux. $M$ is related to the conductance 
of the extended ballistic leads $G_{L}$.
We also define the auxiliary quantities $p=M/N$,
$G_{0} = (e^{2}/2h) \sqrt{MN}$, and $\tilde{G}$ which can 
be estimated from 


\enni \begin{align}
p(B) = &  \frac{h G_{L} \Phi_{0}}{e^{2} S B}, \label{Eq:p} \\
G_{0}(B) = & \sqrt{ \frac{e^{2} G_{L} S B}{4h \Phi_{0}} }, \label{Eq:G0} \\
\tilde{G}(\omega, T, B) = & \frac{G(\omega, T, B)}{ G_{0}(B)} \label{Eq:Dmns_cndc},
\end{align}

\enni where $G(\omega, T, B)$ is the conductance of the junction. As previously mentioned, with $M$ is fixed, 
the ratio $p$ can be tuned via the strength of the transverse field applied 
to the dot. Note that both $G_{0}$ and $p$ are functions of 
the applied field.


\section{Charge transport}
\label{Sec:Results}

The model defined by Hamiltonian $H_I$ in Eq.~(\ref{Eq:Model}) can be
solved analytically using path integral techniques to average over
disorder in the limit of large $N$ and $M$. Specifically, closed form
expressions for fermion propagators can be obtained\cite{Altman2016} in the conformal
regime $(\omega,T)\ll J$. From these, it is possible to evaluate the
conductance of the junction in certain limits, including the linear response regime
(small bias voltage $U$) and the weak tunneling regime (small $V$). This
leads to our main results already given in Introduction as Eqs.\ (\ref{Eq:DC_cndc}) and (\ref{Eq:DC_tun}).

Away from these simple limits and outside the conformal regime
we solve the model in Eq.~(\ref{Eq:Model}) numerically using a large-$N,M$ 
saddle-point approximation and determine the real-time Green's
functions in the Keldysh basis~\cite{Balents2017} in and out of
equilibrium. In practice this amounts to numerically iterating a set of
self-consistent equations, given in Appendix B1, for the fermion
propagators and self energies.
Based on these solutions, we obtain the response to an applied bias
using a variant of the standard Meier-Wingreen formula.\cite{Meir} The numerical results are restricted to finite temperatures and are 
matched to the  analytical results in appropriate limits. 
A detailed discussion of our calculations is given in the Appendices.

\subsection{Linear response AC conductance}

\noindent \begin{figure}[t!]
\includegraphics[width=1\columnwidth]{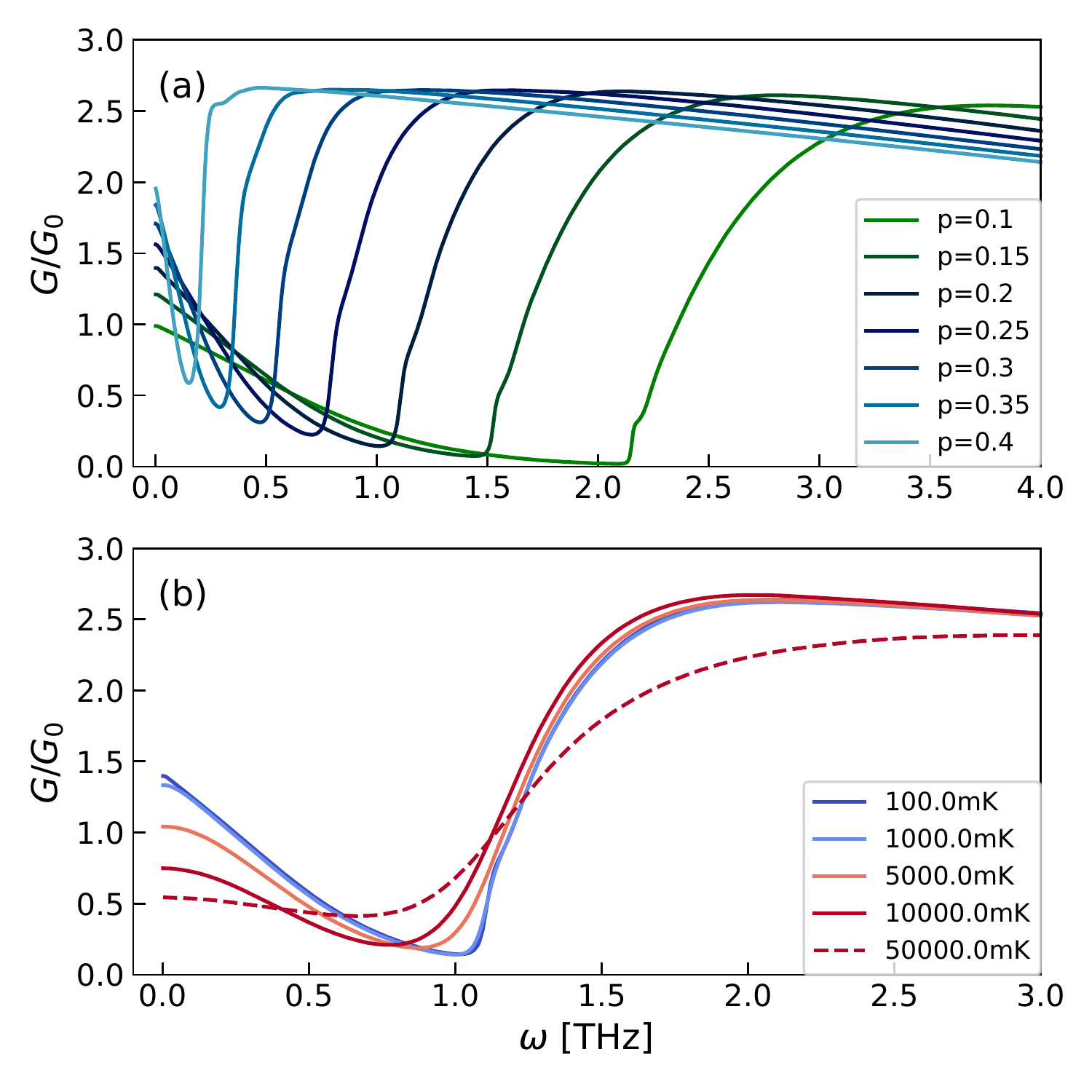}
\caption{(a) Dimensionless tunneling conductance \note{computed using Eq. \ref{LRconductanceformula} and numerical solutions of the saddle point equations } in the NFL 
regime at fixed temperature $T=$ 100 mK and increasing 
values of the tuning parameter $p$ (Eq.~\ref{Eq:p}). 
Note the cross-over from the peak at low energies, 
in the conformal-invariant NFL regime,
to the broad, featureless spectrum beyond a scale $\hbar \omega^{*} \approx k_{B} T^{*}$. 
With increasing $p$, $\omega^{*}$ decreases, while 
the peak is suppressed. The behavior is consistent with 
the approach to a second-order quantum phase transition as sketched in  Fig.~\ref{Fig:Phase_diagram}.
(b) Same as (a), at fixed $p=0.2$ and for several temperatures. 
The height of the peak at lower frequencies decreases with 
increasing temperature indicating a smooth 
cross-over away from the conformal-invariant 
NFL regime. Also note the cross-over to the high-frequency limit which  
occurs at roughly the same frequency. The dashed line corresponds to the high-temperature limit and is included for comparison. 
}
\label{Fig:Qualitative_NFL}
\end{figure}
We first discuss the tunneling conductance obtained via the Kubo formalism~\cite{Mahan} in the presence of a
small oscillating bias applied to the two leads and subsequently present our 
results for current with arbitrarily-large, static biases. 
A detailed account of our calculations is found in  Appendix~\ref{Sec:Appn_lnr}.

Based on dimensional analysis~\cite{Sachdev_trans}, we expect that the dimensionless conductance of the junction (Eq.~\eqref{Eq:Dmns_cndc})
obeys the scaling form 

 \enni \begin{align}
\frac{G}{G_{0}}
= & g \left( \frac{\hbar \omega}{k_{B} T}, \frac{T}{T^{*}}, \frac{\mu}{\mu^{*}}, p \right),
\label{Eq:Scln_NFL}
\end{align}

\enni where $g$ is a dimensionless function which depends on 
the nature of the phases on either side of the transition.
In addition, $\omega$ is the frequency of the driving bias, $T$ is the temperature, $\mu$ is the chemical potential 
common to both leads and dot, while $p \propto \frac{1}{B}$ serves as a tuning parameter. 
$T^{*}(p)$ is given in Eq.\ (\ref{Eq:T*}) and represents cross-over scales associated with the emergence of 
the NFL and FL scaling regimes. It vanishes at the critical point from either side. 
As Ref.~\onlinecite{Altman2016} pointed out, away from particle-hole 
symmetry ($\mu \neq 0$), the NFL and FL phases are separated by an incompressible 
phase for a finite range of $\mu$.  
Since the focus of our work is behavior of the conductance deep within either 
NFL and FL phases, we do not address the intermediate phases. As such, 
we associate $\mu^{*}(p)$ with a scale below and above which 
the conductance follows either NFL/FL scaling. As discussed below, 
for frequencies and temperatures well below $T^{*}$, the conductance 
shows very weak dependence on either $\omega, T$, while it exhibits 
characteristic scaling with $p(B)$ in either phases.

\noindent \begin{figure}[b!]
\includegraphics[width=1\columnwidth]{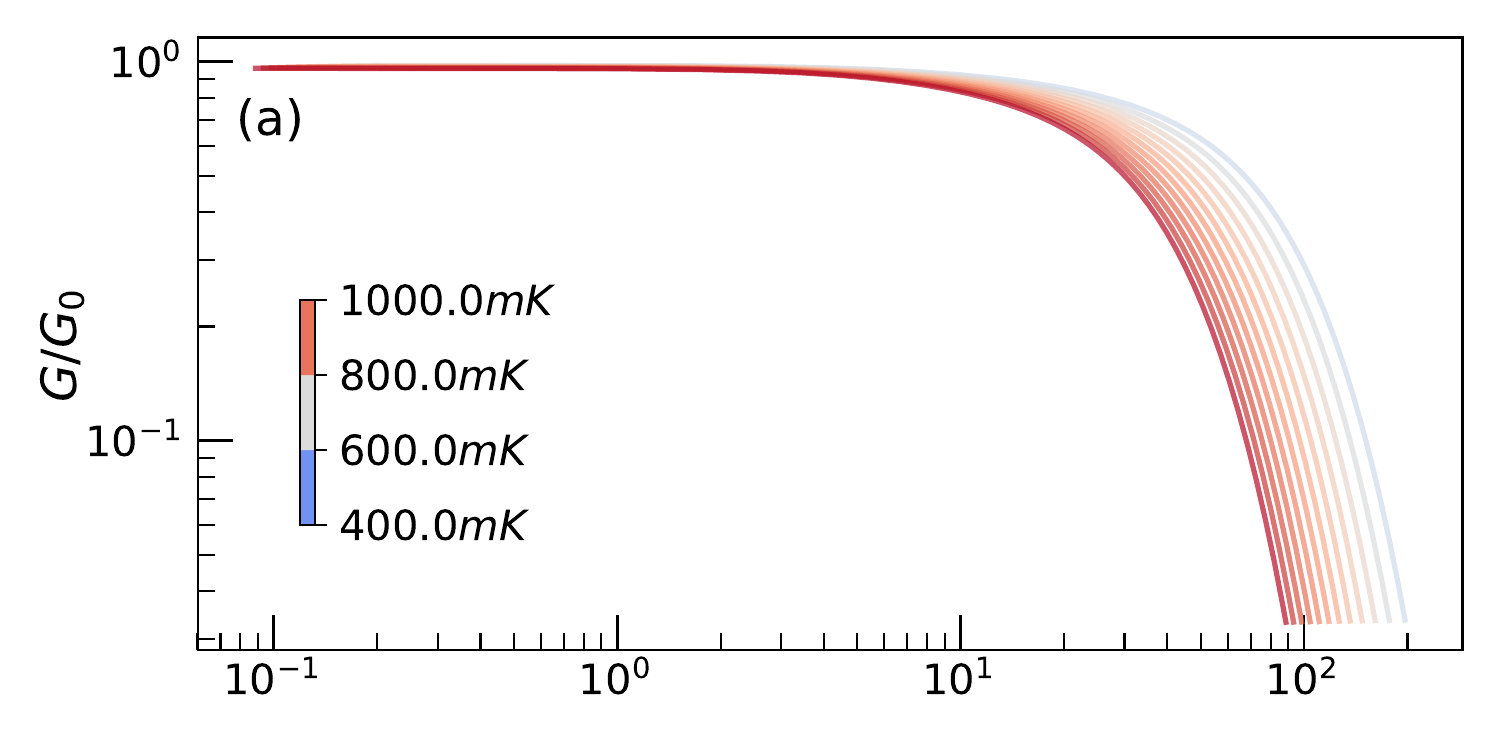}
\includegraphics[width=1\columnwidth]{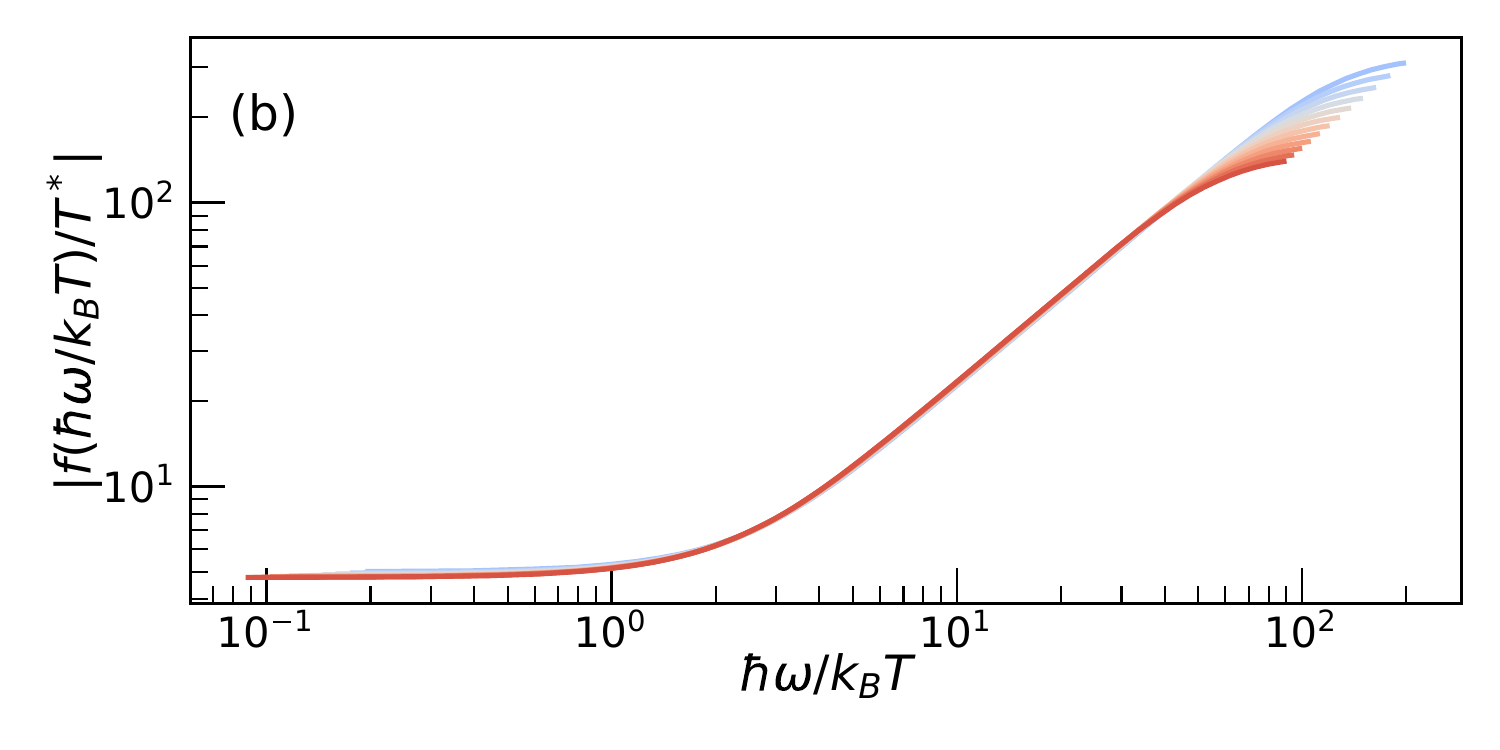}
\caption{(a) Dimensionless tunneling conductance \note{(Eq. \ref{LRconductanceformula})} in the NFL 
regime for $p=0.1$ as a function of $ \hbar \omega/ k_{B} T$ for several temperatures. The conductances 
saturate close to a universal value $g_{0}(p)$ (Eq.~\ref{Eq:Scln_frm})
for vanishing values of the argument. 
Deviations are clearly visible for arguments 
roughly exceeding 1. (b) Scaling collapse 
for the function $f$ which determines the corrections 
to the universal dimensionless conductance $g_{0}(p)$ in 
the conformal-invariant regime (Eq.~\ref{Eq:Scln_frm}). 
It scales linearly for arguments greater than $O(1)$. }
\label{Fig:Scaling_NFL}
\end{figure}
In Fig.~\ref{Fig:Qualitative_NFL}(a), we plot the dimensionless tunneling conductance  $G/G_{0}$ as a function of frequency at combined half-filling $\mu=0$ and 
at a temperature of 100 mK 
in the NFL regime for increasing values of the 
tuning parameter $p < p_{c}=1/2$. Here and below we take $J=25$meV as estimated for the graphene flake in Ref.\ \onlinecite{Chen}. In addition we assume $t=V=J/2$ in the following unless otherwise noted. We distinguish the 
presence of a relatively sharp peak in the 
conformal-invariant NFL regime at low frequencies followed by a cross-over to an essentially featureless
spectrum beyond a scale $\omega^{*}(p)$. Upon increasing $p$, the height of the peak 
increases while the crossover scale tends to zero, as expected for a second-order 
quantum phase transition (Fig.~\ref{Fig:Phase_diagram}). 
In Fig.~\ref{Fig:Qualitative_NFL}(b) we plot the conductance at fixed $p=0.2$ on the NFL side for 
several temperatures. With decreasing $T$, the height of the central peak 
increases while its width remains roughly constant. The broad 
spectrum beyond the cross-over scale shows very little dependence on temperature.

The dimensionless conductance $G/G_{0}(B)$ at half-filling is 
shown in Fig.~\ref{Fig:Scaling_NFL}(a) as a function of $\hbar \omega /k_{B} T$, for $p=0.1$, and
temperatures ranging from 400 to 1000 mK. 
It saturates to a constant for values of the argument below 1. 
A weak temperature dependence in this limit can still be distinguished from the 
offsets of the saturated values. These shifts are due 
to corrections from leading irrelevant operators about the conformal-invariant fixed point value which arise 
with increasing temperature. 
We find that in the $\hbar \omega , k_{B} T \ll k_{B} T^{*}$ limit the dimensionless 
conductance is consistent with the scaling form 

\enni \begin{align}
g \left( \frac{\hbar \omega}{k_{B} T}, \frac{T}{T^{*}} \rightarrow 0, 0, p \right) = & g_{0}(p) - \left( \frac{T}{T^{*}} \right)^{\alpha} f\left( \frac{\hbar \omega}{k_{B} T} \right).
\label{Eq:Scln_frm}
\end{align}

\enni The universal dimensionless conductance $g_{0}(p)$ 
is the contribution in the conformal limit. It varies 
continuously along the line of fixed points associated with the 
stable NFL phases. We estimated $\alpha \approx 1$ for a range of temperatures 
extending over a decade from the lowest numerically-accessible value of 100 mK (see Appendix~\ref{Sec:Appn_crrc}). The exponent also 
holds for higher values of $p$.
In Fig.~\ref{Fig:Scaling_NFL}(b), we plot the universal function $f$ which converges to a constant
for small values of $\hbar \omega / k_{B} T$. For higher values of the argument, 
$f$ scales linearly. This indicates that corrections to the conductance about the conformal-invariant 
NFL fixed point scale linearly with either frequency or temperature. 
Similar behavior emerges for other values of $p$, 
as well as in cases away from particle-hole symmetry. 
\en{We note that the temperature and frequency-dependent corrections to the universal dimensionless conductance $g_{0}$
arise from the sub-leading contributions to the leading spectral densities shown in Eqs.~\eqref{Eq:Scln_frm_c},~\eqref{Eq:Scln_frm_psi} and Fig.~\ref{Fig:rho_scln}.}

\begin{figure}[t!]
\includegraphics[width=1\columnwidth]{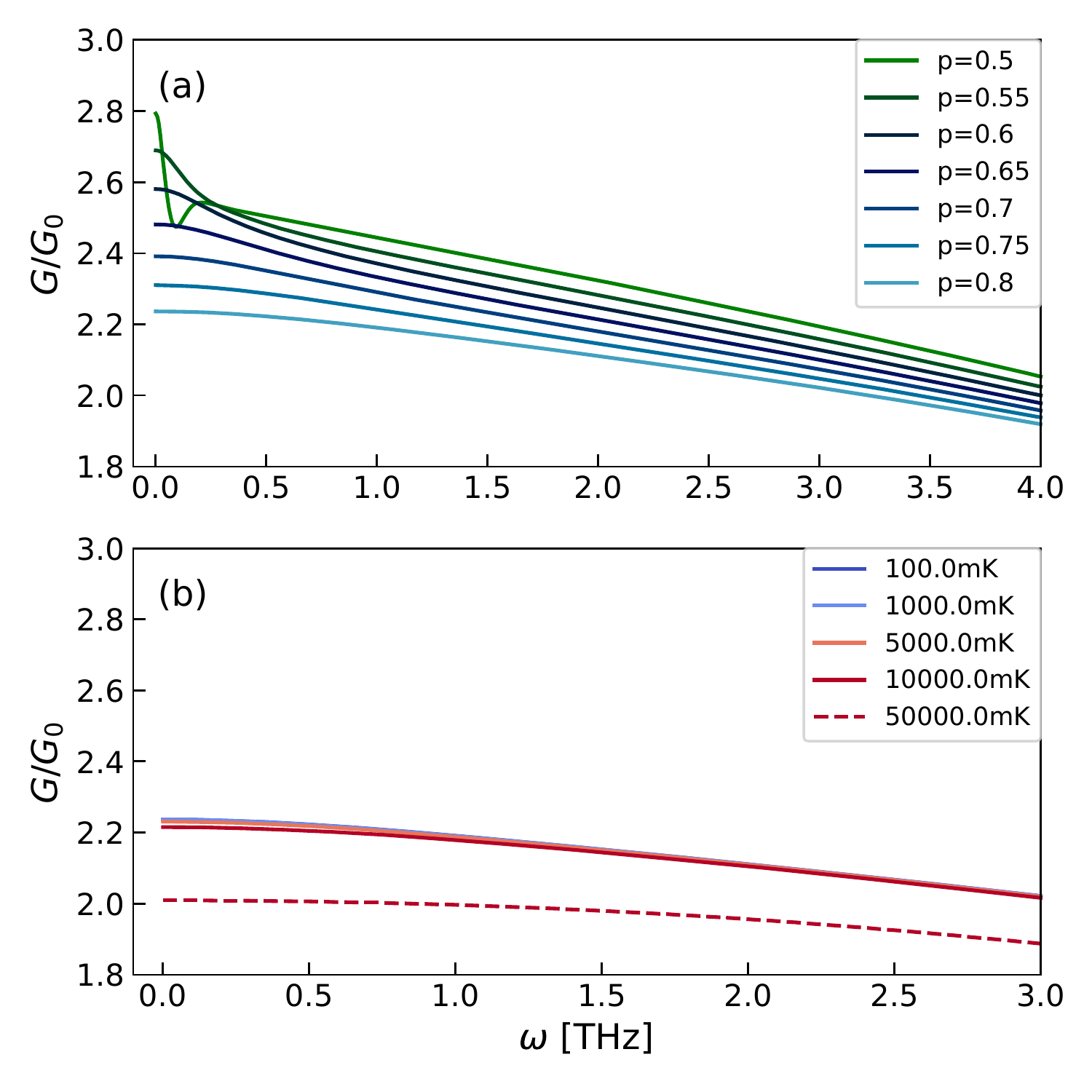}
\caption{(a) Dimensionless tunneling conductance \note{(Eq. \ref{LRconductanceformula})} in the FL 
regime at fixed temperature $T=$ 100 mK and increasing 
values of the tuning parameter $p$.
The peak around zero frequencies rapidly 
broadens and merges with the featureless spectrum 
at higher frequencies. 
(b) Same as (a), at fixed $p=0.8$ and for several temperatures. Note the insensibility to 
variations in temperature relative to the NFL 
regime (Fig.~\ref{Fig:Qualitative_NFL})(b)). 
The dashed line corresponds to the high-temperature limit and is included for comparison.
}
\label{Fig:Qualitative_FL}
\end{figure}

Turning to the FL regime, 
in Fig.~\ref{Fig:Qualitative_FL}(a) we plot 
the dimensionless conductance $G/G_{0}$ as a function 
of frequency, at half-filling, for fixed temperature $T=$ 100 mK and several 
values of $p$. 
Close to the transition, we observe a narrow peak 
which quickly broadens and flattens and becomes 
indistinguishable from the high-energy spectrum.
In addition, as shown in Fig.~\ref{Fig:Qualitative_FL}(b), it
shows a much weaker temperature dependence relative 
to the NFL phase in Fig.~\ref{Fig:Qualitative_NFL}(b).
An analysis similar to the one leading to Eq.~\ref{Eq:Scln_frm} reveals 
a similar scaling form with an exponent $\alpha=2$ which is characteristic of FL regimes~\cite{Hewson} (see 
Appendix~\ref{Sec:Appn_crrc}). 
The relative insensibility to temperature on the FL side is most likely due to the combined 
effect of corrections to the fixed point which scale as $(T/T^{*})^{2}$ and to a relatively large 
cross-over scale, as sketched in Fig.~\ref{Fig:Phase_diagram}.
A similar picture emerges in this regime away from particle-hole symmetry.   

\noindent 

\subsection{Linear response DC conductance}

We now discuss the universal dimensionless conductance $g_{0}(p)$  defined in Eq.~\eqref{Eq:Scln_frm} and compare our analytical and numerical results. Well below the cross-over scales determined by $T^{*}(p)$, $g_{0}(p)$ provides the leading contribution to the dimensionless conductance $G/G_{0}$.  
Appendix~\ref{Sec:Appn_cndc} gives analytical calculation of
$g_{0}(p)$ in the DC ($\omega=0$) and zero-temperature limits. The
linear-response DC conductance is then given 
via the spectral densities for the coupled leads and dot in the conformal
regime. At particle-hole symmetry a simple result already quoted in Eq.\
(\ref{Eq:DC_cndc}) is obtained using BA results for the spectral densities (adapted to two flavors of auxiliary fermions). It shows a universal jump at $p=p_c$.

\begin{figure}[t!]
\includegraphics[width=1\columnwidth]{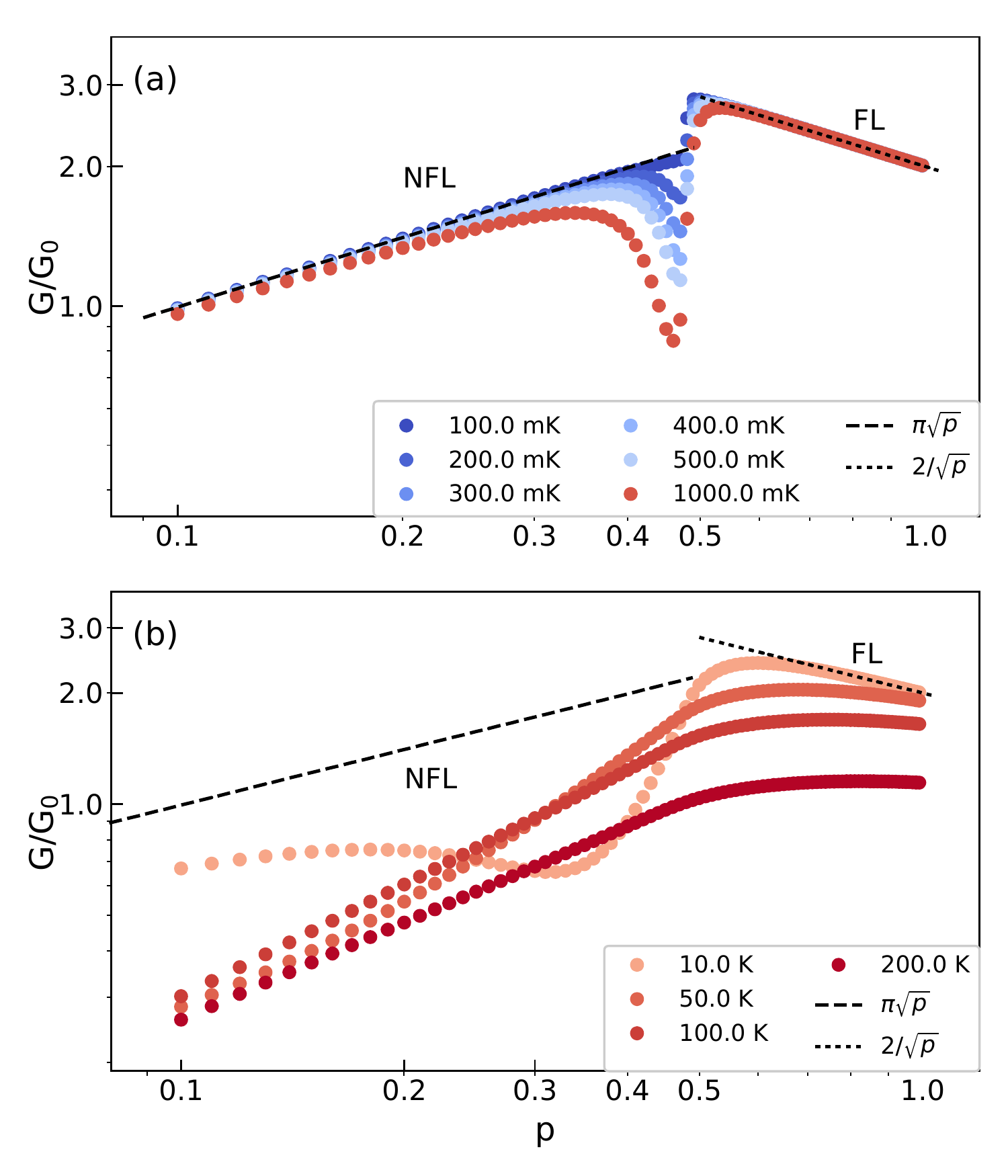}
\caption{Dimensionless DC conductance $G/G_0$ as function of tuning
  parameter $p$, for a range of temperatures, at particle-hole
  symmetry, plotted on a logarithmic scale. The dashed and dotted grey
  lines denote the analytically-determined, universal conductances
  $g_{0}(p)$  given in Eq.~\eqref{Eq:DC_cndc}. The full dots are the numerically-extracted DC conductances \note{obtained by taking the $\omega \rightarrow 0$ limits of AC conductances computed using Eq. \ref{LRconductanceformula}.} ~(a) When $T$ is below $T^{*}(p)$ for almost all values of $p$, the numerical results in each phase are in good agreement with the analytical prediction for $g_{0}(p)$.  Deviations are more pronounced with increasing temperatures, especially on the NFL side, reflecting corrections due to the finite cross-over scale $T^{*}(p)$. ~(b) 
For temperatures exceeding $T^{*}(p)$ in the entire range of $p$ values, we observe a cross-over to a putative quantum-critical regime on the NFL side, as illustrated by the $T = 10$ K curve. 
As the temperature is further increased, we note the onset of a high-temperature regime as indicated by the remaining curves.}
\label{Fig:Conductance_p}
\end{figure}

At nonzero temperature the integrals entering the Kubo formula must be evaluated numerically.   
 Our results for $G/G_{0}$ are shown in Figs.~\ref{Fig:Conductance_p}(a) as a function of $p$, at several lower  temperatures, in the $\mu=0$ case. 
On the NFL side ($p < p_{c}$) we find that the numerically-determined
DC conductance closely follows the analytical prediction for
$g_{0}(p)$, provided that $T$ stays well below $T^{*}(p)$.  We attribute the large deviations observed above 
roughly 500 mK to proximity 
to the quantum-critical regime.  
Similarly, cross-overs to the quantum-critical
regime with increasing $p$ are more pronounced and occur at lower values with increasing 
temperature. This behaviour is completely consistent with the 
presence of a cross-over scale $T^{*}(p)$ which vanishes 
continuously at the critical coupling, as sketched in Fig.~\ref{Fig:Phase_diagram}. 
It is also in agreement with the 
offsets of the saturated values in Fig.~\ref{Fig:Scaling_NFL}. 

Beyond the cross-over to the quantum-critical regime, the DC conductance enters the FL phase, where it exhibits very little dependence on temperature. With increasing temperatures, we find that the dimensionless conductance $G/G_{0}$ exhibits several cross-overs. To illustrate, in Fig.~\ref{Fig:Conductance_p}(b) we plot the dimensionless conductance $G/G_{0}$ as a function of $p$ for temperatures exceeding 1 K. Note the cross-over to a putative quantum-critical regime for $p< 1/2$, as illustrated by the $T=10$ K data. The remaining curves indicate the onset of a distinct, high-temperature regime. We also note that a distinction between the NFL and FL regimes survives in this high-temperature regime. 

\noindent 

\begin{figure}[h]
\includegraphics[width=1\columnwidth]{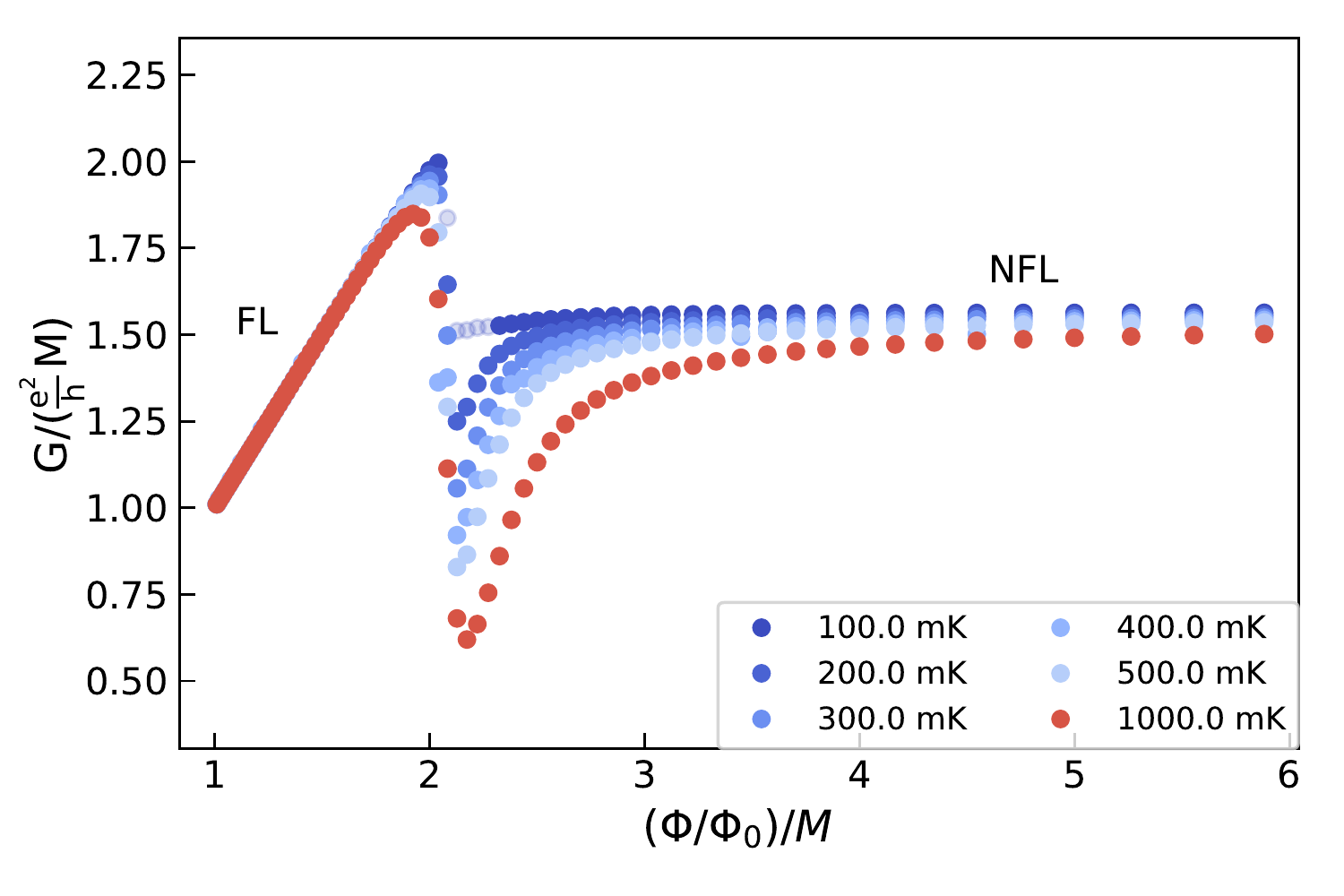}
\caption{Dimensionless tunneling conductance per lead channel as a function of flux threading the dot. Note that $\frac{(\Phi/\Phi_0)}{M} = 1/p$ (see also Eq.~\ref{Eq:p}). The conductance shows linear dependence for low values of external magnetic field in the FL phase below the critical point. With increasing field the sharp cross-over to the NFL behaviour is observed. In the latter regime, the dimensionless conductance is constant in the applied field.}
\label{Fig:Conductance_B}
\end{figure}

\enni In order to illustrate the direct dependence of the response on the applied external magnetic field to the dot, we include Fig.~\ref{Fig:Conductance_B}, which shows the dimensionless conductance per transverse channel as a function of the flux threading the dot. The linear increase in the FL regime reflects the increasing number of channels available for conduction in the dot which is linearly proportional to the Landau level degeneracy $N$. Above the transition, which occurs at total magnetic flux $\Phi=2M\Phi_0$, conductance saturates at a field-independent constant value ${\pi\over 2}{e^2\over h}M$ characteristic of the NFL regime. Observing this remarkable behavior experimentally would constitute an unambiguous evidence of the SYK physics in the system.


\noindent 

Away from exact half filling the DC conductance can still be evaluated
analytically. In the conformal limit on the NFL side we obtain (Appendix~\ref{Sec:Appn_cndc} )

\begin{figure}[ht]
\includegraphics[width=1\columnwidth]{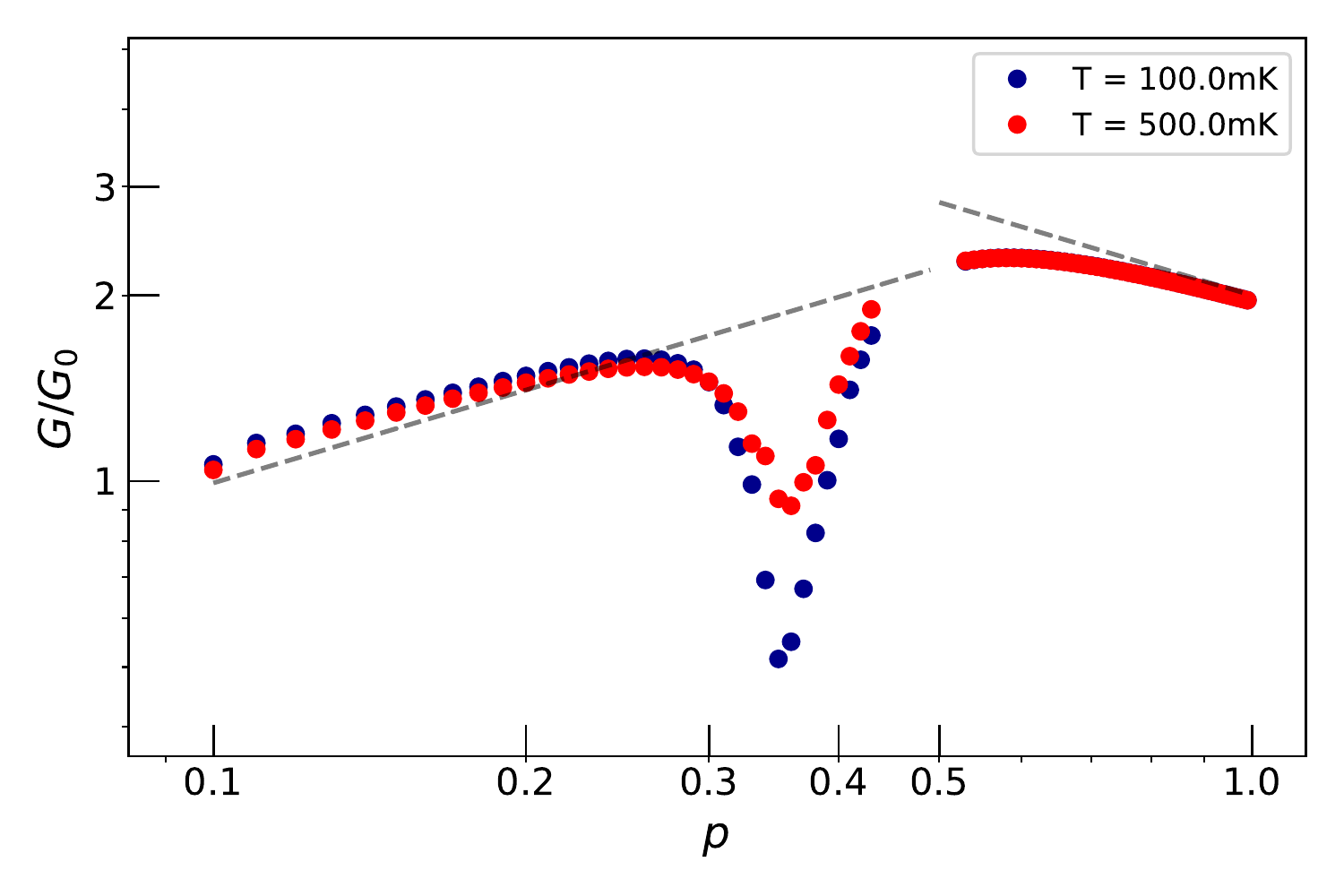}
\caption{Dimensionless tunneling conductance at $T=100, 500$ mK in the dc limit, away from half-filling ($\mu=0.625$ meV). The dashed lines indicate the behavior at particle-hole symmetry. The curves 
closely follow the particle-hole symmetric dependence on $\sqrt{p}$ on the NFL side. On the FL side, we observe stronger deviations in the vicinity of the cross-over, although the $1/\sqrt{p}$ behavior is recovered for $p > 1$.} 
\label{Fig:ph_asmt}
\end{figure}

\enni \begin{align}
& g_{0}(p < 1/2) 
 =  \pi \sin \left( \frac{\pi}{2} + 2 \theta \right) \sqrt{p}.
\end{align}
The phase $\theta \in [-\pi/4, \pi/4]$ is related to the "spectral asymmetry" defined in the context 
of $\text{SYK}_4$ models~\cite{Parcollet_Kondo, Sachdev2015, Altman2016}. As discussed in Ref.~\onlinecite{Altman2016}, 
it is in general a function of the total filling of dot and lead (end points) and of $p = M/N$. 
For particle-hole symmetry, $\theta=0$ for all values of $p$. 
Away from particle-hole symmetry 
$\theta$ must be determined numerically. In Fig.~\ref{Fig:ph_asmt}, we show the DC conductance  
for a chemical potential $\mu=0.625$ meV ($0.025 J$) and at two temperatures as a function of tuning parameter $p$. The dashed line indicates the expected value at particle-hole symmetry extracted via Eq.~\ref{Eq:DC_cndc}.  
In the NFL regime, the dimensionless conductance closely follows the particle-hole symmetric results and shows similar scaling with $p$. The total filling at $p=0$ is 0.42 and undergoes a 10\% increase up to close to the transition. Larger deviations of the conductance with respect to the particle-hole symmetric case are 
observed in the FL regime for $p \ge 1/2$, although the curve approaches the scaling predicted 
for the particle-hole symmetric case for $p>1$. In this regime the total filling varies from 0.46 to 0.48 at $p=1$. As mentioned above, we do not treat the cross-over regimes in great detail in this work. The results indicate that small departures from particle-hole symmetry do not significantly affect the $\sqrt{p}$ scaling determined for $\mu=0$.
\noindent 
\noindent \begin{figure}[t!]
\includegraphics[width=1\columnwidth]{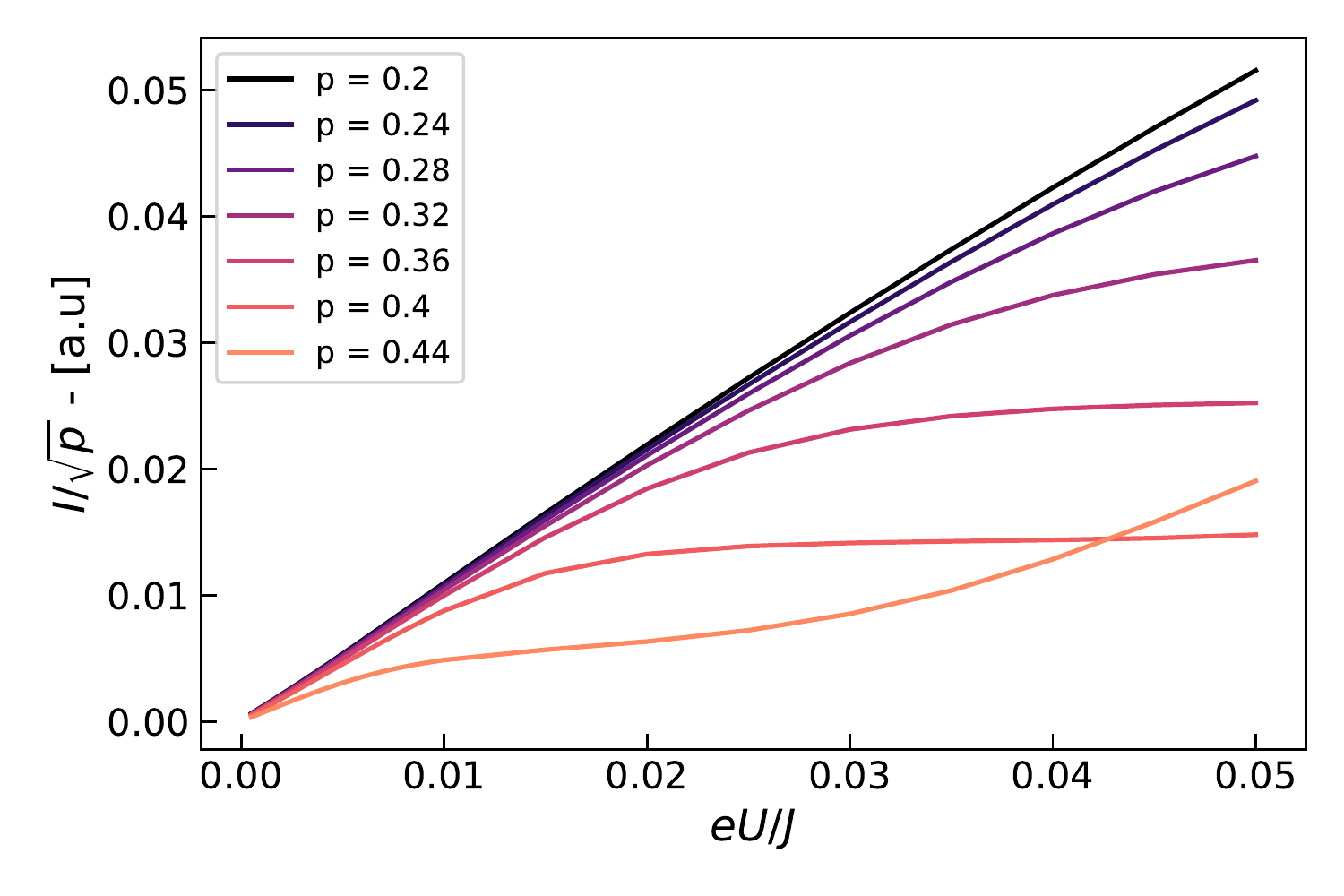}
\caption{Tunneling current \note{(Eq. \ref{Eq:MW_current})} in arbitrary units scaled by $\sqrt{p}$, at a fixed temperature $T=400$ mK ($2.0 \times 10^{-4} J$) 
for the leads, for several values of the tuning parameters $p < p_{c}$ corresponding to the NFL near equilibrium, as a function of bias $eU$ in units of $J$. A cross-over from linear response can be distinguished around $U^{*}(p)$. }
\label{Fig:MW_U}
\end{figure}

\noindent \begin{figure}[t!]
\includegraphics[width=1\columnwidth]{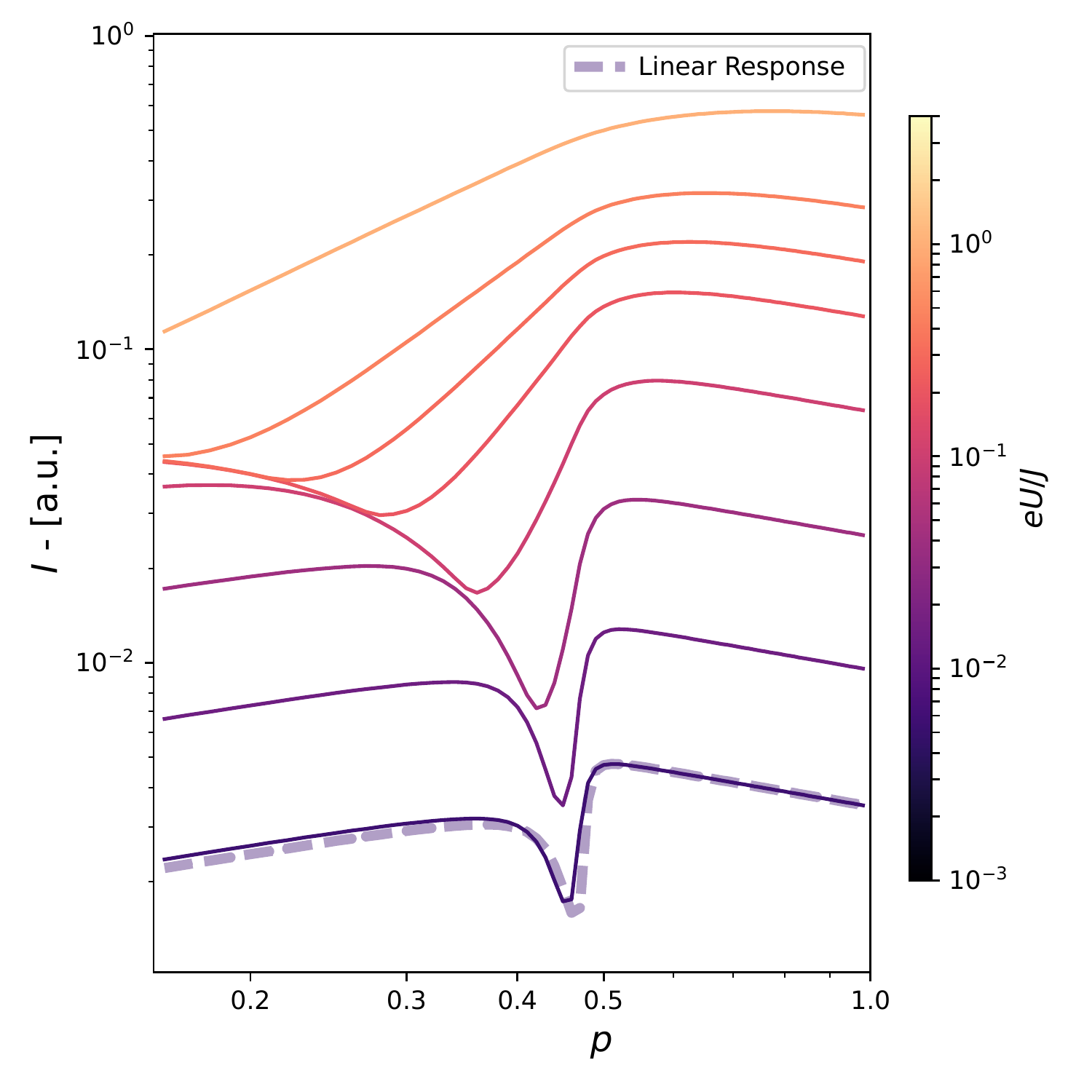}
\caption{Tunneling current \note{(Eq. \ref{Eq:MW_current})} in arbitrary units, at fixed $T=400$ mK ($2.0 \times 10^{-4} J$) for the leads , for a range of biases $eU$ across the dot, as a function of tuning parameter $p$. The results are plotted on a logarithmic scale. The dashed line indicates the prediction based on linear response. For biases up to $O(10^{-2} J)$, the current closely follows the $\sqrt{p}$ and $1/\sqrt{p}$ dependencies encountered in linear response, up to a trivial shift due to near-linear dependence of the current on bias. Increasing bias induces a cross-over to an intermediate regime $O(10^{-1}) < eU/J < O(1)$ analogous to the quantum-critical region near equilibrium. Note the similarity with the temperature-induced cross-overs in Fig.~\ref{Fig:Conductance_p}(a) and (b).}
\label{Fig:MW_p}
\end{figure}


\subsection{Non-linear DC response}

We also calculated the nonlinear current \eqref{Eq:MW_current} for arbitrary \emph{static} 
applied bias across the dot via an approach based on real-time Green's functions in the Keldysh basis~\cite{Meir, Haug}. 
The leads are in thermal equilibrium with reservoirs at chemical potentials shifted by  $\pm eU/2$, where $U$ is the applied bias. 
The details 
of the procedure and implementation are given in Appendix~\ref{Sec:Appn_MW}. In Fig.~\ref{Fig:MW_U} we plot 
 the current in arbitrary units, scaled with $\sqrt{p}$ as a function of the applied bias in units of $J$, at a lead temperature of $T=400$ mK ($2.0 \times 10^{-4} J$), for a range of values of the tuning parameter $p$, corresponding to the NFL in equilibrium. The factor of $\sqrt{p}$ is included to account for variations with tuning parameter already present in linear response. 
The response remains linear up to a bias $U^{*}(p)$ beyond which it saturates to a $p$-dependent constant. The cross-over scale set by the bias $eU^{*}(p)$ decreases and appears to vanish as $p$
approaches the value at which the transition occurs in equilibrium $p_{c}=1/2$.
We note that this behavior persists for applied biases which are well-above the scale set by the temperature in equilibrium. In Fig.~\ref{Fig:MW_p} we show the current, likewise in arbitrary units, at fixed $T=400$  mK for the leads, for a range of applied biases $eU$, as a function of tuning parameter $p$. The dashed line indicates the current predicted from linear response at the same temperature. For vanishingly-small biases, the current closely follows the linear-response prediction with characteristic $\sqrt{p}$ and $1/\sqrt{p}$ scaling 
in the NFL and FL regimes, respectively. Within an expected shift reflecting the linear dependence on bias, the scaling behavior persists for $p$ away from $p_{c}=1/2$ up to biases of $O(10^{-2}) J$, while the cross-over region becomes increasingly broader. Beyond this hallmark bias, the currents for $p<p_{c}$ undergo a clear cross-over to 
an intermediate regime which is no longer well-described by a $\sqrt{p}$ dependence. Finally, 
for large biases approaching $J$, a completely different $p$
dependence is reached, which still maintains a distinction between the
two regimes encountered in linear response. There is a striking
similarity between the cross-overs observed in the non-linear response
with increasing bias and the cross-overs seen with increasing
temperatures in linear response (Fig.~\ref{Fig:Conductance_p}~(a) and
(b)) as a function of $p$.

The numerical saddle-point results are consistent with a lead-dot
coupling which is relevant in the RG sense. Hence, we expect that the
results for the conductance discussed thus far, which imply
renormalized spectral densities for the leads
(Appendix~\ref{Sec:Appn_spct}), are always valid in the $T \ll
T^{*}(p)$ limit. Note that according to Eq.\ (\ref{Eq:T*}) the cross-over scale $T^{*}$ is
expected~\cite{Altman2016} to be of $O(V^{4}/t^{2} J)$ in the lead-dot
coupling $V$. For a weak coupling between leads and dot,
the cross-overs determined by $T^{*}$ are expected to occur at very
low temperatures. Above $T^{*}$ we can estimate the current-bias curve
using a weak-tunnelling approximation discussed below.

\subsection{Weak-tunneling regime}

When the coupling between leads and the dot $V$ is \en{sufficiently small,} one can calculate the tunneling current perturbatively
in this small parameter even when the bias voltage across the two leads is finite. This amounts to the well-known tunneling
Hamiltonian approximation~\cite{Mahan} involving a tunneling rate of
$O(V^2)$ and densities of states corresponding to decoupled leads and
$\text{SYK}_4$ dot in the conformal-invariant regime.
\en{More specifically, we expect that this regime emerges for temperatures well above the cross-over scale $T^{*}$. Below this scale, the contribution from $V$ is non-perturbative, as illustrated by the spectral densities calculated to all orders in $V$ in Eqs.~\eqref{Eq:Scln_frm_c},~~\eqref{Eq:Scln_frm_psi} and Fig.~\ref{Fig:rho_scln}. Since we expect that $T^{*} \propto V^{4}$ (Eq.~\eqref{Eq:T*}) in the vicinity of the transition, a reduction in $V$ will induce a significant decrease in $T^{*}$.}
We found that the weak tunneling current $I_{WT}$ is given by
\begin{equation}\label{eq:lweaktunnelingsummary}\hspace{12pt}
\langle I_{WT} \rangle \propto \begin{cases} 
     eU/\sqrt{ T}  &  (eU \ll k_BT)\\
        \sqrt{ e U} &(eU \gg k_BT)
   \end{cases}
\end{equation} 
Details of the calculation are presented in Appendix~\ref{Sec:Appn_wk}. The weak-tunneling
approximation is expected to be valid in the context of scanning-tunnelling spectroscopy (STM) experiments and in situations when leads are separated from the dot by a thin oxide barrier. 

\begin{figure}[h]
\includegraphics[width=1\columnwidth]{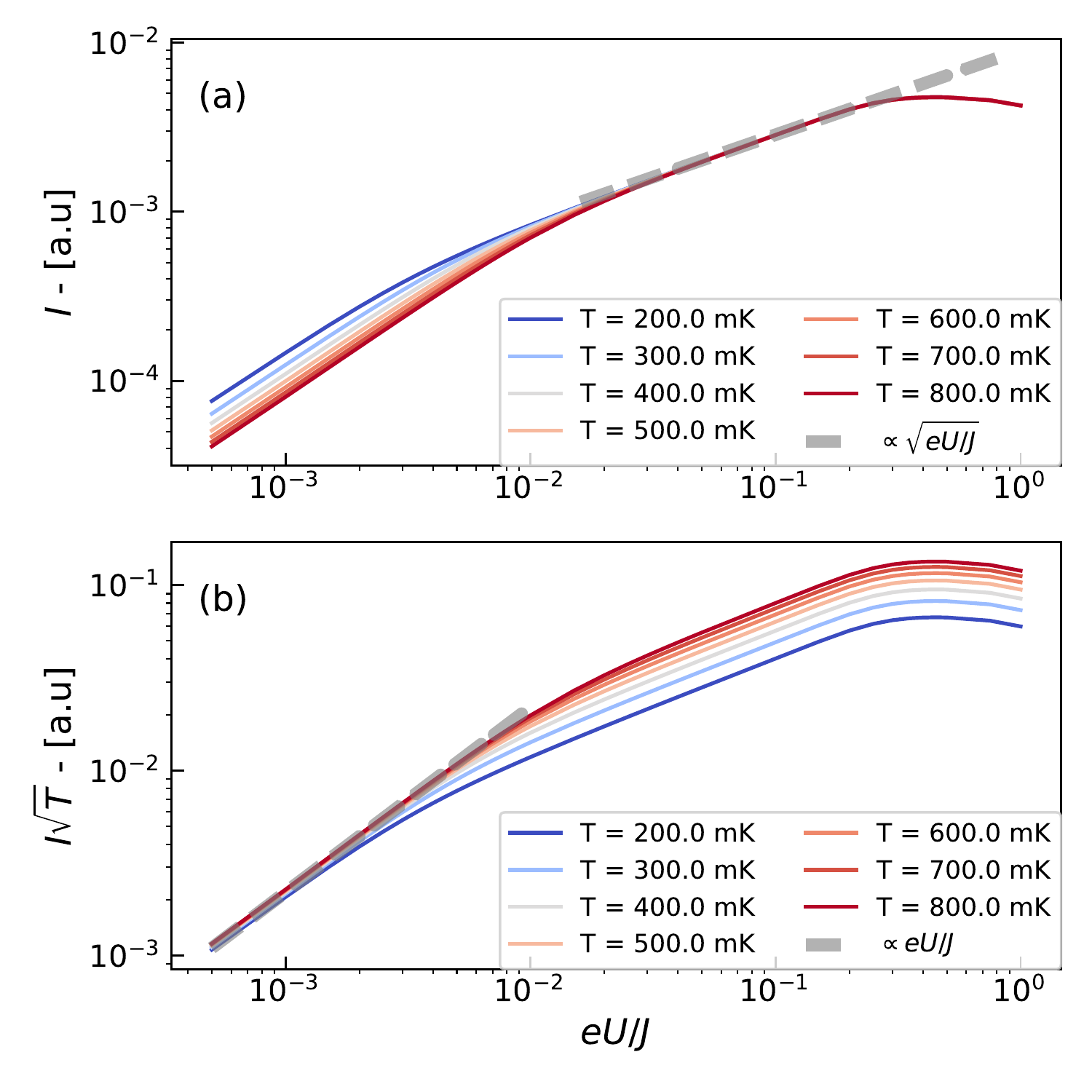}
\caption{Current in weak-tunneling approximation for the NFL phase, plotted in arbitrary units as a function of applied bias $eU$ in units of $J$. Here we take $t=J/2$, $V=0.025J$ and $p=0.3$. (a) In high bias regime $eU \gg k_BT$ we find that the current calculated with \eqref{Eq:MW_current} using numerical solutions of the saddle point equations match weak tunneling analytical prediction \eqref{eq:lweaktunnelingsummary} \note{plotted with dashed lines.} (b)
$I\sqrt{T}$ - $eU/J$ characteristics in the weak tunneling regime for various temperatures. For low bias regime $eU \ll k_BT$ we observe a scaling collapse, confirming the predicted $eU/\sqrt{T}$ dependence. }
\label{Fig:cndt_wk_tnlg}
\end{figure}

We match these analytical predictions to the nonlinear current  \eqref{Eq:MW_current}  \en{which includes contributions to all orders in $V$. To tune the system to the weak-coupling regime we use a lead-dot coupling $V=0.025 J$ which is one order of magnitude smaller than the previously-used value while all remaining parameters, including the temperature range, are kept fixed.} \en{The numerically-determined current} for $p=0.3$, at various temperatures ranging from $T=200$ mK to $T=800$ mK, as a function of applied bias $eU$ \en{is shown} in Fig.~\ref{Fig:cndt_wk_tnlg}.
In high bias regime (Fig.~\ref{Fig:cndt_wk_tnlg}(a)) where $eU \gg k_BT$ \en{the} I-V curves do not depend on temperature and agree with the analytical prediction $I \propto \sqrt{eU/J}$. At low biases, we observe a temperature dependent behaviour which is linear in the applied bias. In Fig. \ref{Fig:cndt_wk_tnlg} (b) we plot $I\sqrt{T}$ versus $eU$ to observe the scaling collapse that occurs for $I \propto eU/\sqrt{T}$ in low bias regime ($eU \ll k_BT$). Once again, this characteristic behavior, if observed experimentally, would furnish strong evidence supporting the SYK state on the dot.

\noindent
\noindent \begin{figure}[t!]
\includegraphics[width=0.9\columnwidth]{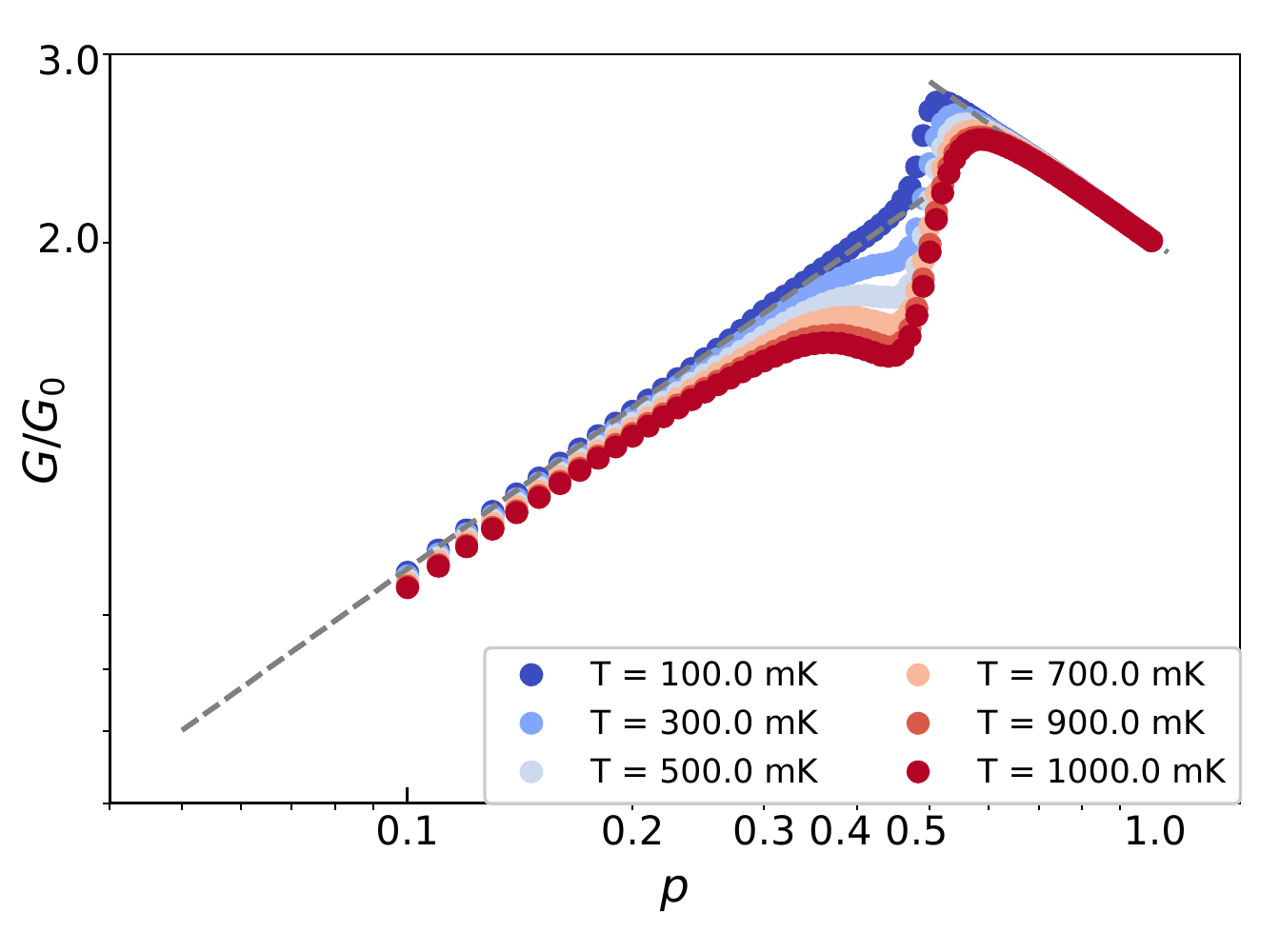}
\caption{Dimensionless conductance for an effective model of the tunneling junction which includes strong coupling of $O(J/2)$ to extended, non-interacting leads  (see Appendix E) in addition to the local disorder on the lead end points, plotted versus tuning
parameter $p$. The same essential features present in Fig.~\ref{Fig:Conductance_p}~(a) for the effective model without local contributions from the lead bulk are also apparent here. The main difference is a narrowing of the cross-over regime.}
\label{Fig:Conductance_wires_dis}
\end{figure}
\noindent 
\begin{figure}[t!]
\includegraphics[width=0.9\columnwidth]{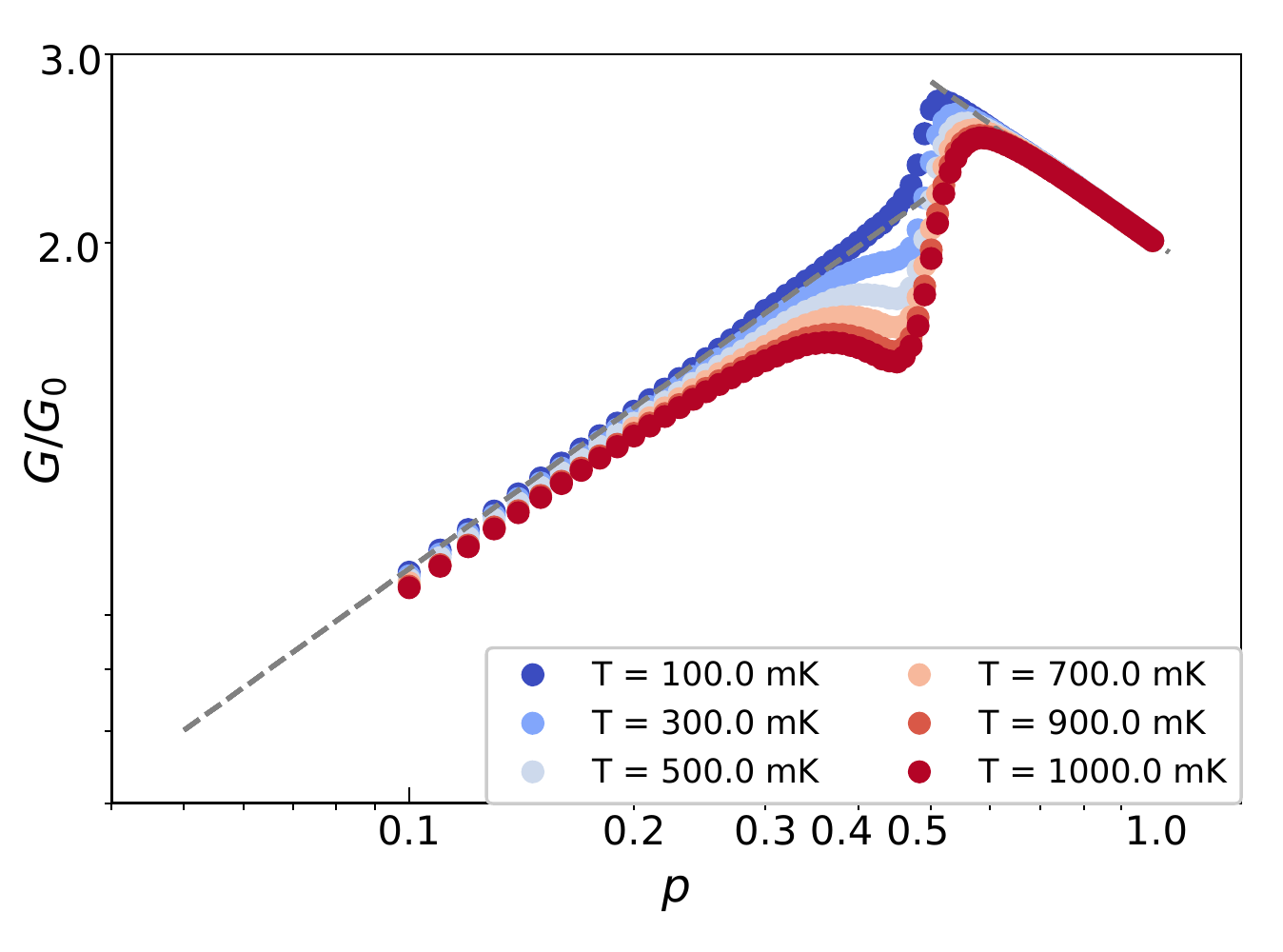}
\caption{Dimensionless conductance for an effective model which includes strong coupling to extended, non-interacting leads of $O(J/2)$ but excludes any local disorder on the lead end points. As in Fig.~\ref{Fig:Conductance_wires_dis}, we observe no essential deviations from the cases shown Fig. ~\ref{Fig:Conductance_p}} \label{Fig:Conductance_wires_no_dis}
\end{figure}

\subsection{Effect of extended leads}

Our results thus far have neglected the effect of the coupling to the bulk of the leads 
on the low-energy and low-temperature spectral densities. Instead, we considered an \emph{effective local 
model} for the junction where disorder-scattering dominated the
low-energy dynamics of the end points of the leads. We now consider an
explicit coupling to extended leads as described by Hamiltonian
(\ref{hE}). A detailed discussion of the modified saddle-point solution is given in Appendix~\ref{Sec:Appn_lead}. 

We find that including a coupling to non-interacting extended leads or ignoring disorder-scattering altogether near the end points has no 
essential effect on the low-temperature tunneling current in either phase.
Consider the effect of coupling to extended leads, which are 
modeled as quasi-one dimensional, ballistic wires, while maintaining 
the disorder at the end points. 
At weak coupling, such additional terms are marginal. A complete numerical 
solution indicates that both phases survive 
for couplings to the bulk of the leads of order $J$. 
Likewise, a transition to the FL occurs 
at the same value of $p_{c}$.  
The DC conductance preserves the same scaling with $p$ in either phase, as shown in Fig.~\ref{Fig:Conductance_wires_dis}(a) for $\mu=0$.
A similar picture emerges upon completely 
neglecting the disorder in the end points of the leads, as shown in
Fig.~\ref{Fig:Conductance_wires_no_dis}. We thus conclude that the
simplified model of the junction studied in the earlier subsections
constitutes a very reasonable approximation for a physical setup with
extended leads.

\section{Summary and conclusions}\label{Sec:Conclusions}
We have characterized the tunneling conductance and current-bias
properties of a graphene quantum-dot realization of the $\text{SYK}$
model coupled to leads with and without disorder in the vicinity of
the junction. The problem is highly non-trivial because the fragile
non-Fermi liquid state on the dot is easily disrupted by coupling to
the leads. We obtained our results using a saddle-point approximation for an effective model of the junction in the limit of large number of transverse modes $M$ for the lead and large degeneracy of the dot zeroth Landau level $N$, with their ratio $p=M/N$ finite. The calculations were carried out analytically in various simple limits and numerically using real-time Green's functions in the Keldysh basis for general parameters. We find clear signatures of distinct emergent conformal-invariant non-Fermi liquid and Fermi-liquid regimes and of the cross-overs associated with a quantum-critical point. The transition can be accessed by tuning the ratio $p$ via the magnetic field applied to the dot.

Deep within the NFL phase, and for temperatures much lower than a
cross-over scale $T^{*}$, we find a universal dimensionless
conductance which shows a $\sqrt{p}$ variation with the tuning
parameter which is directly related to the applied field $B$ through Eq.\ (\ref{Eq:p}). This dependence is intrinsic to the low-energy emergent,
conformal-invariant regime. We also find leading corrections which
scale linearly with temperature and frequency throughout the NFL
regime. Beyond the transition at $p_c=1/2$, we find that the low-temperature FL regime exhibits a $1/\sqrt{p}$ dependence on the tuning parameter and corrections which are quadratic in either temperature or frequency. Results obtained at weak particle-hole asymmetry show a similar scaling with tuning parameter $p$. 

We find that the current is linear with applied bias up to a bias
$U^{*}(p)$ when the coupling to the leads is strong. For larger biases we find cross-overs from the linear to  intermediate and high-bias regimes which are analogous to the quantum-critical and high-temperature regions in linear response. 

In the limit of weak tunneling, relevant for scanning tunneling
spectroscopy and tunnel junction experiments, we find the tunneling conductance proportional to
$\min(1/\sqrt{U},1/\sqrt{T})$. The inverse square root dependence on
the bias in the $T\to 0$ limit reflects the $|\omega|^{-1/2}$ behavior of the electron
spectral function in the NFL regime of the SYK model and has been
noted previously\cite{Chen,Beenakker2018}. Our calculations extend
these results to include the effect of non-zero temperature which is
found to cut off the low-bias divergence of the conductance at a
characteristic value proportional to $1/\sqrt{T}$.

We also find that the similar scaling with $p$ holds in the absence of local disorder on the lead end points.

We note that our results are in some ways similar to those obtained for $SU(K) \otimes SU(N)$ multi-channel Kondo impurity models where $K$ and $N$ refer to the number of conduction electron channels and spin-symmetry group, respectively~\cite{Parcollet_Kondo}. These models host non-trivial emergent conformal invariance at low temperatures and are amenable to saddle-point approximations. It was found that  
the conduction electron scattering rate depends essentially on the ratio between $K$ and $N$.  Corrections at finite temperature or frequency scale with a common non-trivial fixed-point dependent exponent~\cite{Parcollet_Kondo, Affleck}. In our case, we find that the conductance in the NFL phase acquires an analogous $\sqrt{p}$ dependence. However, we find corrections which scale linearly with temperature and frequency throughout the NFL phase for all ratios $p<p_{c}$. 

A closely related aspect involves the small-bias corrections to the differential tunneling conductance (Fig.~\ref{Fig:MW_U}). In the context of two-channel Kondo models, corrections which exhibit $eU/k_{B}T$ scaling for $eU, k_{B}T \ll T_{K}$ have been predicted based on conformal field theory~\cite{Ralph, von_Delft}, and  non-equilibrium Green's functions calculations~\cite{Hettler_1994, Hettler_1998}. These corrections scale as $x^{2}$ and $\sqrt{x}$, for $x \ll 1$ and $x \gg 1$, respectively, where  $x = eU / k_{B} T$. These predictions were subsequently observed in experiment~\cite{Potok}. Based on the analogy with the two-channel Kondo model, and the linear scaling with temperature in equilibrium, we expect a differential conductance with corrections which are linear in the bias, for $k_{B}T \ll eU \ll k_{B}T^{*}$ in our case. Equivalently, we expect corrections to the Ohmic dependence which are quadratic in the bias in this regime. Our results shown in Fig.~\ref{Fig:MW_U} do not show any signatures of this behavior as the current exhibits a linear dependence on bias up to a cross-over scale $eU^{*}$ which is roughly analogous to $k_{B}T^{*}$ in equilibrium. Here, we argue that this is likely due the smallness of these corrections which are expected to be $\sim (eU)^{2}/T^{*} \approx O(10^{-3})J$. We reserve a more detailed analysis of this issue for future work.        

As these results indicate, the SYK model realized in a graphene flake or a similar system
shows a remarkable wealth of experimentally observable transport
phenomena when connected to weakly interacting leads. Perhaps the most
important finding is that the strength of coupling to the leads is
less important than the total number of channels present in the
leads ($2M$ in our notation with two identical leads). When $2M$
exceeds the number $N$ of the active fermion degrees of freedom on the
SYK dot a quantum phase transition is triggered to a Fermi liquid
state. Much of the interesting SYK phenomenology is then lost
(although some signatures may remain in the quantum critical regime at
higher temperatures or frequencies). This result, already contained in the work of Banerjee and
Altman,\cite{Altman2016}  underscores the necessity of designing the
junction with a small number of conduction channels coupled to the
dot. STM tip normally corresponds to a single-channel probe,
which would be ideal to observe properties deep in the NFL regime. It is 
important to remember, however, that a sample in the STM experiment must be
grounded and, according to our results, coupling to the ground must be
carefully controled so that the {\em total} number of channels coupled
to the dot remains small compared to $N$.  

Because $N$ is equal to the number of magnetic flux quanta piercing
the dot the sensitive dependence on the parameter $p=M/N$ affords a
unique opportunity to study the quantum phase transition from NFL to
FL phase by tuning the applied field. At low temperature we predict a
universal jump in dimensionless DC conductance at the transition
accompanied by a characteristic broadening at non-zero $T$. Observing
such a jump would constitute an unambiguous evidence of the phase
transition as well as the SYK state on the high-field side of the transition.

We may thus conclude that transport experiments on a nanoscale
graphene flake with an irregular boundary offer a unique opportunity
to study the iconic SYK model, whose physics cuts across the
boundaries of fields ranging from string theory and quantum gravity to
chaos theory and strongly correlated electron systems.

{\it Acknowledgements:}
We thank Ian Affleck, Josh Folk, \'Etienne Lantagne-Hurtubise and
Chengshu Li for stimulating discussions. Research reported in this
article was supported by NSERC and CIfAR. Final stages of the work
were completed at the Aspen Center for Physics (M.F.). O.C was also supported by QuEST program at the Stewart Blusson Quantum Matter Institute, UBC.

\bibliographystyle{apsrev4-1}
\bibliography{SYK_trans}

\begin{thebibliography}{51}%
\makeatletter
\providecommand \@ifxundefined [1]{%
 \@ifx{#1\undefined}
}%
\providecommand \@ifnum [1]{%
 \ifnum #1\expandafter \@firstoftwo
 \else \expandafter \@secondoftwo
 \fi
}%
\providecommand \@ifx [1]{%
 \ifx #1\expandafter \@firstoftwo
 \else \expandafter \@secondoftwo
 \fi
}%
\providecommand \natexlab [1]{#1}%
\providecommand \enquote  [1]{``#1''}%
\providecommand \bibnamefont  [1]{#1}%
\providecommand \bibfnamefont [1]{#1}%
\providecommand \citenamefont [1]{#1}%
\providecommand \href@noop [0]{\@secondoftwo}%
\providecommand \href [0]{\begingroup \@sanitize@url \@href}%
\providecommand \@href[1]{\@@startlink{#1}\@@href}%
\providecommand \@@href[1]{\endgroup#1\@@endlink}%
\providecommand \@sanitize@url [0]{\catcode `\\12\catcode `\$12\catcode
  `\&12\catcode `\#12\catcode `\^12\catcode `\_12\catcode `\%12\relax}%
\providecommand \@@startlink[1]{}%
\providecommand \@@endlink[0]{}%
\providecommand \url  [0]{\begingroup\@sanitize@url \@url }%
\providecommand \@url [1]{\endgroup\@href {#1}{\urlprefix }}%
\providecommand \urlprefix  [0]{URL }%
\providecommand \Eprint [0]{\href }%
\providecommand \doibase [0]{http://dx.doi.org/}%
\providecommand \selectlanguage [0]{\@gobble}%
\providecommand \bibinfo  [0]{\@secondoftwo}%
\providecommand \bibfield  [0]{\@secondoftwo}%
\providecommand \translation [1]{[#1]}%
\providecommand \BibitemOpen [0]{}%
\providecommand \bibitemStop [0]{}%
\providecommand \bibitemNoStop [0]{.\EOS\space}%
\providecommand \EOS [0]{\spacefactor3000\relax}%
\providecommand \BibitemShut  [1]{\csname bibitem#1\endcsname}%
\let\auto@bib@innerbib\@empty
\bibitem [{\citenamefont {Sachdev}\ and\ \citenamefont {Ye}(1993)}]{SY1996}%
  \BibitemOpen
  \bibfield  {author} {\bibinfo {author} {\bibfnamefont {S.}~\bibnamefont
  {Sachdev}}\ and\ \bibinfo {author} {\bibfnamefont {J.}~\bibnamefont {Ye}},\
  }\href {\doibase 10.1103/PhysRevLett.70.3339} {\bibfield  {journal} {\bibinfo
   {journal} {Phys. Rev. Lett.}\ }\textbf {\bibinfo {volume} {70}},\ \bibinfo
  {pages} {3339} (\bibinfo {year} {1993})}\BibitemShut {NoStop}%
\bibitem [{\citenamefont {Kitaev}(2015)}]{Kitaev2015}%
  \BibitemOpen
  \bibfield  {author} {\bibinfo {author} {\bibfnamefont {A.}~\bibnamefont
  {Kitaev}},\ }\href {http://online.kitp.ucsb.edu/online/entangled15/}
  {\bibfield  {journal} {\bibinfo  {journal} {in KITP Strings Seminar and
  Entanglement 2015 Program}\ } (\bibinfo {year} {2015})}\BibitemShut {NoStop}%
\bibitem [{\citenamefont {Sachdev}(2015)}]{Sachdev2015}%
  \BibitemOpen
  \bibfield  {author} {\bibinfo {author} {\bibfnamefont {S.}~\bibnamefont
  {Sachdev}},\ }\href {\doibase 10.1103/PhysRevX.5.041025} {\bibfield
  {journal} {\bibinfo  {journal} {Phys. Rev. X}\ }\textbf {\bibinfo {volume}
  {5}},\ \bibinfo {pages} {041025} (\bibinfo {year} {2015})}\BibitemShut
  {NoStop}%
\bibitem [{\citenamefont {Maldacena}\ and\ \citenamefont
  {Stanford}(2016)}]{Maldacena2016}%
  \BibitemOpen
  \bibfield  {author} {\bibinfo {author} {\bibfnamefont {J.}~\bibnamefont
  {Maldacena}}\ and\ \bibinfo {author} {\bibfnamefont {D.}~\bibnamefont
  {Stanford}},\ }\href {\doibase 10.1103/PhysRevD.94.106002} {\bibfield
  {journal} {\bibinfo  {journal} {Phys. Rev. D}\ }\textbf {\bibinfo {volume}
  {94}},\ \bibinfo {pages} {106002} (\bibinfo {year} {2016})}\BibitemShut
  {NoStop}%
\bibitem [{\citenamefont {Sachdev}(2010)}]{PhysRevLett.105.151602}%
  \BibitemOpen
  \bibfield  {author} {\bibinfo {author} {\bibfnamefont {S.}~\bibnamefont
  {Sachdev}},\ }\href {\doibase 10.1103/PhysRevLett.105.151602} {\bibfield
  {journal} {\bibinfo  {journal} {Phys. Rev. Lett.}\ }\textbf {\bibinfo
  {volume} {105}},\ \bibinfo {pages} {151602} (\bibinfo {year}
  {2010})}\BibitemShut {NoStop}%
\bibitem [{\citenamefont {Maldacena}\ \emph {et~al.}(2016)\citenamefont
  {Maldacena}, \citenamefont {Shenker},\ and\ \citenamefont
  {Stanford}}]{boundonchaos}%
  \BibitemOpen
  \bibfield  {author} {\bibinfo {author} {\bibfnamefont {J.}~\bibnamefont
  {Maldacena}}, \bibinfo {author} {\bibfnamefont {S.~H.}\ \bibnamefont
  {Shenker}}, \ and\ \bibinfo {author} {\bibfnamefont {D.}~\bibnamefont
  {Stanford}},\ }\href {\doibase 10.1007/JHEP08(2016)106} {\bibfield  {journal}
  {\bibinfo  {journal} {Journal of High Energy Physics}\ }\textbf {\bibinfo
  {volume} {2016}},\ \bibinfo {pages} {106} (\bibinfo {year}
  {2016})}\BibitemShut {NoStop}%
\bibitem [{\citenamefont {You}\ \emph {et~al.}(2017)\citenamefont {You},
  \citenamefont {Ludwig},\ and\ \citenamefont {Xu}}]{Xu2016}%
  \BibitemOpen
  \bibfield  {author} {\bibinfo {author} {\bibfnamefont {Y.-Z.}\ \bibnamefont
  {You}}, \bibinfo {author} {\bibfnamefont {A.~W.~W.}\ \bibnamefont {Ludwig}},
  \ and\ \bibinfo {author} {\bibfnamefont {C.}~\bibnamefont {Xu}},\ }\href
  {\doibase 10.1103/PhysRevB.95.115150} {\bibfield  {journal} {\bibinfo
  {journal} {Phys. Rev. B}\ }\textbf {\bibinfo {volume} {95}},\ \bibinfo
  {pages} {115150} (\bibinfo {year} {2017})}\BibitemShut {NoStop}%
\bibitem [{\citenamefont {Polchinski}\ and\ \citenamefont
  {Rosenhaus}(2016)}]{Polchinski2016}%
  \BibitemOpen
  \bibfield  {author} {\bibinfo {author} {\bibfnamefont {J.}~\bibnamefont
  {Polchinski}}\ and\ \bibinfo {author} {\bibfnamefont {V.}~\bibnamefont
  {Rosenhaus}},\ }\href {\doibase 10.1007/JHEP04(2016)001} {\bibfield
  {journal} {\bibinfo  {journal} {Journal of High Energy Physics}\ }\textbf
  {\bibinfo {volume} {2016}},\ \bibinfo {pages} {1} (\bibinfo {year}
  {2016})}\BibitemShut {NoStop}%
\bibitem [{\citenamefont {Garc\'{\i}a-Garc\'{\i}a}\ and\ \citenamefont
  {Verbaarschot}(2016)}]{Verbaar2016}%
  \BibitemOpen
  \bibfield  {author} {\bibinfo {author} {\bibfnamefont {A.~M.}\ \bibnamefont
  {Garc\'{\i}a-Garc\'{\i}a}}\ and\ \bibinfo {author} {\bibfnamefont {J.~J.~M.}\
  \bibnamefont {Verbaarschot}},\ }\href {\doibase 10.1103/PhysRevD.94.126010}
  {\bibfield  {journal} {\bibinfo  {journal} {Phys. Rev. D}\ }\textbf {\bibinfo
  {volume} {94}},\ \bibinfo {pages} {126010} (\bibinfo {year}
  {2016})}\BibitemShut {NoStop}%
\bibitem [{\citenamefont {Fu}\ \emph {et~al.}(2017)\citenamefont {Fu},
  \citenamefont {Gaiotto}, \citenamefont {Maldacena},\ and\ \citenamefont
  {Sachdev}}]{Fu2016}%
  \BibitemOpen
  \bibfield  {author} {\bibinfo {author} {\bibfnamefont {W.}~\bibnamefont
  {Fu}}, \bibinfo {author} {\bibfnamefont {D.}~\bibnamefont {Gaiotto}},
  \bibinfo {author} {\bibfnamefont {J.}~\bibnamefont {Maldacena}}, \ and\
  \bibinfo {author} {\bibfnamefont {S.}~\bibnamefont {Sachdev}},\ }\href
  {\doibase 10.1103/PhysRevD.95.026009} {\bibfield  {journal} {\bibinfo
  {journal} {Phys. Rev. D}\ }\textbf {\bibinfo {volume} {95}},\ \bibinfo
  {pages} {026009} (\bibinfo {year} {2017})}\BibitemShut {NoStop}%
\bibitem [{\citenamefont {Banerjee}\ and\ \citenamefont
  {Altman}(2017)}]{Altman2016}%
  \BibitemOpen
  \bibfield  {author} {\bibinfo {author} {\bibfnamefont {S.}~\bibnamefont
  {Banerjee}}\ and\ \bibinfo {author} {\bibfnamefont {E.}~\bibnamefont
  {Altman}},\ }\href {\doibase 10.1103/PhysRevB.95.134302} {\bibfield
  {journal} {\bibinfo  {journal} {Phys. Rev. B}\ }\textbf {\bibinfo {volume}
  {95}},\ \bibinfo {pages} {134302} (\bibinfo {year} {2017})}\BibitemShut
  {NoStop}%
\bibitem [{\citenamefont {Gu}\ \emph {et~al.}(2017)\citenamefont {Gu},
  \citenamefont {Qi},\ and\ \citenamefont {Stanford}}]{Gu2016}%
  \BibitemOpen
  \bibfield  {author} {\bibinfo {author} {\bibfnamefont {Y.}~\bibnamefont
  {Gu}}, \bibinfo {author} {\bibfnamefont {X.-L.}\ \bibnamefont {Qi}}, \ and\
  \bibinfo {author} {\bibfnamefont {D.}~\bibnamefont {Stanford}},\ }\href
  {\doibase 10.1007/JHEP05(2017)125} {\bibfield  {journal} {\bibinfo  {journal}
  {Journal of High Energy Physics}\ }\textbf {\bibinfo {volume} {2017}},\
  \bibinfo {pages} {125} (\bibinfo {year} {2017})}\BibitemShut {NoStop}%
\bibitem [{\citenamefont {Berkooz}\ \emph {et~al.}(2017)\citenamefont
  {Berkooz}, \citenamefont {Narayan}, \citenamefont {Rozali},\ and\
  \citenamefont {Sim{\'o}n}}]{Berkooz2016}%
  \BibitemOpen
  \bibfield  {author} {\bibinfo {author} {\bibfnamefont {M.}~\bibnamefont
  {Berkooz}}, \bibinfo {author} {\bibfnamefont {P.}~\bibnamefont {Narayan}},
  \bibinfo {author} {\bibfnamefont {M.}~\bibnamefont {Rozali}}, \ and\ \bibinfo
  {author} {\bibfnamefont {J.}~\bibnamefont {Sim{\'o}n}},\ }\href {\doibase
  10.1007/JHEP01(2017)138} {\bibfield  {journal} {\bibinfo  {journal} {Journal
  of High Energy Physics}\ }\textbf {\bibinfo {volume} {2017}},\ \bibinfo
  {pages} {138} (\bibinfo {year} {2017})}\BibitemShut {NoStop}%
\bibitem [{\citenamefont {Hosur}\ \emph {et~al.}(2016)\citenamefont {Hosur},
  \citenamefont {Qi}, \citenamefont {Roberts},\ and\ \citenamefont
  {Yoshida}}]{Hosur2016}%
  \BibitemOpen
  \bibfield  {author} {\bibinfo {author} {\bibfnamefont {P.}~\bibnamefont
  {Hosur}}, \bibinfo {author} {\bibfnamefont {X.-L.}\ \bibnamefont {Qi}},
  \bibinfo {author} {\bibfnamefont {D.~A.}\ \bibnamefont {Roberts}}, \ and\
  \bibinfo {author} {\bibfnamefont {B.}~\bibnamefont {Yoshida}},\ }\href
  {\doibase 10.1007/JHEP02(2016)004} {\bibfield  {journal} {\bibinfo  {journal}
  {Journal of High Energy Physics}\ }\textbf {\bibinfo {volume} {2016}},\
  \bibinfo {pages} {4} (\bibinfo {year} {2016})}\BibitemShut {NoStop}%
\bibitem [{\citenamefont {Liu}\ \emph {et~al.}(2018)\citenamefont {Liu},
  \citenamefont {Chen},\ and\ \citenamefont {Balents}}]{Liu2017}%
  \BibitemOpen
  \bibfield  {author} {\bibinfo {author} {\bibfnamefont {C.}~\bibnamefont
  {Liu}}, \bibinfo {author} {\bibfnamefont {X.}~\bibnamefont {Chen}}, \ and\
  \bibinfo {author} {\bibfnamefont {L.}~\bibnamefont {Balents}},\ }\href
  {\doibase 10.1103/PhysRevB.97.245126} {\bibfield  {journal} {\bibinfo
  {journal} {Phys. Rev. B}\ }\textbf {\bibinfo {volume} {97}},\ \bibinfo
  {pages} {245126} (\bibinfo {year} {2018})}\BibitemShut {NoStop}%
\bibitem [{\citenamefont {{Huang}}\ and\ \citenamefont
  {{Gu}}(2017)}]{Huang2017}%
  \BibitemOpen
  \bibfield  {author} {\bibinfo {author} {\bibfnamefont {Y.}~\bibnamefont
  {{Huang}}}\ and\ \bibinfo {author} {\bibfnamefont {Y.}~\bibnamefont {{Gu}}},\
  }\href@noop {} {\bibfield  {journal} {\bibinfo  {journal} {ArXiv e-prints}\ }
  (\bibinfo {year} {2017})},\ \Eprint {http://arxiv.org/abs/1709.09160}
  {arXiv:1709.09160 [hep-th]} \BibitemShut {NoStop}%
\bibitem [{\citenamefont {Song}\ \emph {et~al.}(2017)\citenamefont {Song},
  \citenamefont {Jian},\ and\ \citenamefont {Balents}}]{Balents2017}%
  \BibitemOpen
  \bibfield  {author} {\bibinfo {author} {\bibfnamefont {X.-Y.}\ \bibnamefont
  {Song}}, \bibinfo {author} {\bibfnamefont {C.-M.}\ \bibnamefont {Jian}}, \
  and\ \bibinfo {author} {\bibfnamefont {L.}~\bibnamefont {Balents}},\ }\href
  {\doibase 10.1103/PhysRevLett.119.216601} {\bibfield  {journal} {\bibinfo
  {journal} {Phys. Rev. Lett.}\ }\textbf {\bibinfo {volume} {119}},\ \bibinfo
  {pages} {216601} (\bibinfo {year} {2017})}\BibitemShut {NoStop}%
\bibitem [{\citenamefont {Bi}\ \emph {et~al.}(2017)\citenamefont {Bi},
  \citenamefont {Jian}, \citenamefont {You}, \citenamefont {Pawlak},\ and\
  \citenamefont {Xu}}]{Bi2017}%
  \BibitemOpen
  \bibfield  {author} {\bibinfo {author} {\bibfnamefont {Z.}~\bibnamefont
  {Bi}}, \bibinfo {author} {\bibfnamefont {C.-M.}\ \bibnamefont {Jian}},
  \bibinfo {author} {\bibfnamefont {Y.-Z.}\ \bibnamefont {You}}, \bibinfo
  {author} {\bibfnamefont {K.~A.}\ \bibnamefont {Pawlak}}, \ and\ \bibinfo
  {author} {\bibfnamefont {C.}~\bibnamefont {Xu}},\ }\href {\doibase
  10.1103/PhysRevB.95.205105} {\bibfield  {journal} {\bibinfo  {journal} {Phys.
  Rev. B}\ }\textbf {\bibinfo {volume} {95}},\ \bibinfo {pages} {205105}
  (\bibinfo {year} {2017})}\BibitemShut {NoStop}%
\bibitem [{\citenamefont {Lantagne-Hurtubise}\ \emph
  {et~al.}(2018)\citenamefont {Lantagne-Hurtubise}, \citenamefont {Li},\ and\
  \citenamefont {Franz}}]{Lantagne2018}%
  \BibitemOpen
  \bibfield  {author} {\bibinfo {author} {\bibfnamefont {E.}~\bibnamefont
  {Lantagne-Hurtubise}}, \bibinfo {author} {\bibfnamefont {C.}~\bibnamefont
  {Li}}, \ and\ \bibinfo {author} {\bibfnamefont {M.}~\bibnamefont {Franz}},\
  }\href {\doibase 10.1103/PhysRevB.97.235124} {\bibfield  {journal} {\bibinfo
  {journal} {Phys. Rev. B}\ }\textbf {\bibinfo {volume} {97}},\ \bibinfo
  {pages} {235124} (\bibinfo {year} {2018})}\BibitemShut {NoStop}%
\bibitem [{\citenamefont {Franz}\ and\ \citenamefont
  {Rozali}()}]{marcelmoshereview}%
  \BibitemOpen
  \bibfield  {author} {\bibinfo {author} {\bibfnamefont {M.}~\bibnamefont
  {Franz}}\ and\ \bibinfo {author} {\bibfnamefont {M.}~\bibnamefont {Rozali}},\
  }\href@noop {} {\bibfield  {journal} {\bibinfo  {journal} {ArXiv e-prints}\
  }}\Eprint {http://arxiv.org/abs/1808.00541} {arXiv:1808.00541
  [cond-mat.str-el]} \BibitemShut {NoStop}%
\bibitem [{\citenamefont {Danshita}\ \emph {et~al.}(2017)\citenamefont
  {Danshita}, \citenamefont {Hanada},\ and\ \citenamefont
  {Tezuka}}]{ultracoldrealization}%
  \BibitemOpen
  \bibfield  {author} {\bibinfo {author} {\bibfnamefont {I.}~\bibnamefont
  {Danshita}}, \bibinfo {author} {\bibfnamefont {M.}~\bibnamefont {Hanada}}, \
  and\ \bibinfo {author} {\bibfnamefont {M.}~\bibnamefont {Tezuka}},\ }\href
  {\doibase 10.1093/ptep/ptx108} {\bibfield  {journal} {\bibinfo  {journal}
  {Progress of Theoretical and Experimental Physics}\ }\textbf {\bibinfo
  {volume} {2017}},\ \bibinfo {pages} {083I01} (\bibinfo {year}
  {2017})}\BibitemShut {NoStop}%
\bibitem [{\citenamefont {Pikulin}\ and\ \citenamefont {Franz}(2017)}]{mzmsyk}%
  \BibitemOpen
  \bibfield  {author} {\bibinfo {author} {\bibfnamefont {D.~I.}\ \bibnamefont
  {Pikulin}}\ and\ \bibinfo {author} {\bibfnamefont {M.}~\bibnamefont
  {Franz}},\ }\href {\doibase 10.1103/PhysRevX.7.031006} {\bibfield  {journal}
  {\bibinfo  {journal} {Phys. Rev. X}\ }\textbf {\bibinfo {volume} {7}},\
  \bibinfo {pages} {031006} (\bibinfo {year} {2017})}\BibitemShut {NoStop}%
\bibitem [{\citenamefont {Chew}\ \emph {et~al.}(2017)\citenamefont {Chew},
  \citenamefont {Essin},\ and\ \citenamefont {Alicea}}]{majoranawiresyk}%
  \BibitemOpen
  \bibfield  {author} {\bibinfo {author} {\bibfnamefont {A.}~\bibnamefont
  {Chew}}, \bibinfo {author} {\bibfnamefont {A.}~\bibnamefont {Essin}}, \ and\
  \bibinfo {author} {\bibfnamefont {J.}~\bibnamefont {Alicea}},\ }\href
  {\doibase 10.1103/PhysRevB.96.121119} {\bibfield  {journal} {\bibinfo
  {journal} {Phys. Rev. B}\ }\textbf {\bibinfo {volume} {96}},\ \bibinfo
  {pages} {121119} (\bibinfo {year} {2017})}\BibitemShut {NoStop}%
\bibitem [{\citenamefont {Chen}\ \emph {et~al.}(2018)\citenamefont {Chen},
  \citenamefont {Ilan}, \citenamefont {de~Juan}, \citenamefont {Pikulin},\ and\
  \citenamefont {Franz}}]{Chen}%
  \BibitemOpen
  \bibfield  {author} {\bibinfo {author} {\bibfnamefont {A.}~\bibnamefont
  {Chen}}, \bibinfo {author} {\bibfnamefont {R.}~\bibnamefont {Ilan}}, \bibinfo
  {author} {\bibfnamefont {F.}~\bibnamefont {de~Juan}}, \bibinfo {author}
  {\bibfnamefont {D.~I.}\ \bibnamefont {Pikulin}}, \ and\ \bibinfo {author}
  {\bibfnamefont {M.}~\bibnamefont {Franz}},\ }\href {\doibase
  10.1103/PhysRevLett.121.036403} {\bibfield  {journal} {\bibinfo  {journal}
  {Phys. Rev. Lett}\ }\textbf {\bibinfo {volume} {121}},\ \bibinfo {pages}
  {036403} (\bibinfo {year} {2018})}\BibitemShut {NoStop}%
\bibitem [{\citenamefont {Parcollet}\ \emph {et~al.}(1998)\citenamefont
  {Parcollet}, \citenamefont {Georges}, \citenamefont {Kotliar},\ and\
  \citenamefont {Sengupta}}]{Parcollet_Kondo}%
  \BibitemOpen
  \bibfield  {author} {\bibinfo {author} {\bibfnamefont {O.}~\bibnamefont
  {Parcollet}}, \bibinfo {author} {\bibfnamefont {A.}~\bibnamefont {Georges}},
  \bibinfo {author} {\bibfnamefont {G.}~\bibnamefont {Kotliar}}, \ and\
  \bibinfo {author} {\bibfnamefont {A.}~\bibnamefont {Sengupta}},\ }\href
  {\doibase 10.1103/PhysRevLett.121.036403} {\bibfield  {journal} {\bibinfo
  {journal} {Phys. Rev. B}\ }\textbf {\bibinfo {volume} {58}},\ \bibinfo
  {pages} {3794} (\bibinfo {year} {1998})}\BibitemShut {NoStop}%
\bibitem [{\citenamefont {Knezdilov}\ \emph {et~al.}(2018)\citenamefont
  {Knezdilov}, \citenamefont {Hutasoit},\ and\ \citenamefont
  {Beenakker}}]{Beenakker2018}%
  \BibitemOpen
  \bibfield  {author} {\bibinfo {author} {\bibfnamefont {N.~V.}\ \bibnamefont
  {Knezdilov}}, \bibinfo {author} {\bibfnamefont {J.~A.}\ \bibnamefont
  {Hutasoit}}, \ and\ \bibinfo {author} {\bibfnamefont {C.~W.~J.}\ \bibnamefont
  {Beenakker}},\ }\href@noop {} {\bibfield  {journal} {\bibinfo  {journal}
  {ArXiv e-prints}\ } (\bibinfo {year} {2018})},\ \Eprint
  {http://arxiv.org/abs/1807.09099} {arXiv:1807.09099 [cond-mat.mes-hall]}
  \BibitemShut {NoStop}%
\bibitem [{\citenamefont {Affleck}\ and\ \citenamefont
  {Ludwig}(1991)}]{Ludwig}%
  \BibitemOpen
  \bibfield  {author} {\bibinfo {author} {\bibfnamefont {I.}~\bibnamefont
  {Affleck}}\ and\ \bibinfo {author} {\bibfnamefont {A.~W.~W.}\ \bibnamefont
  {Ludwig}},\ }\href@noop {} {\bibfield  {journal} {\bibinfo  {journal} {Nucl.
  Phys. B}\ }\textbf {\bibinfo {volume} {360}},\ \bibinfo {pages} {641}
  (\bibinfo {year} {1991})}\BibitemShut {NoStop}%
\bibitem [{\citenamefont {Meir}\ and\ \citenamefont {Wingreen}(1992)}]{Meir}%
  \BibitemOpen
  \bibfield  {author} {\bibinfo {author} {\bibfnamefont {Y.}~\bibnamefont
  {Meir}}\ and\ \bibinfo {author} {\bibfnamefont {N.~S.~S.}\ \bibnamefont
  {Wingreen}},\ }\href@noop {} {\bibfield  {journal} {\bibinfo  {journal}
  {Phys. Rev. Lett.}\ }\textbf {\bibinfo {volume} {68}},\ \bibinfo {pages}
  {2512} (\bibinfo {year} {1992})}\BibitemShut {NoStop}%
\bibitem [{\citenamefont {Mahan}(2000)}]{Mahan}%
  \BibitemOpen
  \bibfield  {author} {\bibinfo {author} {\bibfnamefont {G.~D.}\ \bibnamefont
  {Mahan}},\ }\href@noop {} {\emph {\bibinfo {title} {Many-particle systems}}}\
  (\bibinfo  {publisher} {Plenum, New York},\ \bibinfo {year}
  {2000})\BibitemShut {NoStop}%
\bibitem [{\citenamefont {Sachdev}(2001)}]{Sachdev_trans}%
  \BibitemOpen
  \bibfield  {author} {\bibinfo {author} {\bibfnamefont {S.}~\bibnamefont
  {Sachdev}},\ }\href@noop {} {\emph {\bibinfo {title} {Quantum phase
  transitions}}}\ (\bibinfo  {publisher} {Cambridge},\ \bibinfo {year}
  {2001})\BibitemShut {NoStop}%
\bibitem [{\citenamefont {Hewson}(1993)}]{Hewson}%
  \BibitemOpen
  \bibfield  {author} {\bibinfo {author} {\bibfnamefont {A.~C.}\ \bibnamefont
  {Hewson}},\ }\href@noop {} {\emph {\bibinfo {title} {The Kondo problem to
  heavy fermions}}}\ (\bibinfo  {publisher} {Cambridge},\ \bibinfo {year}
  {1993})\BibitemShut {NoStop}%
\bibitem [{\citenamefont {Haug}\ and\ \citenamefont {Jauho}(2008)}]{Haug}%
  \BibitemOpen
  \bibfield  {author} {\bibinfo {author} {\bibfnamefont {H.~J.~W.}\
  \bibnamefont {Haug}}\ and\ \bibinfo {author} {\bibfnamefont {A.-P.}\
  \bibnamefont {Jauho}},\ }\href@noop {} {\emph {\bibinfo {title} {Quantum
  Kinetics in Transport and Optics of Semiconductors}}}\ (\bibinfo  {publisher}
  {Springer-Verlag, Berlin},\ \bibinfo {year} {2008})\BibitemShut {NoStop}%
\bibitem [{\citenamefont {Affleck}\ and\ \citenamefont
  {Ludwig}(1993)}]{Affleck}%
  \BibitemOpen
  \bibfield  {author} {\bibinfo {author} {\bibfnamefont {I.}~\bibnamefont
  {Affleck}}\ and\ \bibinfo {author} {\bibfnamefont {A.~W.~W.}\ \bibnamefont
  {Ludwig}},\ }\href@noop {} {\bibfield  {journal} {\bibinfo  {journal} {Phys.
  Rev. B}\ }\textbf {\bibinfo {volume} {48}},\ \bibinfo {pages} {7297}
  (\bibinfo {year} {1993})}\BibitemShut {NoStop}%
\bibitem [{\citenamefont {Ralph}\ \emph {et~al.}(1994)\citenamefont {Ralph},
  \citenamefont {Ludwig}, \citenamefont {Delft},\ and\ \citenamefont
  {Buhrman}}]{Ralph}%
  \BibitemOpen
  \bibfield  {author} {\bibinfo {author} {\bibfnamefont {D.~C.}\ \bibnamefont
  {Ralph}}, \bibinfo {author} {\bibfnamefont {A.~W.~W.}\ \bibnamefont
  {Ludwig}}, \bibinfo {author} {\bibfnamefont {J.~v.}\ \bibnamefont {Delft}}, \
  and\ \bibinfo {author} {\bibfnamefont {R.~A.}\ \bibnamefont {Buhrman}},\
  }\href@noop {} {\bibfield  {journal} {\bibinfo  {journal} {Phys. Rev. Lett.}\
  }\textbf {\bibinfo {volume} {72}},\ \bibinfo {pages} {1064} (\bibinfo {year}
  {1994})}\BibitemShut {NoStop}%
\bibitem [{\citenamefont {Delft}\ \emph {et~al.}(1999)\citenamefont {Delft},
  \citenamefont {Ludwig},\ and\ \citenamefont {Ambegaokar}}]{von_Delft}%
  \BibitemOpen
  \bibfield  {author} {\bibinfo {author} {\bibfnamefont {J.~v.}\ \bibnamefont
  {Delft}}, \bibinfo {author} {\bibfnamefont {A.~W.~W.}\ \bibnamefont
  {Ludwig}}, \ and\ \bibinfo {author} {\bibfnamefont {V.}~\bibnamefont
  {Ambegaokar}},\ }\href@noop {} {\bibfield  {journal} {\bibinfo  {journal}
  {Ann. Phys.}\ }\textbf {\bibinfo {volume} {273}},\ \bibinfo {pages} {175}
  (\bibinfo {year} {1999})}\BibitemShut {NoStop}%
\bibitem [{\citenamefont {Hettler}\ \emph
  {et~al.}(1994{\natexlab{a}})\citenamefont {Hettler}, \citenamefont {Kroha},\
  and\ \citenamefont {Hershfield}}]{Hettler_1994}%
  \BibitemOpen
  \bibfield  {author} {\bibinfo {author} {\bibfnamefont {M.~H.}\ \bibnamefont
  {Hettler}}, \bibinfo {author} {\bibfnamefont {J.}~\bibnamefont {Kroha}}, \
  and\ \bibinfo {author} {\bibfnamefont {S.}~\bibnamefont {Hershfield}},\
  }\href@noop {} {\bibfield  {journal} {\bibinfo  {journal} {Phys. Rev. Lett.}\
  }\textbf {\bibinfo {volume} {73}},\ \bibinfo {pages} {1967} (\bibinfo {year}
  {1994}{\natexlab{a}})}\BibitemShut {NoStop}%
\bibitem [{\citenamefont {Hettler}\ \emph
  {et~al.}(1994{\natexlab{b}})\citenamefont {Hettler}, \citenamefont {Kroha},\
  and\ \citenamefont {Hershfield}}]{Hettler_1998}%
  \BibitemOpen
  \bibfield  {author} {\bibinfo {author} {\bibfnamefont {M.~H.}\ \bibnamefont
  {Hettler}}, \bibinfo {author} {\bibfnamefont {J.}~\bibnamefont {Kroha}}, \
  and\ \bibinfo {author} {\bibfnamefont {S.}~\bibnamefont {Hershfield}},\
  }\href@noop {} {\bibfield  {journal} {\bibinfo  {journal} {Phys. Rev. B}\
  }\textbf {\bibinfo {volume} {58}},\ \bibinfo {pages} {5649} (\bibinfo {year}
  {1994}{\natexlab{b}})}\BibitemShut {NoStop}%
\bibitem [{\citenamefont {Potok}\ \emph {et~al.}(2007)\citenamefont {Potok},
  \citenamefont {Rau}, \citenamefont {Shtrikman}, \citenamefont {Oreg},\ and\
  \citenamefont {Goldhaber-Gordon}}]{Potok}%
  \BibitemOpen
  \bibfield  {author} {\bibinfo {author} {\bibfnamefont {R.~M.}\ \bibnamefont
  {Potok}}, \bibinfo {author} {\bibfnamefont {I.~G.}\ \bibnamefont {Rau}},
  \bibinfo {author} {\bibfnamefont {H.}~\bibnamefont {Shtrikman}}, \bibinfo
  {author} {\bibfnamefont {Y.}~\bibnamefont {Oreg}}, \ and\ \bibinfo {author}
  {\bibfnamefont {D.}~\bibnamefont {Goldhaber-Gordon}},\ }\href@noop {}
  {\bibfield  {journal} {\bibinfo  {journal} {Nature}\ }\textbf {\bibinfo
  {volume} {446}},\ \bibinfo {pages} {167} (\bibinfo {year}
  {2007})}\BibitemShut {NoStop}%
\bibitem [{\citenamefont {Weiss}(2008)}]{Weiss}%
  \BibitemOpen
  \bibfield  {author} {\bibinfo {author} {\bibfnamefont {U.}~\bibnamefont
  {Weiss}},\ }\href@noop {} {\emph {\bibinfo {title} {Quantum Dissipative
  Systems}}}\ (\bibinfo  {publisher} {World Scientific},\ \bibinfo {address}
  {Singapore},\ \bibinfo {year} {2008})\BibitemShut {NoStop}%
\bibitem [{\citenamefont {Negele}\ and\ \citenamefont {Orland}(1998)}]{Negele}%
  \BibitemOpen
  \bibfield  {author} {\bibinfo {author} {\bibfnamefont {J.~W.}\ \bibnamefont
  {Negele}}\ and\ \bibinfo {author} {\bibfnamefont {H.}~\bibnamefont
  {Orland}},\ }\href@noop {} {\emph {\bibinfo {title} {Quantum Many-Particle
  Systems}}}\ (\bibinfo  {publisher} {Westview},\ \bibinfo {address}
  {Westview},\ \bibinfo {year} {1998})\ p.~\bibinfo {pages} {51}\BibitemShut
  {NoStop}%
\bibitem [{\citenamefont {Parcollet}\ and\ \citenamefont
  {Georges}(1999)}]{Parcollet_disorder}%
  \BibitemOpen
  \bibfield  {author} {\bibinfo {author} {\bibfnamefont {O.}~\bibnamefont
  {Parcollet}}\ and\ \bibinfo {author} {\bibfnamefont {A.}~\bibnamefont
  {Georges}},\ }\href@noop {} {\bibfield  {journal} {\bibinfo  {journal} {Phys.
  Rev. B}\ }\textbf {\bibinfo {volume} {59}},\ \bibinfo {pages} {5341}
  (\bibinfo {year} {1999})}\BibitemShut {NoStop}%
\bibitem [{\citenamefont {Gradshteyn}\ and\ \citenamefont
  {Ryzhik}(2000)}]{Gradshteyn}%
  \BibitemOpen
  \bibfield  {author} {\bibinfo {author} {\bibfnamefont {I.~S.}\ \bibnamefont
  {Gradshteyn}}\ and\ \bibinfo {author} {\bibfnamefont {I.~M.}\ \bibnamefont
  {Ryzhik}},\ }\href@noop {} {\emph {\bibinfo {title} {Table of Integrals,
  Series, and Products}}},\ edited by\ \bibinfo {editor} {\bibfnamefont
  {A.}~\bibnamefont {Jeffrey}}\ and\ \bibinfo {editor} {\bibfnamefont
  {D.}~\bibnamefont {Zwilinger}}\ (\bibinfo  {publisher} {Academic},\ \bibinfo
  {year} {2000})\BibitemShut {NoStop}%
\bibitem [{\citenamefont {Abramowitz}\ and\ \citenamefont
  {Stegun}(1972)}]{Abramowitz}%
  \BibitemOpen
  \bibinfo {editor} {\bibfnamefont {M.}~\bibnamefont {Abramowitz}}\ and\
  \bibinfo {editor} {\bibfnamefont {I.~A.}\ \bibnamefont {Stegun}},\ eds.,\
  \href@noop {} {\emph {\bibinfo {title} {Handbook of Mathematical
  Functions}}}\ (\bibinfo  {publisher} {National Bureau Of Standards},\
  \bibinfo {year} {1972})\BibitemShut {NoStop}%
\bibitem [{\citenamefont {Rammer}\ and\ \citenamefont {Smith}(1986)}]{Rammer}%
  \BibitemOpen
  \bibfield  {author} {\bibinfo {author} {\bibfnamefont {J.}~\bibnamefont
  {Rammer}}\ and\ \bibinfo {author} {\bibfnamefont {H.}~\bibnamefont {Smith}},\
  }\href@noop {} {\bibfield  {journal} {\bibinfo  {journal} {Rev. Mod. Phys.}\
  }\textbf {\bibinfo {volume} {58}},\ \bibinfo {pages} {323} (\bibinfo {year}
  {1986})}\BibitemShut {NoStop}%
\bibitem [{\citenamefont {Kamenev}(2004)}]{Kamenev}%
  \BibitemOpen
  \bibfield  {author} {\bibinfo {author} {\bibfnamefont {A.}~\bibnamefont
  {Kamenev}},\ }in\ \href@noop {} {\emph {\bibinfo {booktitle} {Nanophysics:
  Coherence and Transport}}},\ Vol.~\bibinfo {volume} {81},\ \bibinfo {editor}
  {edited by\ \bibinfo {editor} {\bibfnamefont {H.}~\bibnamefont {Bouchiat}},
  \bibinfo {editor} {\bibfnamefont {Y.}~\bibnamefont {Gefen}}, \bibinfo
  {editor} {\bibfnamefont {S.}~\bibnamefont {Gu\'{e}ron}}, \bibinfo {editor}
  {\bibfnamefont {G.}~\bibnamefont {Montambaux}}, \ and\ \bibinfo {editor}
  {\bibfnamefont {J.}~\bibnamefont {Dalibard}}},\ \bibinfo {organization}
  {Lecture Notes of the Les Houches Summer School}\ (\bibinfo  {publisher}
  {Elsevier},\ \bibinfo {year} {2004})\ p.\ \bibinfo {pages} {177}\BibitemShut
  {NoStop}%
\bibitem [{\citenamefont {Vanderplas}(2018)}]{nfft}%
  \BibitemOpen
  \bibfield  {author} {\bibinfo {author} {\bibfnamefont {J.}~\bibnamefont
  {Vanderplas}},\ }\href@noop {} {\bibfield  {journal} {\bibinfo  {journal}
  {non-equispaced fast Fourier transform for Python}\ } (\bibinfo {year}
  {2018})},\ \Eprint {http://arxiv.org/abs/https://github.com/jakevdp/nfft}
  {URL:https://github.com/jakevdp/nfft} \BibitemShut {NoStop}%
\bibitem [{\citenamefont {Langreth}(1976)}]{Langreth}%
  \BibitemOpen
  \bibfield  {author} {\bibinfo {author} {\bibfnamefont {D.~C.}\ \bibnamefont
  {Langreth}},\ }in\ \href@noop {} {\emph {\bibinfo {booktitle} {Linear and
  Nonlinear Electron Transport in Solids}}},\ \bibinfo {series} {8},
  Vol.~\bibinfo {volume} {17},\ \bibinfo {editor} {edited by\ \bibinfo {editor}
  {\bibfnamefont {J.~T.}\ \bibnamefont {Devreese}}\ and\ \bibinfo {editor}
  {\bibfnamefont {E.}~\bibnamefont {van Doren}}},\ \bibinfo {organization}
  {NATO Advanced Study Institute}\ (\bibinfo  {publisher} {Springer},\ \bibinfo
  {address} {New York},\ \bibinfo {year} {1976})\ p.~\bibinfo {pages}
  {9}\BibitemShut {NoStop}%
\bibitem [{\citenamefont {Wingree}\ and\ \citenamefont
  {Meir}(1994)}]{Meir_NCA}%
  \BibitemOpen
  \bibfield  {author} {\bibinfo {author} {\bibfnamefont {N.~S.}\ \bibnamefont
  {Wingree}}\ and\ \bibinfo {author} {\bibfnamefont {Y.}~\bibnamefont {Meir}},\
  }\href@noop {} {\bibfield  {journal} {\bibinfo  {journal} {Phys. Rev. B}\
  }\textbf {\bibinfo {volume} {49}},\ \bibinfo {pages} {11040} (\bibinfo {year}
  {1994})}\BibitemShut {NoStop}%
\bibitem [{\citenamefont {Hershfield}\ \emph {et~al.}(1992)\citenamefont
  {Hershfield}, \citenamefont {Davies},\ and\ \citenamefont
  {Wilkins}}]{Hershfield}%
  \BibitemOpen
  \bibfield  {author} {\bibinfo {author} {\bibfnamefont {S.}~\bibnamefont
  {Hershfield}}, \bibinfo {author} {\bibfnamefont {J.~H.}\ \bibnamefont
  {Davies}}, \ and\ \bibinfo {author} {\bibfnamefont {J.}~\bibnamefont
  {Wilkins}},\ }\href@noop {} {\bibfield  {journal} {\bibinfo  {journal} {Phys.
  Rev. B}\ }\textbf {\bibinfo {volume} {46}},\ \bibinfo {pages} {7046}
  (\bibinfo {year} {1992})}\BibitemShut {NoStop}%
\bibitem [{\citenamefont {Kadanoff}\ and\ \citenamefont
  {Baym}(1962)}]{Baym_book}%
  \BibitemOpen
  \bibfield  {author} {\bibinfo {author} {\bibfnamefont {L.~P.}\ \bibnamefont
  {Kadanoff}}\ and\ \bibinfo {author} {\bibfnamefont {G.}~\bibnamefont
  {Baym}},\ }\href@noop {} {\emph {\bibinfo {title} {Quantum Statistical
  Mechanics}}}\ (\bibinfo  {publisher} {W. A. Benjamin},\ \bibinfo {address}
  {New York},\ \bibinfo {year} {1962})\BibitemShut {NoStop}%
\bibitem [{\citenamefont {Giamarchi}(2003)}]{Giamarchi}%
  \BibitemOpen
  \bibfield  {author} {\bibinfo {author} {\bibfnamefont {T.}~\bibnamefont
  {Giamarchi}},\ }in\ \href@noop {} {\emph {\bibinfo {booktitle} {Quantum
  Physics in One Dimension}}}\ (\bibinfo  {publisher} {Clarendon},\ \bibinfo
  {address} {Oxford},\ \bibinfo {year} {2003})\ p.\ \bibinfo {pages}
  {306}\BibitemShut {NoStop}%
\end{thebibliography}%


\appendix

\section{Linear response}
\label{Sec:Appn_lnr}

\subsection{Tunneling conductance in the saddle-point approximation}

We calculate the tunneling current as a function of 
applied oscillating potentials on the lead (end points) included via the terms   

\enni \begin{align}
H \rightarrow & H + H_{U, L} + H_{U, R} \\
H_{U, L/R} = & \pm \left( \frac{eU}{2} \right) \cos(\omega_{0}t) \sum_{\alpha} \psi^{\dagger}_{\alpha, L/R} \psi_{\alpha, L/R},
\label{Eq:bias}
\end{align}

\enni where $U$ is the amplitude of the scalar potential and $H$ is the Hamiltonian for 
the junction (Eq.~\eqref{Eq:Model}).

We eliminate the scalar potential via a temporal gauge transformation  which introduces time-dependent phases (Sec.3.4 of Ref.~\onlinecite{Weiss})~

\enni \begin{align}
\phi(t) = & \left(\frac{eU}{2\hbar} \right) \frac{\sin(\omega_{0}t)}{\omega_{0}} \\
\hbar \frac{d \phi}{dt} = & \left(\frac{eU}{2}\right) \cos(\omega_{0}t). 
\end{align}

\enni This amounts to the  gauge transformation $H_{U, R/L} \rightarrow  0 $ and 

\enni \begin{align} 
H_{L/RD} \rightarrow & H_{LR/D}(t) = \sum_{i \alpha} \frac{V_{ i \alpha}}{(NM)^{1/4}} c^{\dagger}_{i} \psi_{L/R \alpha}e^{ \mp i\phi(t)}  \notag \\
& + \sum_{i \alpha} \frac{V^{*}_{i \alpha}}{(NM)^{1/4}} \psi^{\dagger}_{L/R \alpha}e^{\pm i \phi(t) } c_{i}.
\end{align} 

\enni  We remind the reader that the  
tunneling coefficients $V_{L,R}$ connecting left/right lead and dot are chosen to be complex, random, Gaussian-distributed 
variables of identical variance $V^{2}$. As such, we suppress $L/R$ indices.
 We expand the coupling between left/right lead and dot to linear order in the phase 

\enni \begin{align}
 H_{L/RD}(t) \approx & \sum_{i \alpha} \frac{V_{ i \alpha}}{(NM)^{1/4}} c^{\dagger}_{i} \psi_{L/R \alpha}  + \sum_{i \alpha} \frac{V^{*}_{ i \alpha}}{(NM)^{1/4}} \psi^{\dagger}_{L/R \alpha} c_{i} \notag \notag \\
& \mp i \frac{eU \sin(\omega_{0}t)}{2 \hbar \omega_{0}} \sum_{i \alpha} \frac{V_{ i \alpha}}{(NM)^{1/4}} c^{\dagger}_{i} \psi_{L/R \alpha} \notag \\
& \pm i \frac{eU \sin(\omega_{0}t)}{2\hbar \omega_{0}} \sum_{i \alpha} \frac{V^{*}_{i \alpha}}{(NM)^{1/4}} \psi^{\dagger}_{L/R \alpha} c_{i}  \notag \\
= & H_{L/RD}(U =0 ) \pm A(t) I_{L/RD}.
\end{align}

\enni The currents \emph{out of left and right leads} are obtained from $\braket{ I_{L/R} }= (ie/ \hbar) \braket{ \left[ N_{L/R}, H \right] }$

\enni \begin{align}
\braket{ I_{L/R} } 
= & \frac{ie}{\hbar} \sum_{i, \alpha} \left\{ \frac{V_{Li\alpha}}{(NM)^{1/4}} \braket{\psi^{\dagger}_{L\alpha} c_{i}} - \frac{V^{*}_{Li\alpha}}{(NM)^{1/4}} \braket{c^{\dagger}_{i} \psi_{L \alpha}} \right\},
\label{Eq:J_L}
\end{align}

\enni where $N_{L/R}$ is the total number operator for left and right leads, respectively, and 

\enni \begin{align}
A(t) = \frac{U \sin(\omega_{0}t)}{2 \omega_{0}}.
\end{align}

Following the standard linear-response formalism~\cite{Negele}, the current is 

\enni \begin{align}
\overline{\braket{I_{L/R}(t)}}  
= &   \int^{\infty}_{-\infty} dt' A(t') C^{R}(t-t'),
\end{align}

\enni where we defined the disorder-averaged, retarded current-current correlator 

\enni \begin{align}
C^{R}(t-t') = -i \theta(t-t') \overline{\braket{\left[ I_{L/R}(t), I_{L/R}(t') \right]}}.
\end{align}
 
\enni After a Fourier transform, we obtain 

\enni \begin{align} \label{readofftheconductance}
\overline{\braket{I_{L/R}(t)}} = & \frac{\text{Im}C^{R}( \omega_{0})}{ \omega_{0}}  \left( \pm \frac{U}{2} \cos(\omega_{0}t) \right) \notag \\ 
& - i \frac{\text{Re}C^{R}( \omega_{0})}{  \omega_{0}} \left( \pm \frac{U}{2} \sin(\omega_{0}t) \right),
\end{align}

\enni where $\text{Im}C^{R}$ and $\text{Re}C^{R}$ are odd and even functions, respectively. 

We identify the tunneling conductance 

\enni \begin{align}
G(\omega, T, \mu, p) = \frac{\text{Im}C^{R}(\omega)}{ 2\omega}.
\end{align}

\enni Note the additional factor of $1/2$. Recall that we assume symmetric leads implying equal conductance for the left and right junctions. Furthermore, $U$ is the total potential difference between the two leads, as opposed to $U/2$ across each left/right junction. The factor of $1/2$ then yields the conductance of the entire system. 

The retarded, disorder-averaged, current-current correlator $C^{R}$ can be determined by considering its 
imaginary time-ordered analogue:

\begin{widetext}

\enni \begin{align}
 C^{T}(\tau - \tau')  
= & \frac{e^{2}}{\hbar^{2}(NM)^{1/4} } \sum_{i, j, \alpha, \beta} \left\{  \overline{ V_{i\alpha} V^{*}_{j\beta} \bigg \langle T  \psi_{\alpha}^{\dagger}(\tau) c_{i}(\tau)  
 c^{\dagger}_{j}(\tau') \psi_{ \beta}(\tau')  \bigg \rangle }
+ \overline{V^{*}_{i\alpha} V_{j\beta} \bigg \langle  T  c^{\dagger}_{i}(\tau) \psi_{ \alpha}(\tau)\psi_{\beta}^{\dagger}(\tau') c_{j}(\tau') \bigg \rangle}
\right\} \\
= & \frac{- e^{2}V^{2}\sqrt{NM}}{\hbar^{2} }  \left\{ G_{c}(\tau - \tau') G_{\psi}(\tau'-\tau) + G_{\psi}(\tau - \tau') G_{c}(\tau' - \tau) \right\},
\end{align}

\enni where the bar indicates disorder-averaging. We also temporarily suppressed $L/R$ indices for clarity. In obtaining the last line, we used the fact that at saddle-point the lead and 
dot electrons are decoupled~\cite{Altman2016} with  single-particle Green's functions $G_{c}$ and $G_{\psi}$ which are diagonal in the
$\alpha$ and $i$ indices. Taking into account the definition of $V$ (Eq.~\eqref{Eq:Model}), the summation over the indices produces the overall factor of $\sqrt{NM}$.

Straightforward Fourier transform, change to the Lehmann representation, Matsubara frequency summation~\cite{Mahan}, and subsequent analytical 
continuation lead to the expression for the tunneling conductance

\enni \begin{align}\label{LRconductanceformula}
G(\omega, T, \mu, p) = &  \frac{ e^{2}V^{2}\sqrt{NM} \pi}{ 2\hbar}   \left\{ \int^{\infty}_{-\infty} d\epsilon \rho_{c}(\epsilon, T) 
\rho_{\psi}(\epsilon + \omega, T) \left[ \frac{ f(\epsilon) - f(\epsilon + \omega)}{\omega} \right] + (\psi \leftrightarrow c) \right\},
\end{align}

\enni where we introduced the spectral densities 
 $\rho_{c, \psi}(\omega, T) = - (1/\pi) \text{Im}G_{c,\pi}(\omega, T)$, and the standard Fermi-Dirac function $f(\omega)$. We also suppressed the explicit dependence 
of the spectral densities on $\mu$ and $p$ for simplicity.


\subsection{DC and zero-temperature limits}
\label{Sec:Appn_cndc}

In the DC limit the conductance is given by 

\enni \begin{align}
\lim_{\omega \rightarrow 0} G =  &  \frac{  e^{2}V^{2}\sqrt{NM} \pi }{2 k_{B} T  }  \int^{\infty}_{-\infty} d \epsilon \rho_{c}(\epsilon, T) 
\rho_{\psi}(\epsilon, T) \left( \frac{1}{ \cosh^{2}(\frac{\hbar \epsilon}{4k_{B}T})} \right). 
\label{Eq:dc_lmt}
\end{align}

\enni In the emergent conformal-invariant regime on the NFL side, $\hbar \omega, k_{B}T \ll k_{B} T^{*}$, we approximate the spectral densities by the forms given
in Appendix A of Ref.~\onlinecite{Altman2016}

\enni \begin{align}
\rho_{c}(\epsilon, T) =   \frac{B_{c}}{J} \left( \frac{k_{B}T}{J} \right)^{-1/2} \frac{e^{-\alpha/2}}{\sqrt{2} \pi^{2}} \cosh{ \left( \frac{\hbar \epsilon}{2 k_{B}T} \right)} \frac{\Gamma(\frac{1}{4} - i \frac{\alpha}{2\pi} + i \frac{\hbar \epsilon}{\pi 2k_{B}T }) \Gamma(\frac{1}{4} + i \frac{\alpha}{2\pi} - i \frac{\hbar \epsilon}{\pi 2k_{B}T }) }{\Gamma(\frac{1}{2})} 
\label{Eq:Scln_frm_c} \\
\rho_{\psi}(\epsilon) =  \frac{B_{\psi}}{J} \left( \frac{k_{B}T}{J} \right)^{1/2} \frac{e^{-\alpha/2} \sqrt{2}}{ \pi^{2}} \cosh{ \left( \frac{\hbar \epsilon}{2k_{B}T} \right)} \frac{\Gamma(\frac{3}{4} - i \frac{\alpha}{2\pi} + i \frac{\hbar \epsilon}{\pi 2k_{B}T }) \Gamma(\frac{3}{4} + i \frac{\alpha}{2\pi} - i \frac{\hbar \epsilon}{\pi 2k_{B}T }) }{\Gamma(\frac{3}{2})} 
\label{Eq:Scln_frm_psi},
\end{align}

\end{widetext}

\enni where the 
dimensionless constants are

\enni \begin{align}
B_{c} = \frac{\Lambda}{\sqrt{1+ e^{-2\alpha}}} &,~~ B_{\psi} =  \frac{\sqrt{p} \pi J^{2}}{2V^{2} \Lambda \sqrt{1+ e^{-2\alpha}}} \\
\Lambda = \left( \frac{(1-2p)}{\cos 2\theta} \right)^{\frac{1}{4}} &,~~ 
\alpha = \ln \left( \tan\left( \frac{\pi}{4} + \theta \right)\right).
\end{align}

\enni These
are obtained from the general solution in Ref.~\onlinecite{Altman2016} by rescaling $p$ and $V$ in the saddle-point equations~(Eq.~\ref{Eq:rscl}). Upon substituting the low-energy forms of the spectral densities in Eq.~\ref{Eq:dc_lmt}, the explicit dependence on temperature cancels and the expression reduces to a dimensionless integral over four Gamma functions. 
As discussed in Ref.~\onlinecite{Parcollet_disorder} , where a similar calculation was considered, this integral can be evaluated as

\enni \begin{align}
 & \int^{i \infty}_{-i\infty} ds \Gamma(\alpha + s) \Gamma(\beta + s) \Gamma(\gamma -s) \Gamma(\delta - s) \notag \\
& =  2\pi i \frac{\Gamma(\alpha + \gamma)\Gamma(\alpha+\delta)\Gamma(\beta+\gamma)\Gamma(\beta+\delta)}{\Gamma(\alpha+\beta+\gamma+\delta)},
\end{align}

\enni where $\text{Re}(\alpha, \beta, \gamma, \delta) > 0$ according to Eq.~6.412 of Ref.~\onlinecite{Gradshteyn}. After some 
straightforward algebra, the DC conductance reduces
to 

\enni \begin{align}
G(\omega \rightarrow 0, T, \mu) = & \left( \frac{e^{2}}{2h} \sqrt{NM} \right) \left[ \pi \sin \left( \frac{\pi}{2} + 2 \theta \right) \right] \sqrt{p} 
\label{Eq:anlt_DC}, 
\end{align}

\enni valid for $p<p_{c}=1/2$. As initially discussed in the context of overscreened Kondo impurities~\cite{Parcollet_Kondo}  and subsequently 
in SYK models~\cite{Altman2016, Sachdev2015}, the 
phase $\theta$ is related to $p$ and the total 
filling on dot and lead (end points) via a form of Luttinger's theorem . At particle-hole symmetry, $\theta=0$ for all $p$.

We follow a similar procedure to determine the
zero-temperature tunneling conductance from

\enni \begin{align}
 & \lim_{T \rightarrow 0} G = \frac{ e^{2}V^{2}\sqrt{NM}  \pi^{2}}{ h \omega} \times \notag \\ & \int^{0}_{-|\omega|} d \epsilon \left[ \rho_{c}(\epsilon) 
\rho_{\psi}(\epsilon + \omega) +  \rho_{\psi}(\epsilon) \rho_{c}(\epsilon + \omega)  \right], \omega \ge 0.
\label{Eq:zr_T_cndc}
\end{align}

\enni The spectral densities in the conformal regime $\hbar \omega \ll k_{B} T^{*}$ on the NFL side can be obtained by using Sterling's formula~\cite{Abramowitz} for the Gamma functions in Eqs.~\eqref{Eq:Scln_frm_c},~\eqref{Eq:Scln_frm_psi} or by using an analytical continuation from the ansatz in Ref.~\onlinecite{Sachdev2015}:

\enni \begin{align}
\rho_{c}(\epsilon, T=0) = & \frac{1}{\pi} \frac{\Lambda}{\sqrt{J \hbar |\epsilon|}} L(\epsilon),  \\
\rho_{\psi}(\epsilon, T=0) = & \frac{1}{\pi} \frac{ \sqrt{p} \sqrt{J \hbar |\epsilon|}}{V^{2} \Lambda} L(\epsilon), 
\end{align}

\enni where 

\enni \begin{align}
L(\epsilon) = 
\begin{cases}
\sin(\pi/4 + \theta), & \epsilon \ge 0 \\
\cos(\pi/4 + \theta), & \epsilon <0.
\end{cases}
\end{align}

\enni Substitution into the zero-temperature expression for the conductance and use of  Eq.~3.192 in Ref.~\onlinecite{Gradshteyn} gives 

\enni \begin{align}
G(\omega, T \rightarrow 0, \mu) 
= & \left( \frac{e^{2}}{2h} \sqrt{NM} \right) \left[ \pi \sin \left( \frac{\pi}{2} + 2 \theta \right) \right] \sqrt{p}, 
\end{align}

\enni valid for $p < p_{c}=1/2$. This is identical to the DC conductance in Eq.~\eqref{Eq:anlt_DC}. The dimensionless conductance $g_{0}(p)$ discussed in the main text follows from these expressions.

Note that these results correspond to the \emph{leading} 
contribution in the emergent conformal-invariant regime on the NFL side. We ignored corrections $\sim (T/T^{*}), (\hbar \omega/ k_{B}T^{*} )$ due to leading irrelevant terms which break this symmetry~\cite{Sachdev2015}.

In the FL regime at particle-hole symmetry, we determine the leading contribution to the conductance by substituting Eqs.~\eqref{Eq:FL_spect_c} and \eqref{Eq:FL_spect_psi} into Eq.~\eqref{Eq:zr_T_cndc}:

\enni \begin{align}
G(\omega, T \rightarrow 0, \mu =0 , p > 1/2) 
= & \left( \frac{e^{2}}{2h} \sqrt{NM} \right) \frac{2}{\sqrt{p}}. 
\end{align}

\enni The same expression holds in the $\omega \rightarrow 0$ limit as obtained by substituting the forms in Eqs.~\eqref{Eq:FL_spect_c},~\eqref{Eq:FL_spect_psi} into Eq.~\eqref{Eq:dc_lmt}.

\subsection{Corrections to the universal conductance}
\label{Sec:Appn_crrc}

In the main text, we discussed corrections to the conformal-invariant NFL and FL solutions which are linear and quadratic in temperature, respectively. Here we support these statements with numerical results. 

In Fig.~\ref{Fig:T_crrc}~(a), we plot deviations from the universal conductance $g_{0}(p < 1/2)$ (Eq.~\eqref{Eq:Scln_frm}) in the NFL regime, for $p=0.1$, versus temperature. The linear dependence is apparent. 
Corrections to the universal conductance $g_{0}(p > 1/2)$ in the FL regime, for $p=0.8$, versus temperature are shown in Fig.~\ref{Fig:T_crrc}~(b). We see that they scale quadratically with temperature.

\enni \begin{widetext}
\noindent \begin{figure*}[ht!]
\subfloat{\includegraphics[width=0.5\columnwidth]{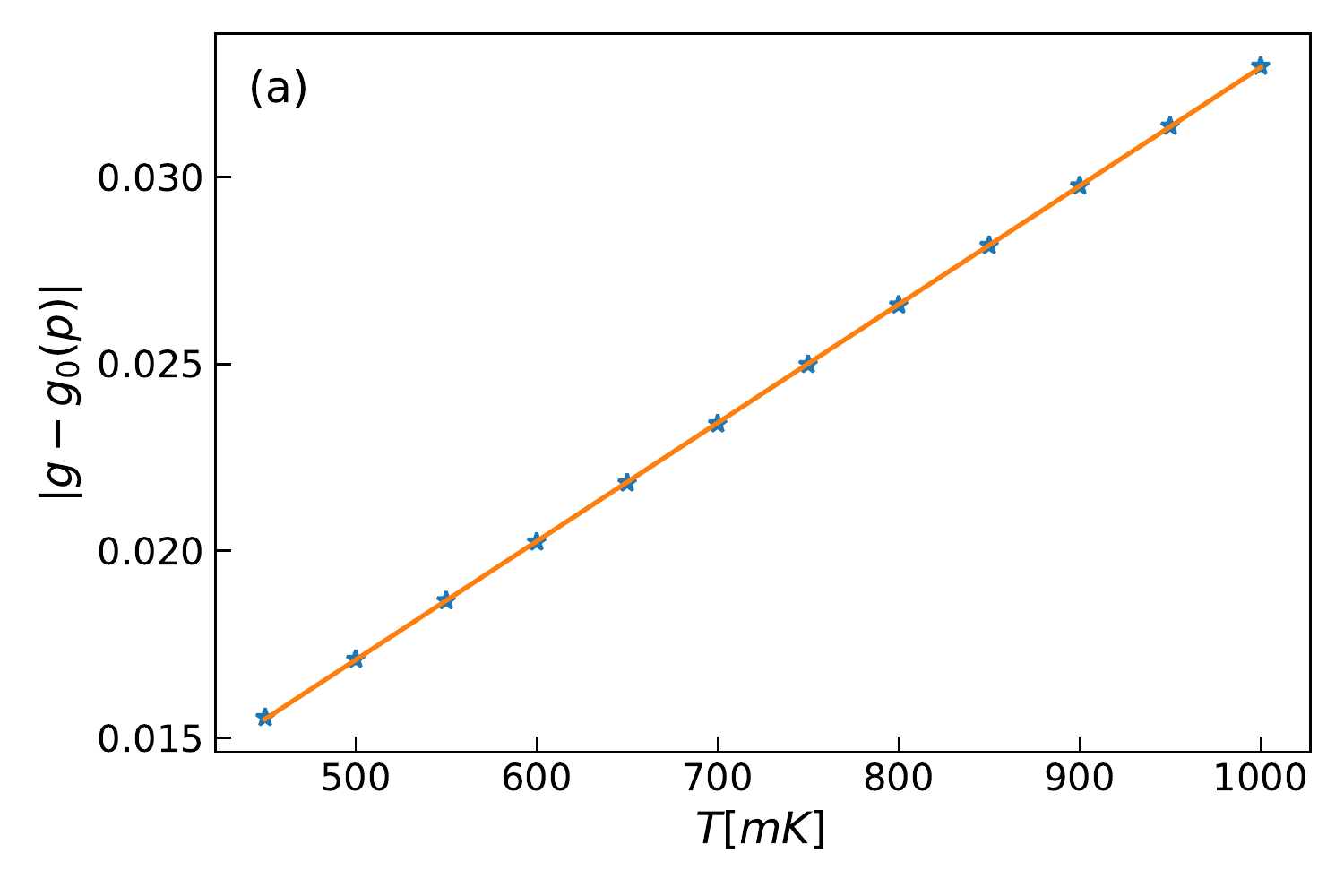}}
\subfloat{\includegraphics[width=0.5\columnwidth]{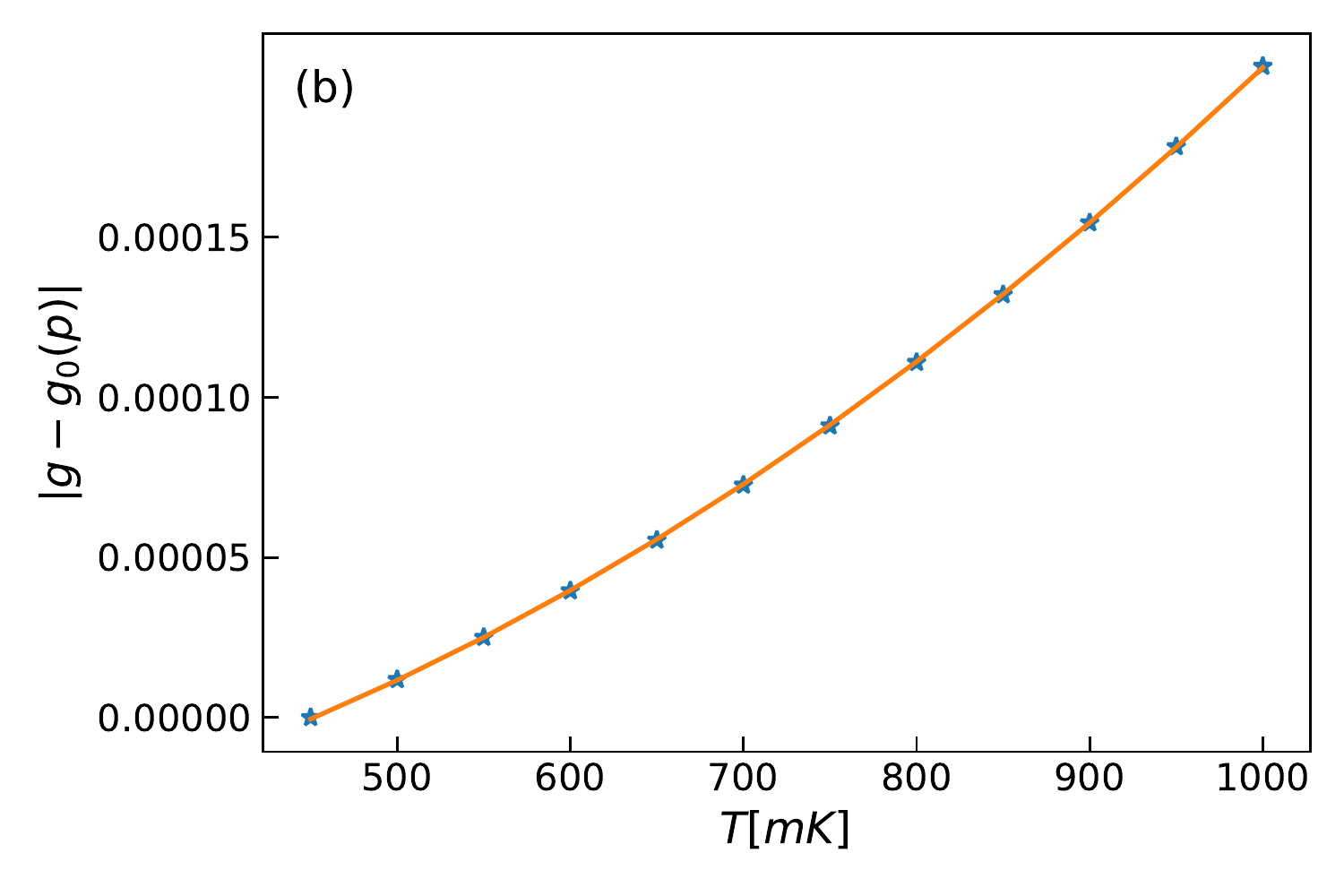}}
\caption{(a) Corrections to the universal \note{DC} conductance $g_{0}(p)$ in the NFL regime (Eq.~ \eqref{Eq:Scln_frm} in the main text, \note{for $\omega \rightarrow 0$}), for $p=0.1$, versus temperature. The linear dependence is apparent. (b) Same as (a) in the FL regime, for $p=0.8$, versus temperature. A quadratic dependence is obtained. 
}
\label{Fig:T_crrc}
\end{figure*}

\section{Large N, M saddle-point solutions}
\subsection{Saddle-point equations}

Following Ref.~\onlinecite{Balents2017}, we write down the path integral for our model, ignoring the extended leads, (Eq.~\eqref{Eq:Model}) and obtain an effective action after disorder averaging

\enni \begin{align}
\overline{Z} = \int \mathcal{D}\left[\psi_{L/R} ,\overline{\psi}_{L/R},c,\overline{c} \right]
e^{iS},
\end{align}

\enni where the Grassmann fields correspond to left and right lead end points and dot, respectively. 
The real-time action is defined on the Keldysh contour~\cite{Rammer} and can be written as a sum of
contributions from the left/right leads, dot, and coupling between them:

\enni \begin{align}
S = & S_{L} + S_{R} + S_{D} + S_{LD} + S_{RD}, \\   
S_{L/R} = & \sum_{s} \sum_{\alpha} \int dt  \left\{ \overline{\psi}_{\alpha s}(t) s \left[i \partial_{t} + \mu \right]\psi_{\alpha s}(t) \right\}
- \sum_{ss'} \int \int dt dt' 
\left\{ ss' \frac{it^2}{2M}\left( \sum_{\alpha} \overline{\psi}_{\alpha s}(t) \psi_{\alpha s'}(t') \right)\left(\sum_{\beta} \overline{\psi}_{\beta s'}(t') \psi_{\beta s}(t) \right) \right\} \\
S_{D} = & \sum_{s} \sum_{i} \int dt  \left\{ \overline{c}_{i s}(t) s \left[i \partial_{t} + \mu \right] c_{i s}(t) \right\} + \sum_{ss'} \int \int dt dt' 
\left\{ss'\frac{iJ^2}{4N^3}
\left(\sum_i\overline{c}_{is}(t) c_{is'}(t')\right)^2 \left(\sum_j \overline{c}_{js'}(t')  c_{js}(t) \right)^2  \right\} \\
S_{L/RD} = & - \sum_{ss'} \int \int dt dt' 
\left\{ss'\frac{iV^2}{\sqrt{NM}}
 \left( \sum_i \overline{c}_{is}(t) c_{is'}(t') \right) \left( \sum_{\alpha}   \overline{\psi}_{\alpha s'}(t') \psi_{\alpha s}(t) \right) \right\}.
\end{align}

\enni We suppressed the $L/R$ indices on the lead fields for clarity. The integrals run from $-\infty$ to $\infty$ and the index $s=\pm 1$ labels the forward and backward direction on the Keldysh contour~\cite{Balents2017}. We introduce the fields $G_{c,\psi}$ together with the Lagrange multipliers $\Sigma_{c, \psi}$:

$$
\int \mathcal{D}[G_{c},\Sigma_{c}] \exp\left(N\sum_{ss'}\int \int dt dt' \left\{\Sigma_{c, ss'}(t,t') \left[G_{c, s's}(t',t)-\frac{i}{N}\sum_i \overline{c}_{is}(t)  c_{is'}(t') \right] \right\} \right)=1
$$

$$
\int \mathcal{D}[G_{\psi},\Sigma_{\psi}] \exp\left(M\sum_{ss'}\int \int dt dt' \left\{\Sigma_{\psi, ss'}(t,t') \left[G_{\psi, s's}(t',t)-\frac{i}{M}\sum_i \overline{\psi}_{is}(t)  \psi_{is'}(t') \right] \right\} \right)=1.
$$

\enni The resulting action is
\begin{multline}
S_{L/R} = \sum_{ss'} \sum_{\alpha} \int\int  dt dt'  \left\{\overline{\psi}_{\alpha s}(t) \left[ \sigma^z_{ss'}\delta_{tt'}\left(i\partial_{t} + \mu \right) - \Sigma_{\psi, ss'}(t, t') \right]\psi_{\alpha s'}(t') \right\} \\
+ \sum_{ss'}\int \int dt dt' 
\bigg\{i ss'\frac{M t^2}{2} G_{\psi, s's}(t',t)G_{\psi, ss'}(t,t') -i M \Sigma_{\psi, ss'}(t, t') G_{\psi, s's}(t',t) \bigg\}
\end{multline}

\begin{multline}
S_{D} = \sum_{ss'} \sum_{i} \int\int  dt dt'  \left\{\overline{c}_{is}(t) \left[ \sigma^z_{ss'}\delta_{tt'}\left(i\partial_{t} + \mu \right) - \Sigma_{c, ss'}(t, t') \right]c_{i s'}(t') \right\} \\
+ \sum_{ss'} \int \int dt dt' 
\bigg\{i ss'\frac{NJ^2}{4}
 G^{2}_{c,s's}(t',t) G^{2}_{c, ss'}(t, t')  -i N \Sigma_{c, ss'}(t, t') G_{c, s's}(t',t) \bigg\}
\end{multline}

\begin{multline}
S_{L/RD} =   \sum_{ss'} \int \int dt dt' \left\{i ss' \sqrt{NM} V^2
  G_{c, s's}(t',t) G_{\psi, ss'}(t,t') \right\} \\
\end{multline}

\enni After integrating out the fermions, we find the saddle point of the action

\enni \begin{align}\label{spwrtG}
\frac{\delta S}{\delta G_{a, ss'}(t, t')} =  0 , \ \ \  
\frac{\delta S}{\delta \Sigma_{a, ss'}(t, t')} =  0, 
\end{align}

\enni where $a$ stands for $c$ and $\psi$ indices for dot and leads, respectively. We drop the dependence on two time indices and obtain the saddle-point equations that follow from equation \eqref{spwrtG}

\enni \begin{align}
\Sigma_{c, ss'}(t) = & ss'J^2
G^{2}_{c, ss'}(t) G_{c, s's}(-t) +  ss'\sqrt{p} V^2
  G_{\psi, L, ss'}(t) + ss'\sqrt{p}V^{2}
  G_{\psi, R, ss'}(t) \label{Eq:sddl_c}\\
  \Sigma_{\psi, L/R, ss'}(t) = & ss't^2    G_{\psi, L/R,ss'}(t)  + ss'\frac{V^{2}}{\sqrt{p}}
  G_{c,ss'}(t) \label{Eq:sddl_psi},
\end{align}
\end{widetext}

\enni where $p=M/N$. These are supplemented by 
the (matrix) Dyson equation for the frequency-dependent Green's functions which we obtain from \eqref{spwrtG}

\enni \begin{align}
G_{a, ss'}(\omega) = \left[ \sigma^z\left( \omega + \mu \right) - \Sigma \right]_{a, ss'}^{-1} \label{Eq:Dyson}.    
\end{align}

\enni The matrix equations are cast in a Keldysh basis for retarded, advanced, and Keldysh components via the standard transformation~\cite{Rammer}. In equilibrium,  a "fluctuation-dissipation" relation~\cite{Kamenev} is imposed.

\enni \begin{align}
G^{K}(\omega) = 2i \tanh \left( \frac{\beta \omega}{2} \right) \text{Im} G^{R}(\omega).    
\end{align}

Recall that we consider identical left and right leads. In this case, $G_{\psi, L}= G_{\psi, R}$ the saddle-point equations (Eq.~\eqref{Eq:sddl_c}, \eqref{Eq:sddl_psi}) 
are formally identical to those in Ref.~\onlinecite{Altman2016} after a simple re-scaling of $V$ and $p$ 

\enni \begin{align} \label{Eq:rscl}
V^{2} \rightarrow  \sqrt{2} V^{2},~~ 
p \rightarrow  2 p 
\end{align}

\enni Consequently, the phase transition occurs at $p_{c}=1/2$. 

\subsection{Numerical Solution} 

The saddle-point equations (\ref{Eq:sddl_c}-\ref{Eq:Dyson}) and are solved by direct numerical iteration with Green's functions defined on a discrete set of $2^{16}$ time points, with an ultraviolet cutoff of $10J$ in the frequency domain. Since the saddle point equations have a simpler form in time while the frequency representation is more natural for Dyson's equation, we used nfft \cite{nfft} library for Python to switch between the time and frequency representations of the Green's functions at each iteration. Using the nfft with non-equispaced frequencies allows for an effective sampling the of the spectral weights near zero and shorter computation times. The plots shown in the main text are determined for fixed $V = t = \frac{J}{2}$ unless stated otherwise.

\subsection{Spectral densities in equilibrium}
\label{Sec:Appn_spct}

\begin{figure}[ht!]
\includegraphics[width=1\columnwidth]{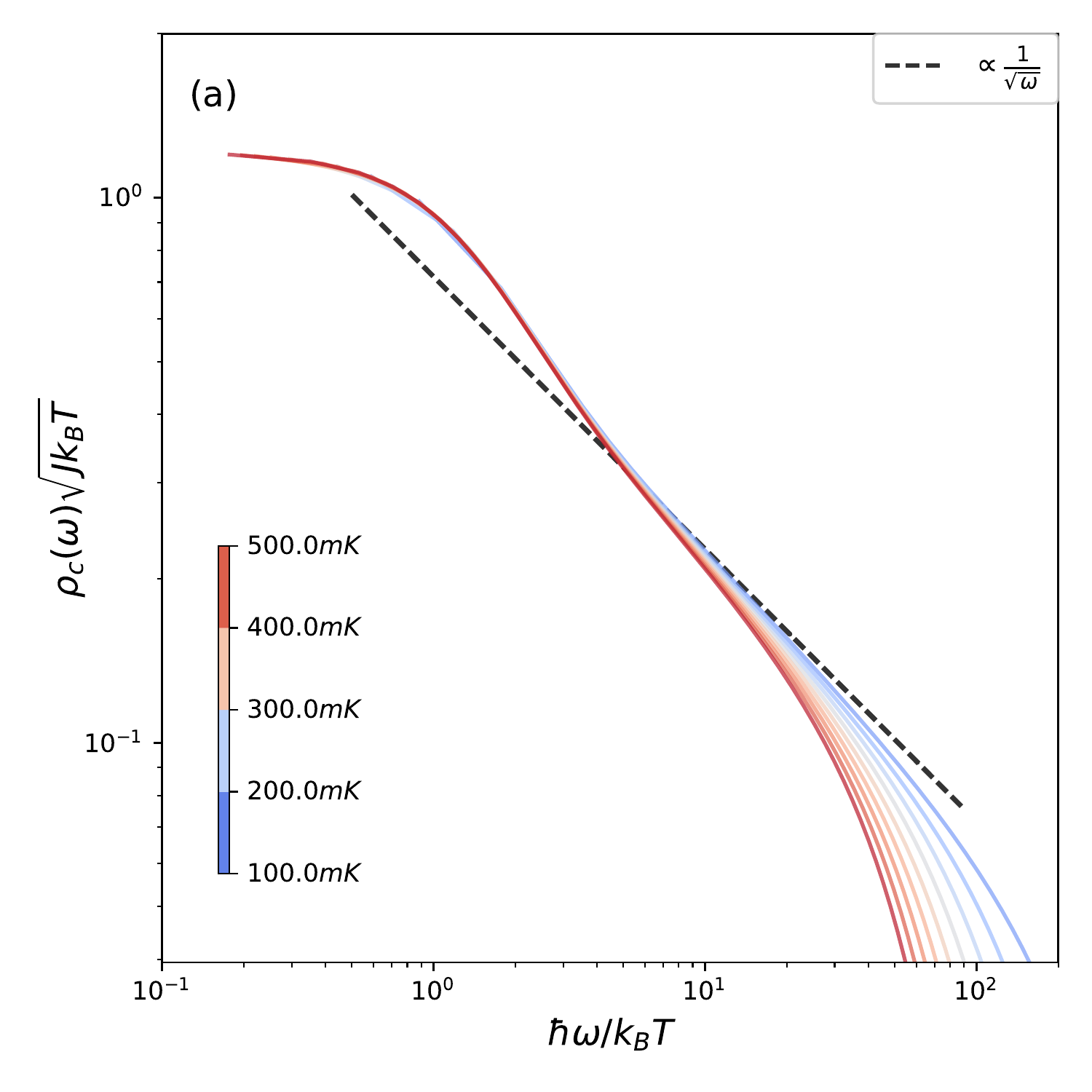}
\includegraphics[width=1\columnwidth]{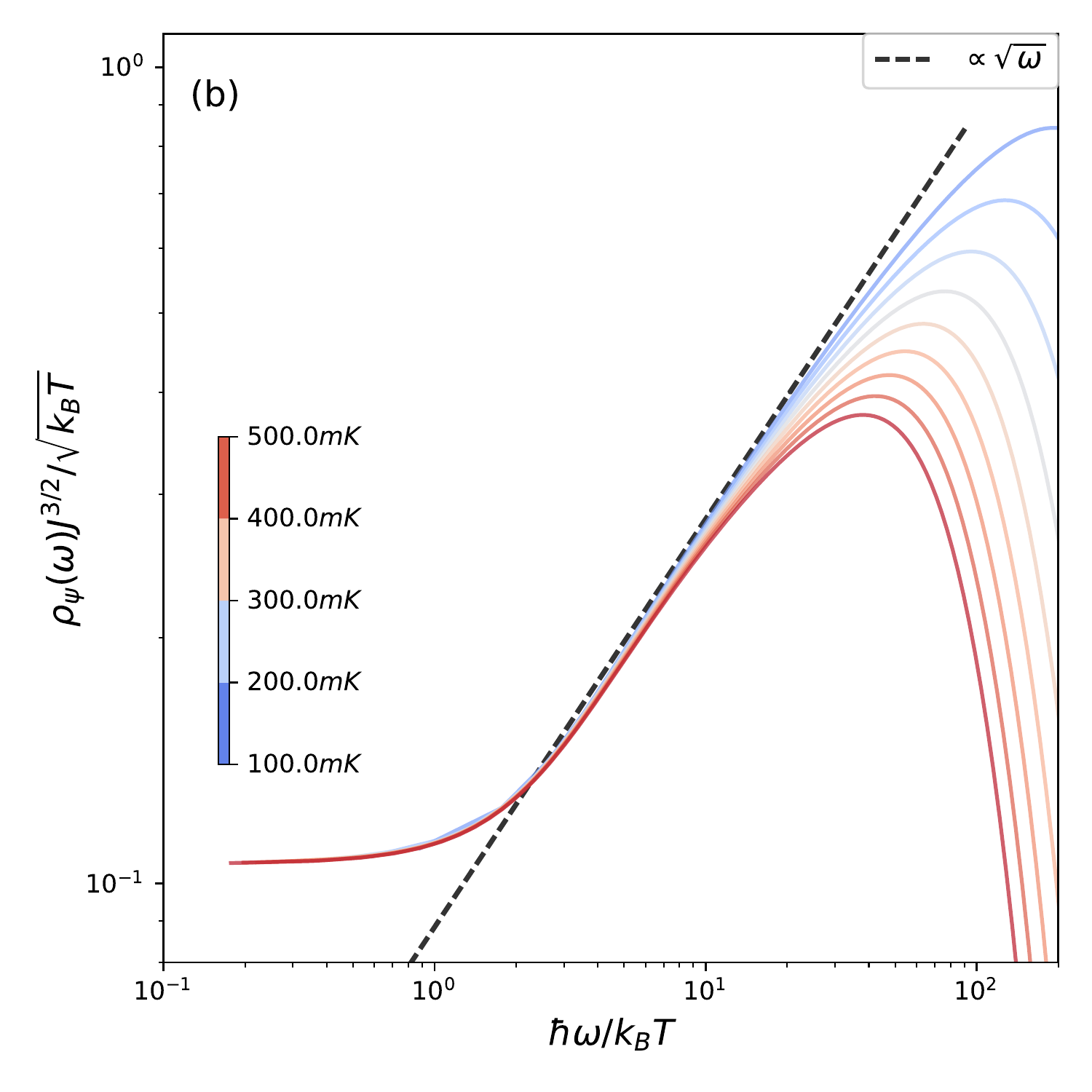}
\caption{(a) Spectral density $\rho_{c}$ for the dot electrons scaled by $\sqrt{k_{B} T}$, for several temperatures, at $p=0.1 < p_{c}$ and particle-hole symmetry, as a function of the dimensionless parameter $\hbar \omega / k_{B} T$.  The curves show scaling collapse below values of the abscissa of $O(10^{2})$. This regime corresponds to the leading behavior for conformal-invariant solutions in Eq.~\eqref{Eq:Scln_frm_c}, as indicated by the dashed line. For lower temperatures, the divergence is cut-off by a finite peak of width $\sim T$. (b) Same as (a)
for the lead end point spectral functions in Eq.~\eqref{Eq:Scln_frm_psi}.}    
\label{Fig:rho_scln}
\end{figure}

In equilibrium, the spectral densities in the conformal-invariant NFL regime were shown in Eqs.~\eqref{Eq:Scln_frm_c},~\eqref{Eq:Scln_frm_psi}. Our numerical solutions are consistent with these forms. In Fig.~\ref{Fig:rho_scln}~(a)  we plot the 
spectral densities for the dot electrons $\rho_{c}$
scaled by $\sqrt{J k_{B}T}$, for $p=0.1 < p_{c}$, at particle-hole symmetry, for several temperatures, versus the dimensionless parameter $\hbar \omega / k_{B} T$.
We see scaling collapse for values of the abscissa below roughly $10^{2}$. Above this cutoff, clear departures from scaling associated with the cross-over scale $k_{B}T^{*}$ are apparent. Immediately below it, the curves follow a $1/\sqrt{x}$ dependence corresponding to the high-frequency limit of Eq.~\ref{Eq:Scln_frm_c} in the conformal regime, as can be checked by using Sterling's formula (Eq. 6.3.17 in Ref.~\onlinecite{Abramowitz}). 
For $\hbar \omega / k_{B} T \ll 1$, the divergence in the high-frequency regime  is cut off by a 
 peak of width $\sim{T}$. 
 A similar scaling holds for the lead end point spectral densities $\rho_{\psi}$ as shown in Fig.~\ref{Fig:rho_scln}~(b).
In either cases, we also 
observe slight departures from the ideal scaling of the conformal-invariant solutions due to corrections $\sim T/T^{*}$. 

\noindent 

At particle-hole symmetry the leading spectral densities in the FL regime can be obtained from Ref.~\onlinecite{Altman2016} via the transformation defined in Eq.~\eqref{Eq:rscl}:

\enni \begin{align}
\rho_{c}= & \frac{1}{\pi} \frac{1}{\sqrt{2p-1}} \frac{t}{\sqrt{2} V^{2}} \label{Eq:FL_spect_c},\\
\rho_{\psi}= & \frac{1}{\pi} \sqrt{\frac{2p-1}{2p}} \frac{1}{t}.
\label{Eq:FL_spect_psi}
\end{align}


\section{Tunneling current for static bias beyond linear response}
\label{Sec:Appn_MW}

We determine the steady-state current by allowing the biases defined in Eq.~\ref{Eq:bias}  
to be arbitrarily large.  The disorder-averaged current \emph{out} 
of the left and right leads, respectively, are obtained from Eq.~\eqref{Eq:J_L}.
Recall that we consider couplings to the $L/R$ leads which are statistically identical with equal 
variance $V^{2}$. We determine the current for arbitrary applied bias $U$ and to all 
orders in the coupling constants $t, V, J$ by keeping contributions to leading order 
in $N, M$. 
The diagrammatic expansion is evaluated using a contour-ordered formalism, 
followed by an analytic continuation to real times. For an in-depth 
discussion of this we refer the reader to Refs.~\onlinecite{Langreth, Haug}.
More specifically, we allow for non-interacting, disorder-free leads at $t \rightarrow -\infty$ 
which are in equilibrium with large reservoirs at shifted chemical
potentials $\mu \pm eU/2$ for 
left and right leads, respectively. We subsequently turn on all couplings adiabatically. 
In practice, we ignore the initial state of the dot. This is a commonly-employed  approximation when calculating steady-state currents~\cite{Meir_NCA, Haug}. In addition, we assume that for  sufficiently long times, the leads reach equilibrium with the large reservoirs.   
Likewise, we ignore the time dependence of the current.

To calculate the steady-state current we require the 
$<$ Green's functions

\enni \begin{align}
G^{<}_{I}(t, t') = & i \braket{\psi^{\dagger}_{L\alpha}(t) c_{i}(t')} \\
G^{<}_{II}(t, t') = &  i \braket{c^{\dagger}_{i}(t) \psi_{L\alpha}(t') }.
\end{align} 

\enni Consider the related, contour-ordered Green's functions

\enni \begin{align}
G_{C, I}(\tau, \tau') = &  -i \overline{ \frac{V^{*}_{Li\alpha}}{(NM)^{1/4}} \braket{T_{C}c_{\tau}(t) \psi^{\dagger}_{L\alpha}(\tau') }} \label{Eq:GC_II} \\
G_{C,II}(\tau, \tau') = & -i \overline{ \frac{V_{Li\alpha}}{(NM)^{1/4}} \braket{T_{C}\psi_{L\alpha}(\tau) c^{\dagger}_{i}(\tau')} }, \label{Eq:GC_I}
\end{align}

\begin{figure}[h!]
\includegraphics[width=1\columnwidth]{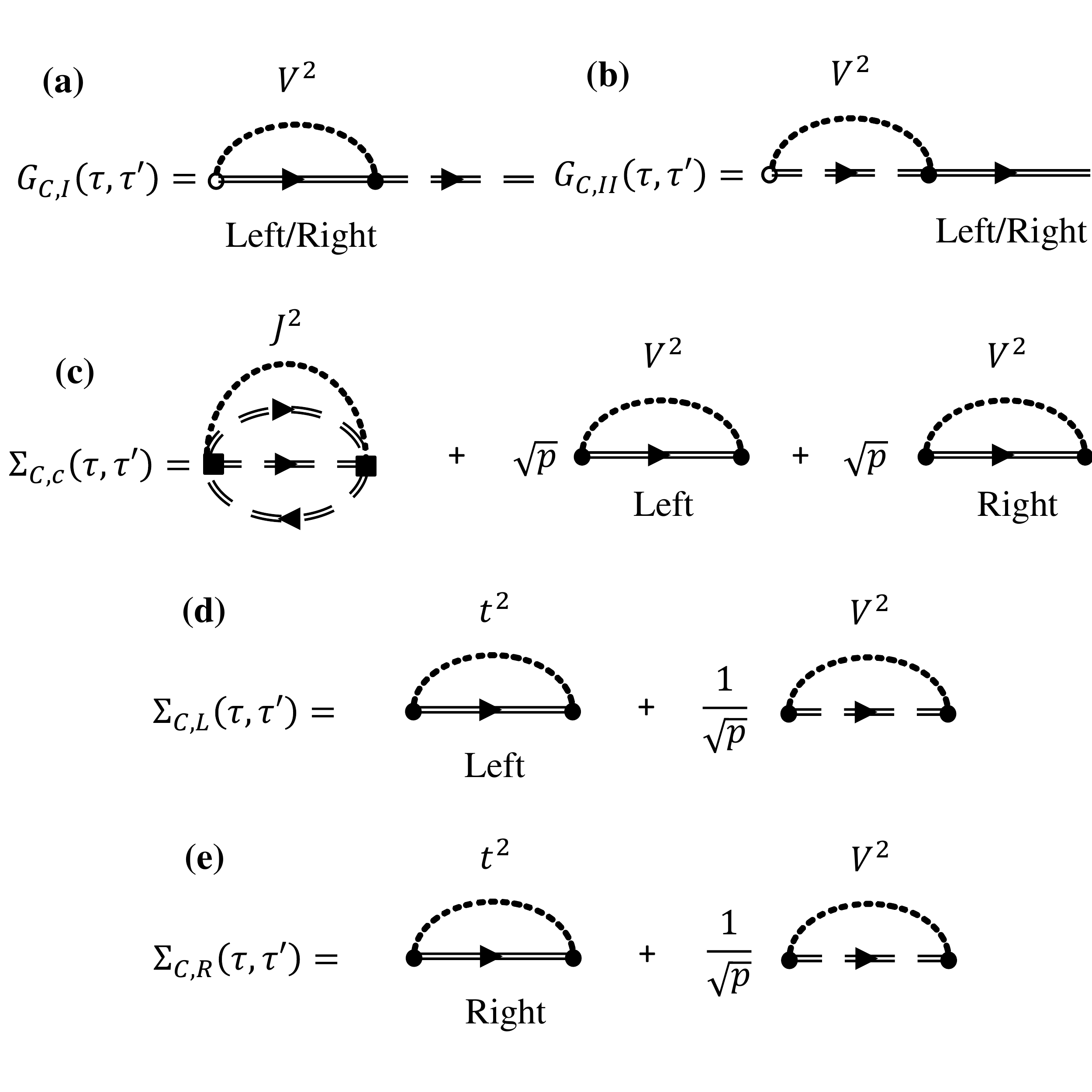}
\caption{Contour-ordered diagrammatic expansions for the disorder-averaged tunneling current to leading 
order in $N, M$ and to all orders in $t, V, J$.  
The dashed, double line represents the fully-dressed dot propagator, while the 
continuous, double lines stand for the fully-dressed left/right lead propagators.    
Internal vertices are indicated by filled symbols, while the external vertex for the current is denoted 
an empty circle. Dotted lines connecting vertices represent disorder averages. 
(a) The contour-ordered Green's function $G_{C,I}$ in Eq.~\ref{Eq:GC_I} 
for either left or right lead. (b) 
Same as (a) for $G_{C,II}$ in Eq.~\ref{Eq:GC_II}. (c) Diagrams which determine 
the self-energy for the dot to leading order in $N, M$. (d) and (e) Same as (c) 
for the left and right leads. Note that the saddle-point equations in Eqs.~\ref{Eq:sddl_c}, \ref{Eq:sddl_psi} can also be obtained via analytical continuation~\cite{Haug}.  
The effect of the bias is included  
via the unperturbed lead propagators.  
}
\label{Fig:MW_expansion}
\end{figure}

\enni defined on the contour extending from $-\infty$ and back, passing through $\tau$ and $\tau'$ once~\cite{Haug}. 
$T_{C}$ stands for ordering along the contour. These functions 
coincide with the real-time $<$ propagators when $\tau$ and $\tau'$ are on the upper- and lower-branch
of the Keldysh contour~\cite{Rammer}, respectively.
As discussed in a number of references~\cite{Langreth, Haug, Rammer}, the contour-ordered Green's functions 
allow for a straightforward application of Wick's theorem. 
We note that this procedure entails no difficulties with respect to disorder averaging. 
One can check that the diagrammatic expansion in Fig.~\ref{Fig:MW_expansion}
, followed by analytical continuation~\cite{Haug} to the real axis reproduces
the saddle-point equations in equilibrium (Eqs.~\eqref{Eq:sddl_c}, \eqref{Eq:sddl_psi}), within minus signs for $\Sigma^{<,>}_{c, \psi}$.
The difference between the two approaches is due to a convention in defining the self-energies which appear in the saddle-point derivation~\cite{Balents2017}, and is otherwise innocuous. 
In order to keep the sign convention common in the literature on transport through quantum dots~\cite{Mahan, Rammer, Haug}, in this section we use the more typical convention for the signs of the $<,>$ self-energies.   
 
Once the functions $G^{C}_{I, II}(\tau, \tau')$ are known, 
we proceed to determine their real-time counterparts via the same analytical continuation.

\noindent 

Using the diagrammatic expansion in Fig.~\ref{Fig:MW_expansion}, we determine 

\enni \begin{align}
& G_{C,I}(\tau, \tau') = \notag \\ 
& \frac{V^{2}}{\sqrt{NM}} \delta_{\alpha \beta} \delta_{ij} \int_{C} d\tau_{1} G_{C, c, ij}(\tau, \tau_{1}) G_{C, \psi L/R, \alpha \beta}(\tau_{1}, \tau') \\
& G_{C,II}(\tau, \tau') = \notag \\ 
& \frac{V^{2}}{\sqrt{NM}} \delta_{\alpha \beta} \delta_{ij} \int_{C} d\tau_{1} G_{C, \psi L/R, \alpha \beta}(\tau, \tau_{1}) G_{C, c, ij}(\tau_{1}, \tau'),
\end{align}

\enni where factors of $1/\sqrt{NM}$ are due to the definition of the hybridization $V$ (Eq.~\eqref{Eq:Model}). We analytically continue these expressions onto the real axis according to the following rule~\cite{Haug, Langreth}:

\enni \begin{align}
C = \int_{C} d\tau_{1} AB \rightarrow C^{<} = \int_{t} dt_{1} \left[ A^{R}B^{<} + A^{<} B^{A} \right] ,
\end{align}

\enni where the $R,A$ indices stand for the retarded and advanced components. The corresponding expression for the current from the left or right leads into the dot is

\enni \begin{align}
& \braket{I_{L/R}}  = \frac{e}{\hbar} V^{2} \sqrt{NM} \int^{\infty}_{-\infty} dt_{1} \notag \\
& \bigg[ G^{R}_{c}(t, t_{1}) G^{<}_{\psi L/R}(t_{1}, t^{+}) + G^{<}_{c}(t, t_{1}) G^{A}_{\psi L/R}(t_{1}, t^{+})  \notag \\
& - G^{R}_{\psi L/R} (t, t_{1}) G^{<}_{c}(t_{1} , t^{+}) 
 -G^{<}_{\psi L/R} (t, t_{1}) G^{A}_{c}(t_{1}, t^{+}) \bigg].
\end{align}

\enni Note that this expression can also be obtained by using diagram rules directly 
in the matrix formulation~\cite{Rammer}. The factor of $\sqrt{NM}$ is due to the summation over the $\alpha, i$ indices (Eq.~\eqref{Eq:J_L}). As we are considering the steady-state current $t \gg t_{0} = -\infty$,
for a static bias, and since
the interactions are turned on adiabatically, we can consider only the difference between the time arguments. We obtain the steady-state 
current 

\enni \begin{align}
\braket{I_{L/R}} = & \frac{e}{\hbar} V^{2} \sqrt{NM} \int \frac{d\omega}{2\pi} 
\bigg\{ G^{<}_{\psi L/R} (\omega) \left[ G^{R}_{c}(\omega) - G^{A}_{c}(\omega) \right] \notag \\
& -  G^{<}_{c}(\omega)  \left[G^{R}_{\psi L/R} (\omega) -G^{A}_{\psi L/R} (\omega) \right]  \bigg\}.
\label{Eq:MW_current}
\end{align} 

\enni This is the central result of this section. 

Note that our expression for the tunneling current   Eq.~\ref{Eq:MW_current} is analogous to cases involving an interacting dot coupled to \emph{non-interacting leads}~\cite{Meir, Hershfield, Haug}.
The important difference is due to the disordered coupling between leads and dot $V_{i, \alpha}$ which implies
that both dot and lead Green's functions must be determined self-consistently. 
Indeed, the single-particle propagators which enter our expression for the current are determined 
via the same set of saddle-point equations encountered in equilibrium (Eqs.~\eqref{Eq:sddl_c}, \eqref{Eq:sddl_psi} , see Fig.~\ref{Fig:MW_expansion}), 
with the additional 
contribution due to the biases. 

%
%
%
%

Before proceeding to a discussion of the numerical implementation, we note a number of important 
points. First, the current vanishes in equilibrium, as expected. This can be seen by 
using the equilibrium forms for the $<$ Green's functions (Eq. 2.160 of Ref.~\onlinecite{Mahan}), which are given by

\enni \begin{align}
G^{<} = & 2\pi i f(\omega) \rho(\omega) \label{Eq:Eqlb_frms}.
\end{align}

\enni The components which enter this expression are the Fermi-Dirac function $f(\omega)$
 and the spectral density $\rho(\omega)$.
 When the bias is set to zero, the left and right leads have the same 
chemical potential. Substitution of the equilibrium forms into the expression for the 
current ensures that the latter vanishes, as expected. 

Second, we consider the condition for charge conservation on the dot in the steady state regime:

\enni \begin{align}
I_{L} + I_{R} = 0.
\end{align}

\enni Using the well-known analytical property~\cite{Rammer, Mahan} $G^{>} - G^{<} = G^{R} - G^{A}$, the conservation of charge is equivalent to 

\begin{widetext}
\enni \begin{align}
\int d\omega
& \left\{  G^{>}_{c}(\omega) \left[ V^{2} G^{<}_{\psi L} (\omega) + V^{2} G^{<}_{\psi R} (\omega) \right] 
- G^{<}_{c}(\omega) \left[ V^{2} G^{>}_{\psi L} (\omega) + V^{2} G^{>}_{\psi R} (\omega) \right]  \right\} = 0.
\end{align}
\end{widetext}

\enni From the diagrams shown in Fig.~\ref{Fig:MW_expansion} we obtain for the 
dot self-energy

\enni \begin{align}
\Sigma^{<,>}_{c} = \sqrt{p} V^{2} \left[ G^{<,>}_{L} + G^{<,>}_{R}\right] + \Sigma^{<,>}_{int},
\end{align}

\enni where $\Sigma_{int} \sim J^{2}$ is the proper self-energy of the dot due to interactions. 
Following Eq.~12.28 in Ref.~\onlinecite{Haug}, 
we solve for $V^{2} \left[G^{<,>}_{L} +  G^{<,>}_{R}\right] $ in terms of the self-energies, substitute into the charge conservation condition, and obtain

\enni \begin{align}
& \int d\omega \left\{  G^{>}_{c} \left[ \Sigma^{<}_{c} - \Sigma^{<}_{int} \right] 
- G^{<}_{c} \left[ \Sigma^{>}_{c} - \Sigma^{>}_{int} \right]  \right\} = 0.
\end{align}

\enni In our case,  the Keldysh equation for $G^{<}_{c}$ reads~\cite{Haug}

\enni \begin{align}
G^{<}_{c} = G^{R} \Sigma^{<}_{c} G^{A}_{c}.
\end{align}

\enni This differs from the full expression (Eq.~2.159 in Ref.~\onlinecite{Mahan}) 
by terms proportional to $G^{<, (0)}_{c}$.  
These functions represent the initial correlations at $t \rightarrow -\infty$ which are ignored 
in our calculations (see for example comment 33 in Ref.~\onlinecite{Meir_NCA}). This is a standard approximation in the context of transport through 
interacting quantum dots~\cite{Haug}. The simplified Keldysh equation implies that

\enni \begin{align}
G^{>}_{c} \Sigma^{<}_{c} - G^{<}_{c} \Sigma^{>}_{c} =0.
\end{align}

\enni Therefore, conservation of charge reduces to the condition involving the 
self-energy of the dot due to interactions

\enni \begin{align}
& \int d\omega \left\{  G^{<}_{c} \Sigma^{>}_{int} 
- G^{>}_{c} \Sigma^{<}_{int}  \right\} = 0,
\label{Eq:Chrg_cnsr}
\end{align}

\enni which is also well-known in the context of transport through interacting Anderson impurity models~\cite{Hershfield}. 

We show that our saddle-point approximation, as given by the self-consistent diagrams in Fig.~\ref{Fig:MW_expansion}, 
ensures that this condition is satisfied \emph{for each frequency}. At saddle-point, the self-energy due to interactions is 

\enni \begin{align}
\Sigma^{<,>}_{J}(t) = J^{2} \left[ G^{<,>}_{c}(t) \right]^{2} G^{>,<}_{c}(-t). 
\end{align}
 
\enni Fourier transforming and substituting into the condition for charge conservation (Eq.~\eqref{Eq:Chrg_cnsr}) 
we obtain 

\begin{widetext}
 \enni \begin{align}
\int d\omega  \int d\omega_{1,2,3} \delta(\omega_{1} + \omega_{2} - \omega_{3} - \omega ) 
\left[ G^{<}_{c}(\omega) G^{>}_{c}(\omega_{1}) G^{>}_{c}(\omega_{2}) G^{<}_{c}(\omega_{3}) -
 G^{>}_{c}(\omega) G^{<}_{c}(\omega_{1}) G^{<}_{c}(\omega_{2}) G^{>}_{c}(\omega_{3})\right]=0.
\end{align}
\end{widetext}

\enni We can re-label the indices in the second term as $
(\omega_{1} , \omega_{2}) \leftrightarrow (\omega, \omega_{3})$. The even $\delta$ function of the same term 
is invariant under the transformation. 
Thus, our approximation satisfies the conserving condition for each frequency $\omega$.  

This important point allows us to determine an effective distribution function $F(\omega)$ for the 
dot out of equilibrium. To do so, we re-write the charge conservation as a sum of left and right 
currents (Eq~\eqref{Eq:MW_current})

\begin{widetext}
\enni \begin{align}
\braket{I_{L} + I_{R}} =  & \frac{e}{\hbar} V^{2}_{L} \sqrt{NM} \int \frac{d\omega}{2\pi} 
\bigg\{ \left[ G^{R}_{c}(\omega) - G^{A}_{c}(\omega) \right]  \left[ G^{<}_{\psi L} (\omega) + G^{<}_{\psi R} (\omega) \right]  
- G^{<}_{c}(\omega) \left[ G^{R}_{\psi L} (\omega) -G^{A}_{\psi L} (\omega) + G^{R}_{\psi R} (\omega) -G^{A}_{\psi R} (\omega)\right] 
\bigg\}.
\end{align}
\end{widetext}

\enni Since the integrand vanishes for all frequencies, we can readily solve for 

\enni \begin{align}
G^{<}_{c}(\omega) = & \frac{G^{<}_{\psi L} (\omega) + G^{<}_{\psi R} (\omega)}{G^{R}_{\psi L} (\omega) -G^{A}_{\psi L} (\omega) + G^{R}_{\psi R} (\omega) -G^{A}_{\psi R} (\omega)} \times \notag \\
& \left[ G^{R}_{c}(\omega) - G^{A}_{c}(\omega) \right]. 
\end{align}

We assume that the leads are maintained in thermal equilibrium at shifted chemical 
potentials throughout the temporal evolution, implying the equilibrium forms (Eq.~\eqref{Eq:Eqlb_frms}) 
for either left/right leads with Fermi-Dirac distributions  

\enni \begin{align}
f_{L/R}(\omega) = & \frac{1}{e^{\beta(\hbar \omega - \mu \pm eU/2)}+1}, \label{Eq:L_eqlb}
\end{align}

\enni where $U$ is the applied bias. We obtain the distribution function for the dot out of equilibrium according to 

\enni \begin{align}
G^{<}_{c}(\omega) = & 2 \pi i F_{c}(\omega) \rho_{c}(\omega) \label{Eq:Dot_form} \\
F_{c}(\omega) = & \frac{ f_{L}(\omega) \rho_{L}(\omega) +  f_{R}(\omega) \rho_{R}(\omega)}{\rho_{L}(\omega) + \rho_{R}(\omega)}. 
\label{Eq:dstr_fnct}
\end{align}

To summarize, we employ the following procedure:\\

1) The lead end points are kept in thermal equilibrium with large reservoirs at shifted 
chemical potentials according to the distribution in  Eq.~\eqref{Eq:L_eqlb}. 
Likewise, we include the bias terms $\sim \pm eU/2$ in the Hamiltonian for left and right lead respectively, 
in accordance with the standard procedure for systems under an applied constant field
(Eq. 10-7 in Ref.~\onlinecite{Baym_book}). \\

2) The saddle-point equations (Eqs.~\eqref{Eq:sddl_c}, \eqref{Eq:sddl_psi}) in the presence of the biases are solved 
numerically by imposing the form in Eqs.~\eqref{Eq:Dot_form},~\eqref{Eq:dstr_fnct}. 
This step determines the local spectral densities for the lead end points $\rho_{L,R}$, and the spectral 
density for the dot $\rho_{c}$ together with the distribution function $F_{c}$. \\

3) The tunneling current is determined according to Eq.~\eqref{Eq:MW_current} via the 
functions $G^{<}_{L/R/c}$ and the spectral functions.

\section{Weak-tunneling approximation}
\label{Sec:Appn_wk}

In order to determine the disorder-averaged current in the weak tunneling approximation, we employ the standard tunneling conductance formula\cite{Mahan}. The latter gives the current as a convolution of the spectral densities of the lead and the dot calculated with their mutual coupling set to zero. After Gaussian averaging over the couplings $V_{ij}$ we obtain the following formula for our serup
\begin{multline}\label{eq:weaktunneling}
\langle I_{WT} \rangle =  4\pi^2\frac{e}{h}V^2\sqrt{NM} \\ \times \int_{-\infty}^\infty \rho_\psi\left(\epsilon+ \frac{eU}{\hbar} \right)\rho_c(\epsilon)\left[f( \epsilon)-f\left(\epsilon+\frac{eU}{\hbar} \right)\right]d\epsilon.
\end{multline}
We assumed that both leads and dot are kept in thermal equilibrium at the same temperature with separate, large reservoirs. However, we assume that the chemical potentials for the leads and dot are shifted due to a finite bias. 
Furthermore, we assume that the current from left lead to dot is equal to the current from dot to right lead, as in the previous sections. 

The spectral functions are obtained from the retarded Green's function of the $\text{SYK}_4$ model.\cite{Sachdev2015} At particle-hole symmetry and non-zero temperature the dot Green's function  is given by
$$
G^R = \frac{-iC}{\sqrt{2\pi T}}\frac{\Gamma(1/4-i\beta \hbar \omega/2\pi)}{\Gamma(3/4-i\beta \hbar \omega/2\pi)},
$$
which gives the spectral density $\rho_c=-\frac{1}{\pi}\text{Im}G^R$, 
$$
\rho_c \propto \frac{1}{\sqrt{ T}} \lvert \Gamma(1/4+i\beta \hbar \omega/2\pi) \rvert^2 \cosh{\left(\frac{\beta \hbar \omega}{2}\right)}
$$
The Green's function for the lead can be obtained by setting $V=0$ in saddle point equations given in Appendix B1 and solving for the lead Green's function. One obtains
$$
\rho_\psi=\frac{1}{\pi t}\text{Re}\sqrt{1-\left(\frac{\hbar \omega}{2t}\right)^2}.
$$

Substituting these expressions into equation \eqref{eq:weaktunneling} we find
\begin{multline*}
\langle I_{WT} \rangle \simeq  e\frac{V^2}{t}\sqrt{NM}\frac{1}{\sqrt{ T}}\\ \times \int_{-\infty}^\infty  \lvert \Gamma(1/4+i\beta \hbar \epsilon/2\pi) \rvert^2 \cosh{\left(\frac{\beta \hbar \epsilon}{2}\right)} \times \notag \\
\left[f( \epsilon)-f\left(\epsilon+\frac{eU}{\hbar} \right)\right]d\epsilon
\end{multline*} 
where we assumed that the lead spectral density $\rho_\psi \approx \frac{1}{\pi t} $ is flat. This will be valid when the range of integration $|\epsilon| \ll t$ which we expect to be true at reasonable bias voltages. 
We estimate this integral in two limits:
\\
\paragraph{$eU \ll k_BT$:}
\noindent In this case Fermi factors reduce effectively to a derivative and we have 
$$
\lim_{e U \rightarrow0}{f( \epsilon)-f\left(\epsilon+\frac{eU}{\hbar}\right)\over \frac{e U}{\hbar}} = \frac{\hbar \beta }{4\cosh^2{(\beta \hbar \epsilon/2)}}.
$$
The integral above becomes
\begin{multline*}
\langle I_{WT} \rangle \simeq  e\frac{V^2}{t}\sqrt{NM}\frac{1}{\sqrt{ T}}\\ \times e U\int_{-\infty}^\infty\frac{\lvert \Gamma(1/4+i\beta \hbar \epsilon/2\pi) \rvert^2}{4\cosh{(\beta \hbar \epsilon/2)}} d(\beta \hbar \epsilon)
\end{multline*}
Substitution $y=\beta \hbar \epsilon$ reduces the integral to a dimensionless constant. From this expression we can easily extract the dependence on the bias voltage and temperature
\begin{equation}
\langle I_{WT} \rangle   \propto \frac{eU}{\sqrt{ T}}, \hspace{12pt} (eU \ll k_BT)
\end{equation}
\paragraph{$eU \gg k_BT$:}
In this case Fermi factors introduce limits to the integral, as they are effectively step functions. 
We have
\begin{multline*}
\langle I_{WT} \rangle \simeq  e\frac{V^2}{t}\sqrt{NM}\frac{1}{\sqrt{ T}}\\ \times \frac{1}{\beta}\int_{-\beta e U}^0  \lvert \Gamma(1/4+i\beta \epsilon/2\pi) \rvert^2 \cosh{\left(\frac{\beta \epsilon}{2}\right)} d(\beta \epsilon).
\end{multline*}
For $\beta \epsilon \gg 1 $ the integrand can be approximated as 
$$
\lvert \Gamma(1/4+i\beta \epsilon/2\pi) \rvert^2 \cosh{\left(\frac{\beta \epsilon}{2}\right)} \simeq \frac{1}{\sqrt{\lvert \beta\epsilon \rvert}}.
$$ 
The integral can be estimated as
$$
\langle I_{WT} \rangle \propto  e\frac{V^2}{t}\sqrt{NM} \frac{1}{\sqrt{\beta T}}\sqrt{ e U},
$$ 
from which we can extract the dependence of the average tunneling current on the external parameters
\begin{equation}
\langle I_{WT} \rangle \propto  \sqrt{ e U}, \hspace{12pt}(eU \gg k_BT).
\end{equation}
It is important to recall that we assumed $\epsilon \ll t$ above, which implies that the results will be valid only when the temperature and the bias are much smaller than $t$. 

To summarize we found that the weak tunneling current $I_{WT}$ is given by
\begin{equation}\label{eq:lweaktunnelingsummary}\hspace{12pt}
\langle I_{WT} \rangle \propto \begin{cases} 
     eU/\sqrt{ T},  &  (eU \ll k_BT),\\
        \sqrt{ e U}, &(eU \gg k_BT).
   \end{cases}
\end{equation} 


\section{Saddle-point equations in the presence of 
explicit coupling to extended leads}
\label{Sec:Appn_lead}

The discussion in the main text involved an \emph{effective local model} (Eq.~\eqref{Eq:Model}) for the junction between lead end points and graphene dot. We assumed that the lead \emph{end points} are  dominated by local disorder scattering. Consequently, they were described by a local SYK$_{2}$ Hamiltonian. As such, we effectively ignored a coupling to the bulk of the leads. This approximation is expected to be valid provided that the phase diagram in Fig.~\ref{Fig:Phase_diagram} is essentially unchanged when a coupling to the bulk of the leads is included in the effective model for the junction. We find that this is indeed the case for quasi-one-dimensional ballistic leads. 

We model the extended leads as a set of $M$ independent, semi-infinite non-interacting chains, labeled by an index $\alpha$. Each of the chains includes simple nearest-neighbor hopping and is coupled to  
\emph{a single end point state $\psi_{L/R, \alpha}$}, as indicated by Hamiltonian Eq.\ (\ref{hdot-lead}). In the absence of interactions, the coupling to the leads can included in the effective local model for the junction to all orders by a redefinition of the bare lead end point propagator 

\enni \begin{align}
G^{0}_{\psi, L/R}(i\omega_{n}) \rightarrow 
\tilde{G}^{0}_{\psi, L/R}(i\omega_{n}) = {1\over i\omega_{n} + \mu - \Sigma_{EL/R}(i \omega_{n})},
\end{align}

\enni while the saddle-point equations (Eqs.~\eqref{Eq:sddl_c},~\eqref{Eq:sddl_psi}) preserve their form. 
This can be seen via expanding the diagrams in Fig.~\ref{Fig:MW_expansion}~(b) and (c) and 
inserting all corrections due to $H_{EL/R}$ in all of the bare lead propagators (full lines). 

The self-energies due to the additional coupling to the extended leads are given by 

\enni \begin{align}
\Sigma_{EL/R, \alpha}(i \omega_{n}) = & \left(t_{-1/1, \alpha} \right)^{2} \int d \epsilon \frac{\rho_{loc, L/R,\alpha}(\epsilon)}{i \omega_{n} - \epsilon}. 
\label{Eq:Slf_enrg}
\end{align} 

\enni These depend on the local density of states at site $-1,1$, $
\rho_{loc, L/R, \alpha}(\epsilon) = -(1 / \pi) \text{Im}G^{R}_{\tilde{i}= -\mp 1}(\epsilon).$

Our goal is to account for the effect of the bulk of the leads in an effective model for the junction. In a manner analogous to treatments of Anderson impurity models coupled to a bath of conduction electrons~\cite{Hewson}, we approximate $\rho_{loc}$ by 
 a constant density of states near the end of a semi-infinite chain~\cite{Giamarchi}. Furthermore, we assume 
 that the chains are identical and ignore the $\alpha$ index. The local density of states is then given by

\enni \begin{align}
\rho_{loc, L/R ,\alpha}(\epsilon) = \rho_{E}, \ \ |\epsilon| \ll D
\end{align}

\enni where $D \gg V, J$ is a cutoff of the order of the 
bandwidth of the extended leads. 

For simplicity we re-label $t_{-1/1}=t_{E}$. 
By substituting $\rho_{E}$ in Eq.~\ref{Eq:Slf_enrg}, and continuing to real frequencies we obtain the retarded self-energy due to coupling to the extended leads

\enni \begin{align}
\text{Re} \Sigma^{R}_{E}(\omega)= & \rho_{E} t^{2}_{E} \ln \left| \frac{\omega+D}{\omega-D} \right| \\
\text{Im} \Sigma^{R}_{E}(\omega) = & - \pi \rho_{E} t^{2}_{E}.
\end{align}

In equilibrium, the Keldysh component is given by 
~\cite{Rammer}

\enni \begin{align}
\Sigma^{K}_{E}(\omega) = 2i \tanh\left(\frac{\beta \omega}{2} \right) \text{Im}\Sigma^{R}_{E}(\omega).
\end{align}

\enni These components are added to the self-energies in the matrix Dyson equation (Eq.~\ref{Eq:Dyson}) and solved together with the saddle-point equations  numerically.



%
%

\end{document}